%% file: Euclid-Margalef2025.tex
\documentclass[longauth]{aaEC}

\usepackage{graphicx}
\usepackage{natbib}
\usepackage{scalerel}
\usepackage{lipsum}  
\usepackage[switch]{lineno}

\usepackage[table]{xcolor}

\renewcommand{\textbf}{\normalfont}

\usepackage{txfonts}
\usepackage[pdfencoding=auto,psdextra]{hyperref}
\hypersetup{
    colorlinks=true,
    linkcolor=blue,
    filecolor=magenta,      
    urlcolor=blue,
    citecolor=blue
}
\makeatletter
\renewcommand*\aa@pageof{, page \thepage{} of \pageref*{LastPage}}
\makeatother

\usepackage[utf8]{inputenc}

\usepackage{euclid}

\begin{document}
%
% Put the title and authors of your (Standard Project) paper here
%

\title{Euclid Quick Data Release (Q1)} \subtitle{First \Euclid\ statistical study of the active galactic nuclei contribution fraction}

%\title{Euclid Quick Data Release (Q1)} \subtitle{First Euclid statistical study of the active galactic nuclei contribution fraction}

\include{authors.tex}

%
% For a Key Project paperm please use instead:
%
% \title{\Euclid\/ preparation. TBD. Your title}
%
% \author{Euclid Collaboration: F.~Author, .....}
%

% 
% Put your abstract here:
%
\abstract{
 
Active galactic nuclei (AGN) are an important phase in galaxy evolution. However, they can be difficult to identify due to their varied observational signatures. Furthermore, to understand the impact of an AGN on its host galaxy, it is important to quantify the strength of the AGN with respect to the host galaxy.
 We developed a deep learning (DL) model for identifying AGN in imaging data by deriving the contribution of the central point source. 
  %Our method consists of training a DL model to calculate the AGN contribution ($f_{\rm PSF}$) to a galaxy. 
The model was trained with Euclidised mock galaxy images in which we artificially injected different levels of AGN, in the form of varying contributions of the point-spread function (PSF). Our DL-based method can precisely and accurately recover the injected AGN \textbf{contribution} fraction $f_{\rm PSF}$, with a mean difference between the predicted and true $f_{\rm PSF}$ of $-0.0078$ and an overall root mean square error (RMSE) of 0.051. \textbf{With this new method, we move beyond the simplistic AGN versus non-AGN classification, allowing us to precisely quantify the AGN contribution and study galaxy evolution across a continuous spectrum of AGN activity.}
%, while the relative absolute error (RAE) is 0.3. 
We apply our method to \textbf{a stellar-mass-limited sample (with $M_{\ast} \ge 10^{9.8} M_{\odot}$ and $0.5 \le z \le 2.0$ ) selected from} the first \Euclid quick data release (Q1) and, using a threshold of $f_{\rm PSF} > 0.2$, we identify \textbf{$48\,840 \pm 78$} AGN over 63.1\,deg$^2$ ($7.8\pm0.1$\% of our sample). We compare these DL-selected AGN with AGN selected in the X-ray, mid-infrared (MIR), and via optical spectroscopy \textbf{and investigate their overlapping fractions depending on different thresholds on the PSF contribution. We find that the overlap increases with increasing X-ray or bolometric AGN luminosity.} We observe a positive correlation between the luminosity \textbf{in the $I_{\rm E}$ filter} of the AGN and the host galaxy stellar mass, suggesting that supermassive black holes (SMBHs) generally grow faster in more massive galaxies. Moreover, the mean relative contribution of the AGN is higher in the quiescent galaxy population than in the star-forming population. In terms of absolute power, starburst galaxies, as well as the most massive galaxies (across the star-formation main sequence), tend to host the most luminous AGN, indicating concomitant assembly of the SMBH and the host galaxy. 
%Similarly the fraction of galaxies with $f_{\rm PSF} > 0.2$ is higher in the quiescent galaxies.
}

%
% Provide up to five key words:
%
%    from the list in
%     https://www.aanda.org/for-authors/latex-issues/information-files#pop}
%
% Add short versions of title and author list for page headings
%
   \titlerunning{First \Euclid statistical study of AGN contribution fraction}
   \authorrunning{Euclid collaboration: B.~Margalef-Bentabol et al.}
   
   \maketitle
%
%-------------------------------------------------------------------
%
%
%   Start the main text of your paper here
%
   
\section{\label{sc:Intro} Introduction}
%
% Background
% Topic importance
% Existing Knowledge: 
% Knowledge Gap: 
% Rationale
% Research Question
% Aim/Objective
% Paper structure

Active \textbf{galactic} nuclei (AGN) are widely considered to be a crucial phase in the evolution of massive galaxies \textbf{\citep{Zhuang2023}, and may also play a role in the growth and regulation of low-mass galaxies \citep{Mezcua2019, Greene2020}.} They are powered by accretion of matter onto the supermassive black holes (\textbf{SMBHs}) at the centres of galaxies and can emit radiation across the whole electromagnetic spectrum \citep{Ueda2003,Padovani2017, Bianchi2022}. AGN can be broadly divided into categories, such as type I and type II AGN \citep{Antonucci1993, Urry1995}, depending on their observational characteristics. Type I are unobscured AGN which are luminous in the ultraviolet and optical. Type II are obscured AGN, in which the dust and gas torus surrounding the central engine can conceal their emission at certain wavelengths. Therefore, they need to be selected using different methods \citep{Hickox2018}, such as X-ray luminosity or colour criteria. While methods based on optical colours are affected by dust obscuration, and therefore will be biased against dust-obscured sources, methods based on mid-infrared (MIR) colours will tend to select dust-obscured AGN. AGN selected based on X-ray emission can include both obscured and unobscured AGN, although the soft X-ray selection tends to be biased toward unobscured sources. However, X-ray selection can be biased towards more massive galaxies \textbf{\citep{Aird2012, Mendez2013, Azadi2017}}. %\textcolor{violet}{What do these different AGN types correspond to? How do they overlap/differ?}

There exists a close connection between the assembly of the SMBH and the formation and evolution of its host galaxy \citep{Kormendy2013}. This is seen by various \textbf{scaling relations} between galaxy physical properties (such as stellar velocity dispersion, bulge luminosity, and bulge mass) and the mass of the SMBH \citep{Gultekin2009, Beifiori2012, Graham2013, McConnell2013, Lasker2014, Shankar2016}. To better measure these correlations and trace their evolution with time, there is clearly a need to accurately separate the light contribution from the accreting SMBH and its host galaxy. Traditionally, one way to achieve this is by performing a \textbf{two-dimensional (2D)} decomposition of the observed surface brightness, in a way that the galaxy is often modelled by a parameterised model, typically a S\'ersic profile \citep{Li2021, Toba2022}, while the AGN component can be assumed to be a point source described by the point-spread function (PSF) of the relevant observational instrument. This method can be further customised with different profiles to describe more complex light distributions of the host galaxy and different PSF models to account for any temporal and spatial variations. However, in addition to making simplified assumptions regarding galaxy morphology and structure (which may be particularly problematic for certain galaxy types), the surface brightness fitting approach is very time-consuming, making it unfeasible for large surveys. These fitting methods, \textbf{based on surface brightness fitting by codes such as \texttt{GALFIT} \citep{Peng2002},} often fail when the galaxy cannot be easily described by a parameterised profile, which can be the case for highly irregular galaxies or merging galaxies, leading to a high failure rate \textbf{\citep{Ribeiro2016, Ghosh2023, Margalef2024b}}. Another practical difficulty is that traditional methods usually do not have a built-in mechanism to easily take into account systematic effects such as variations in the PSF, which can fundamentally limit the level of precision in the decomposition. 

With the advent of the \Euclid space telescope \citep{Laureijs11} and its uniquely transformative power in high spatial resolution, sensitivity, and survey volume, we have an unprecedented opportunity to trace the co-evolution of the SMBHs and their host galaxies \textbf{in statistically large samples} across cosmic history. However, as explained above, traditional methods increasingly struggle to cope with the data complexity and volume and often fail to meet the \textbf{higher} requirements on precision and accuracy. To overcome these issues, in this work, we use a deep-learning (DL) based method developed in \citet{Margalef2024b} to
determine the AGN contribution \textbf{fraction} to the total observed light of a galaxy in imaging data. Specifically, we train a DL model with realistic mock images generated from cosmological hydro-dynamical simulations in which we inject AGN at different levels by adjusting the relative contribution of the PSF. Consequently, the trained model can be used to estimate the fraction of the light originating from a central point source, which we can then interpret as the AGN contribution \textbf{fraction}. \textbf{This method enables a more nuanced study of AGN, moving beyond a binary classification of AGN presence or absence. Some galaxies, while not classified as AGN by traditional selection methods, may still exhibit AGN activity that influences their host galaxies. By quantifying the AGN contribution fraction, we can better assess the role of AGN across a continuum of activity levels. Moreover, for comparisons with other selection techniques (generally binary selection) such as those based on X-ray luminosity, MIR colours, and optical spectroscopy, we can still apply specific thresholds on the PSF contribution fraction to classify AGN candidates. In this work, we present samples of AGN candidates identified with our method and compare it to these alternative selection approaches.}

Due to the relatively short timescale of the AGN activity and its diverse observational signatures, it is often challenging to construct sufficiently large samples of AGN to perform robust statistical and multi-dimensional analyses of the AGN population and co-evolution with the host galaxies. With \Euclid's large survey area this problem can be overcome. Furthermore, \Euclid's high spatial resolution and sensitivity make it the perfect survey to which our DL-based method can be applied to analyse the AGN contribution in imaging data. 
In this paper, we present the first study of identifying AGN using DL-based image decomposition technique in the first Quick Data Release of the \Euclid mission \citep[Q1,][]{Q1cite}. Throughout the paper, we use the terms PSF contribution and AGN contribution interchangeably.

The paper is organised as follows. 
In Sect.\,\ref{sc:Data}, we first briefly introduce the \Euclid imaging data used in this work and the key characteristics. Then we explain our galaxy sample selection and the various AGN-selection methods which are used to compare with our DL-based AGN identification method. In Sect.\,\ref{sc:Method}, we describe the \Euclid mock observations generated from the cosmological hydrodynamic simulations and how they are used to train our DL model.
In Sect.\,\ref{sc:Results}, 
Finally, using the AGN identified by our model, we examine the relation between the growth of the SMBHs (as traced by the AGN \textbf{contribution} fraction and AGN luminosity) and their host galaxy properties. 
In Sect.\,\ref{sc:Conc}, we present our main conclusions and future directions.
Throughout the paper we assume a flat $\Lambda$CDM universe with $\Omega_{\rm m}=0.3089$, $\Omega_{\Lambda}=0.6911$, and $H_0=67.74\,{\rm km\,s}^{-1}\,{\rm Mpc}^{-1}$ \citep{Plank2016}.

\section{\label{sc:Data} Data}

\begin{figure}
\centering
   \includegraphics[width=0.49\textwidth]{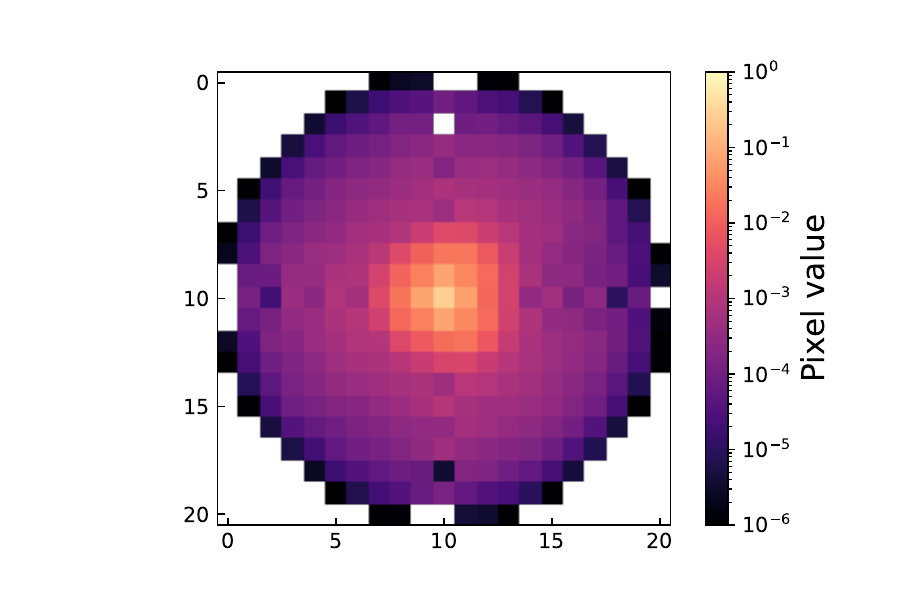}
    \includegraphics[width=0.49\textwidth]{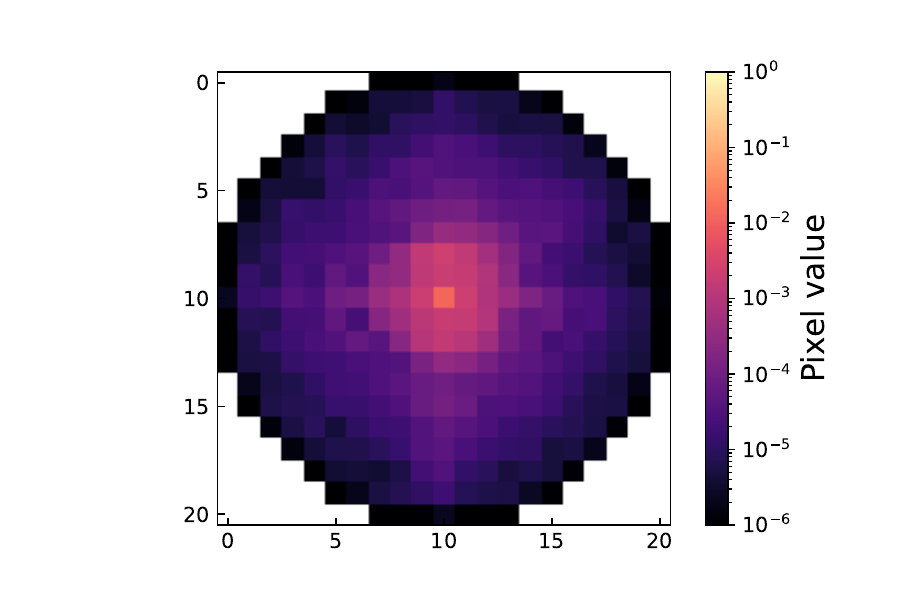}
    \includegraphics[width=0.49\textwidth]{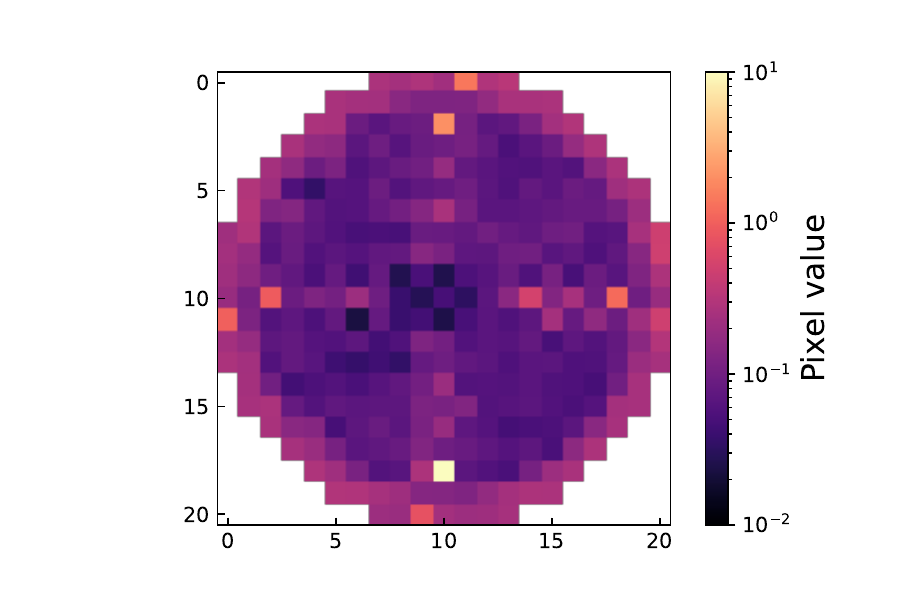}
    \caption{\Euclid VIS PSF. We stacked 500 random empirical PSFs and show the mean PSF (top panel), standard deviation (central panel), and the coefficient of variation (bottom panel), calculated pixel by pixel. The pixel resolution is \mbox{\ang{;;0.1}}\,pixel$^{-1}$. The axes show the number of pixels. The colour bar shows the value of each pixel.}
    \label{fig.psf}
\end{figure}

In this section, we first briefly describe the \Euclid data products used in this paper. Then we explain the selection criteria imposed to construct a stellar-mass-limited galaxy sample. Finally, we present various commonly used AGN-selection methods that will be used to compare with our AGN identification method.

\subsection{\label{subsc:data.imaging}\Euclid data products}

For this work, we exploit the first Euclid Quick Data Release \citep[Q1,][]{Q1cite}, which constitutes the first public data release. A detailed description of the mission and a summary of the scientific objectives are presented in~\cite{EuclidSkyOverview} and a description of the DR can be found in~\cite{Q1-TP001}, \cite{Q1-TP002}, \cite{Q1-TP003}, and \cite{Q1-TP004}. Q1 covers $63.1\,{\rm deg}^2$  in total in the Euclid Deep Fields (EDFs): North (EDF-N), South (EDF-S) and Fornax (EDF-F), observed by a single visit with the Visible Camera (VIS) in a single broadband filter $I_{\rm E}$ ~\citep{EuclidSkyVIS} and the Near Infrared Spectrometer and Photometer (NISP) instrument in three bands $Y_{\rm E}$, $J_{\rm E}$, and $H_{\rm E}$~\citep{EuclidSkyNISP}. 
%A detailed description of the DR is presented in~\cite{Q1-TP001}.  areas observed with both the VIS~\citep{EuclidSkyVIS} and NISP~\citep{EuclidSkyNISP} instruments.

%the visible imaging instrument (VIS; Sect. 3.4); and the Near Infrared Spectrometer and Photometer (NISP; Sect. 3.5).

We make use of a number of data products from Q1, including imaging data and associated catalogues (Altieri et al., in prep.), photometric measurements, and galaxy physical properties, such as stellar masses and star-formation rates (SFRs). \textbf{For a description of the photometric catalogues, see \cite{Q1-TP004}.} For a complete description of how photometric redshifts (photo-$z$) and physical properties are inferred in the Q1 data (\citealt{Q1-TP005}, Tucci et al. in prep.). \textbf{We note that photometric redshifts and physical properties are currently estimated without accounting for AGN contribution. As a result, some findings (especially for sources with significant AGN contribution) may be less reliable. Future analyses should incorporate more refined measurements to improve accuracy.} For this work, we use the VIS imaging data \textbf{due to its high spatial resolution, with a pixel resolution of \mbox{$\ang{;;0.1}\,\text{pixel}^{-1}$} and a depth of 24.7 AB mag ($10\,\sigma$ observed depth)}, observed in the visible filter $I_{\rm E}$. For each galaxy in our selected sample (see Sect.\,\ref{subsc:data.sample}), we made a cutout of $4\arcsec \times 4\arcsec$ ($40\,{\rm pixels} \times 40\,{\rm pixels}$), with the source at the centre. This size corresponds to a physical size between  $25\,{\rm kpc}\times 25\,{\rm kpc}$  and $35\,{\rm kpc}\times 35\,{\rm kpc}$ in the redshift range  $0.5<z<2$. \textbf{This redshift range is chosen to ensure similar physical size across cutouts.} Along with the VIS images, we use the empirical VIS PSFs \citep{EuclidSkyVIS}. Each source is accompanied by a PSF and, from all available PSFs, we choose a random sample of approximately $500\,000$ PSFs distributed within all three EDFs. In Fig.\,\ref{fig.psf} we show the stacked PSFs (a randomly selected subsample of 500) and display the mean, standard deviation, and coefficient of variation (i.e., the standard deviation divided by the mean). The typical VIS PSF full width at half maximum (FWHM) is \ang{;;0.13}. The mean coefficient of variation is around $13\%$. This level of intrinsic variation in the observed PSF will be later compared to the precision of our method in recovering the contribution of the PSF in the VIS imaging data (see Sect.\,\ref{subsc:results.performance}). \textbf{It is worth noticing that potential differences in the spectral energy distribution (SED) of stars used to generate the PSFs and those of AGN could lead to variations in the shape and size of the PSF. Further work is needed to investigate how these variations depend on the choice and colours of stars, and to assess their impact on our method.}

\subsection{\label{subsc:data.sample}Galaxy sample selection}

We construct our sample of galaxies from the \Euclid catalogues by first applying several conditions to remove possible contaminants and ensure \textbf{a high S/N detection}. The conditions applied are the following.
\begin{itemize}
    \item $\texttt{VIS}\_\texttt{DET} = 1$ to ensure \textbf{detection in the $I_{\rm E}$ band}.
    \item $\texttt{DET}\_\texttt{QUALITY}\_\texttt{FLAG} < 4$. This criterion ensures that we remove contaminants in the form of close neighbours, sources blended with another source, saturated sources or sources too close to the border, within the VIS or NIR bright star masks or within an extended source area, and bad pixels. 
    \item $\texttt{SPURIOUS}\_\texttt{FLAG} = 0$ to remove spurious sources.
    \item $\texttt{MUMAX}\_\texttt{MINUS}\_\texttt{MAG} > -2.6$, to remove \textbf{sources that have a high probability of being point-like \citep{Q1-TP004}}.
    \item \textbf{$I_{\rm E} < 24.5$, corresponding to a $10\,\sigma$ detection.}
\end{itemize} 
Furthermore, we imposed the following additional criteria to ensure good-quality photo-$z$ ($z_{\rm ph}$) and physical parameter estimations:
\begin{itemize}
    \item $\texttt{PHZ}\_\texttt{FLAGS} = 0$;
    \item $\texttt{PHYS}\_\texttt{PARAM}\_\texttt{FLAGS} = 0$;
    \item $\texttt{QUALITY}\_\texttt{FLAG} = 0$.
\end{itemize} 
Finally, we select galaxies in the redshift range $0.5 \le z_{\rm ph} \le 2.0$ and with stellar mass $M_{\ast} \ge 10^{9.8} M_{\odot}$. 
%The stellar mass limit is necessary to ensure that our sample is over 90\% complete in stellar mass out to redshift $z=2$. 
The stellar mass limit chosen here is motivated by \citet{Q1-SP031}, who using a similar multi-wavelength sample of galaxies find that at $z = 2$ and for galaxies with $M_{\ast} \ge 10^{9.8} M_{\odot}$, the sample is more than $90\%$ complete. After all these selection criteria, we are left with a final stellar-mass-limited sample of $624\,153$ galaxies.

\begin{figure}
\includegraphics[width=0.48
\textwidth]{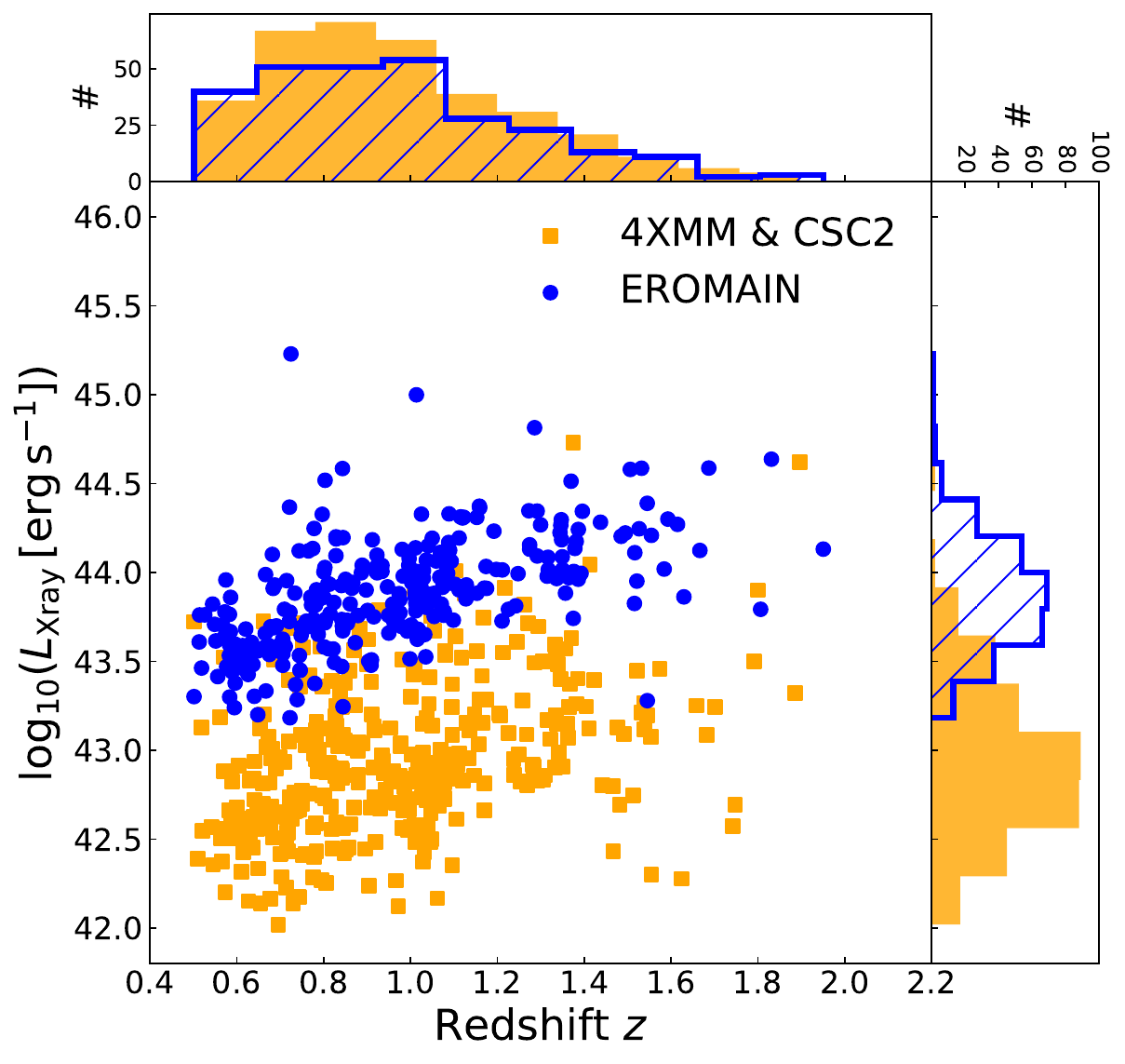}
\caption{X-ray luminosity of the \textbf{X-ray detected \Euclid sources} as a function of redshift. In blue we show the sources detected in EROMAIN and in yellow those detected in 4XMM and CSC2. The top histogram shows the redshift distributions of the different samples and the right histogram shows the distributions in the X-ray luminosity.}\label{fig.xray_sources}
\end{figure}

%We use the \Euclid catalogue (particularly photometric redshifts and stellar masses) to select our sample of galaxies. %We have applied other cuts to ensure a good sample.. 

\subsection{\label{subsc:data.agn_selection}AGN selections}

%Explain the different ways that we use to select AGN: X-ray luminosity, MIR colours, and optical spectroscopy. 
AGN can exhibit various observational features across the entire electromagnetic spectrum. Consequently, different selection techniques are used to identify different flavours of AGN (e.g., observed with different viewing angles, dust obscuration levels and/or at evolution stages). We use the following three widely used AGN-selection techniques to compare with our DL-based methodology.
\begin{enumerate}

    \item X-ray AGN: Galaxies are classified as AGN if they have a counterpart identified in \citet{Q1-SP003}. \citet{Q1-SP003} constructed a catalogue of Q1 galaxies with counterparts in any of the three 
    X-ray surveys that overlap with the EDFs: the XMM-{\it Newton} 4XMM-DR14 survey \citep[4XMM;][]{webbXMMNewtonSerendipitousSurvey2020}; the {\it Chandra} Source Catalog v.2.0 \citep[CSC2;][]{evansChandraSourceCatalog2024}; and the eROSITA  DR1 Main sample \citep[EROMAIN;][]{predehlEROSITAXrayTelescope2021a,merloniSRGEROSITAAllsky2024}. \textbf{All three surveys (4XMN, CSC2, and EOROMAIN have soft X-ray energy range of 0.5--2\,keV)}. However, EROMAIN does not cover EDF-N. The catalogue includes public spectroscopic redshifts (spec-$z$s), if available. In those cases, we use the spec-$z$ instead of the photo-$z$ from the Q1 catalogue. To select a \textbf{high-purity}  sample of X-ray AGN \textbf{and minimise contaminants}, we select only sources that satisfy the following criteria:
    \begin{itemize}
        \item \texttt{match\_flag} equal to 1,  \textbf{to select sources with the highest individual probability of being the correct counterpart to the X-ray sources, in any of the surveys};
        \item low Galactic probability, $\texttt{Gal\_proba}<0.5$, to ensure selecting extragalactic sources;
        \item X-ray signal-to-noise ratio $\rm S/N\geq2$;
        \item X-ray luminosity $L_{\rm X\,[0.5\--2\,\rm{keV}]}\geq 10^{42}\,{\rm erg\,s}^{-1}$. %Note that $L_{X\,[0.5\--2\,keV]}$ has not been $K-$corrected.
    \end{itemize}
     There are a total of 335 sources in 4XMM, 14 in CSC2 and 276 in EROMAIN. Figure\,\ref{fig.xray_sources} shows the different sources identified in the $L_{\rm X\,[0.5\--2\,keV]}$ versus redshift plane. Due to small number statistics and similar sensitivities, we decided to combine 4XMM and CSC2. At a given redshift, galaxies detected in EROMAIN tend to have higher $L_{\rm X\,[0.5\--2\,\rm{keV}]}$. This is expected given the flux limits of the three X-ray surveys. 
    \item MIR AGN: MIR colour-selected AGN \textbf{selection was done by \citep{Q1-SP027}. They followed} the criteria from \citet{Assef2018} \textbf{to find} sources with counterparts in the AllWISE Data Release 6 \citep[DR6,][]{Wright2010, Mainzer2011}, which integrates data from both the WISE cryogenic and NEOWISE post-cryogenic survey \citep{Mainzer2011} phases, providing the most complete mid-infrared sky coverage available to date. The AllWISE DR6 mapped the entire sky in the four bands, W1, W2, W3, W4 (centred at 3.4, 4.6, 12, and $22\,\mu {\rm m}$, respectively), detecting over 747 million sources. \textbf{\cite{Q1-SP027} used two diagnostics, defined in \cite{Assef2018},} to select AGN. The first diagnostic, C75, focusing on achieving $75\%$ completeness for the selected AGN candidates  \textbf{(while achieving $51\%$ reliability)}, is defined by
    \begin{equation}
        \rm W1 - \rm W2 > 0.71\;,
        \label{eq:C75}
    \end{equation}
    where $\rm W1$ and $\rm W2$ \textbf{are given in the Vega magnitude system.}
    The second diagnostic R90, focusing on obtaining a sample with 90\% reliability \textbf{(and achieving 17\% completeness)}, is defined as follows:
    \begin{equation}
        \rm W1 - \rm W2 > \begin{cases}
          0.65 \exp[0.153(\rm W2-13.86)^2],  & \rm if\   \rm W2>13.86\;,\\
          0.65, & \rm if\  \rm W2 \leq 13.86\;.\\
        \end{cases}
    \end{equation}
    These two diagnostics must also satisfy the following extra conditions:
\begin{itemize}
    \item $\rm W1>8$ and $\rm W2>7$, \textbf{with \textbf{$\rm S/N_{W2}\geq5$,}} to only consider sources with W1 and W2 magnitudes fainter than the saturation limits of the survey;
    \item $\verb|cc_flags| = 0$ to ensure that the sources are not artefacts or affected by artefacts \citep{Q1-SP027}.
\end{itemize}

    According to the C75 and R90 diagnostic, there are 9052 and 835  AGN, respectively.
    
    \item \textbf{DESI spectroscopic AGN: This AGN selection was done by \cite{Q1-SP027} by selecting the counterparts with the} spectroscopically identified AGN in the DESI \textbf{Early Data Release \citep{DESI_EDR}, with emission line fluxes, widths and equivalent widths measured with \texttt{FastSpecFit} \citep{2023ascl.soft08005M}}, including:
    \begin{itemize}
        \item QSO classification based on DESI \texttt{SPECTYPE} \textbf{\citep{Siudek2024}};
        \item AGN classification based on the detection of broad H$\,\alpha$, H$\,\beta$, \ion{Mg}{II}, or \ion{C}{IV} emission lines with a FWHM $\geq$ 1200\,\kms;
        \item AGN classification for DESI sources with a spectroscopic $z \ge 0.5$ based on the KEx diagnostic diagram of \cite{ZhangHao2018}, which makes use of the [\ion{O}{III}]\,$\lambda$5007 emission line width, the  BLUE diagram of \cite{Lamareille2010}, which makes use of the equivalent width of the H$\,\beta$ and [\ion{O}{II}]\,$\lambda$3727 emission lines\textbf{, or the WHAN diagram of \citep{Fernandes2010}, which makes use of the equivalent width of the H$\alpha$ emission line.}
    \end{itemize}
    For these sources, a catalogue with spec-$z$ and estimates of the AGN bolometric luminosity exist \citep{Siudek2024}. 
    %In those cases we adopt the spec-$z$ instead of photo-$z$. 
    In total, there are 229 AGN spectroscopically confirmed within our parent sample\textbf{, with 47 QSO, 64 AGN with broad line emission, and 134 AGN confirmed through the different diagrams.}
\end{enumerate}

\begin{table}[]
    \centering
    \caption{Number of AGN for each selection method used \textbf{for comparing with our method} in this paper.}
    \begin{tabular}{lcc}
        \hline\hline
        \noalign{\vskip 1pt}
        AGN-selection method & AGN  \\
        \hline
        \noalign{\vskip 1pt}
        X-ray detection (4XMM \& CSC2) & 349 \\
        X-ray detection (EROMAIN) & 276 \\
        DESI optical spectroscopy & 229 \\
        MIR colours (C75, AllWISE) & 9052 \\
        MIR colours (R90, AllWISE) & 835 \\
        \hline
    \end{tabular}
    \label{tab.agn_counts}
\end{table}

Table\,\ref{tab.agn_counts} shows a summary of the number of AGN depending on the selection method. \textbf{We construct a sample of galaxies with no clear AGN signatures for comparison with the different AGN samples, which we call `non-AGN'. This sample is selected in the EDF-S, where the X-ray and MIR coverage overlap, and includes galaxies that have no X-ray detection and do not satisfy either of the two MIR AGN diagnostics. While some AGN may still be present due to the sensitivity limits of the X-ray and MIR data sets available and the lack of optical spectroscopic AGN identification, AGN are relatively rare, so statistically, this `non-AGN' sample should provide a reasonable representation of galaxies without AGN.}
%\textcolor{violet}{You can refer to La Marca et al. for a detailed comparison of the various AGN selections.}

\section{\label{sc:Method} Methodology}
In this section, we first describe the construction of the mock host galaxy \Euclid VIS images with different injected levels of the AGN contribution, which is approximated by varying contributions of the PSF. Then we briefly explain our DL model used to retrieve the PSF contribution in imaging data and the training process.

\subsection{\label{subsc:method.mock_data}Mock \Euclid VIS data}

The IllustrisTNG project \citep{ Naiman2018, Nelson2018, Marinacci2018, Pillepich2018, Springel2018, Nelson2019} is a series of cosmological hydrodynamical simulations of galaxy formation and evolution, with different runs that differ in volume and resolution. The initial conditions are drawn from {\it Planck} results \citep{Plank2016}. For this work, we used TNG100 and TNG300, which have comoving length sizes of 100, and 300\,Mpc\,$h^{-1}$, respectively. TNG100 contains $1820^3$ dark matter (DM) particles with a mass resolution of $7.5 \times10^6 M_{\odot}$, while TNG300 contains $2500^3$ dark matter (DM) particles with a mass resolution of $6 \times10^7\,M_{\odot}$. The baryonic particle resolution is $1.4\times10^6\,M_{\odot}$ for TNG100 and $1.1\times10^7\,M_{\odot}$ for TNG300. More details on IllustrisTNG can be found in \cite{Pillepich2018}. 

We selected galaxies from simulation snapshots corresponding to redshifts between $z=0.5$ and $2$ (snapshot number between 67 and 25). The time step between each snapshot is around $150\,\text{Myr}$. For TNG100, we selected galaxies with stellar mass $M_{*}>10^9\,M_{\odot}$, while for TNG300 the lower mass limit for our sample is $M_{*}>8\times10^9M_{\odot}$. \textbf{These limits ensure that most galaxies have a sufficient number of stellar particles in each simulation (hence are reasonably well resolved)}. To construct our training data set we used a sample of $150\,000$ galaxies chosen to cover the redshift range and stellar mass range uniformly. We do this so that massive galaxies or high-redshift galaxies are not under-represented in our training sample. We also limit the number of galaxies for computational reasons.

For each galaxy, we generated a synthetic \Euclid VIS observation from the simulations at the same pixel resolution (\mbox{\ang{;;0.1}\,pixel$^{-1}$}) and in the photometric filter, following these steps.
\begin{itemize}
\item Each stellar particle contributes its SED that depends on mass, age, and metallicity and is derived from stellar population synthesis models of \cite{Bruzual2003}. The sum of all stars' contributions is passed through the \Euclid $I_{\rm E}$ filter to create the smoothed 2D projected map \citep{Rodriguez-Gomez2019, Martin2022}, \textbf{with the galaxy at its centre}. The image is cut to a size of $4''\times4''$, which corresponds to approximately $25\,{\rm kpc}\times 25\,{\rm kpc}$  to $35\,{\rm kpc}\times 35\,{\rm kpc}$ in the redshift range of this work.
\item After that, each image was convolved with a randomly chosen \Euclid VIS PSF (to account for the spatial and temporal variations of the PSF).
\item To account for the statistical variation of a source's photon emissions over time, Poisson noise was added to each image.
\item Finally, each image was injected into cutouts of real \Euclid VIS data, to ensure that our training data include realistic \Euclid background and noise. The cutouts of $4''\times4''$ size were obtained randomly within the Q1 area, with the condition that \textbf{there} are no invalid pixels within the cutouts and that there is no source in the centre \textbf{(within a $9\arcsec$ radius, derived from the estimated source density of the \Euclid Q1 deep fields.)}, where we inject the simulated galaxy. The real \Euclid VIS sky cutouts were retrieved and processed from ESA Datalabs \citep{Navarro2024}. %Datalabscite
\end{itemize}

\begin{figure}
    \centering
    \includegraphics[width=1\linewidth]{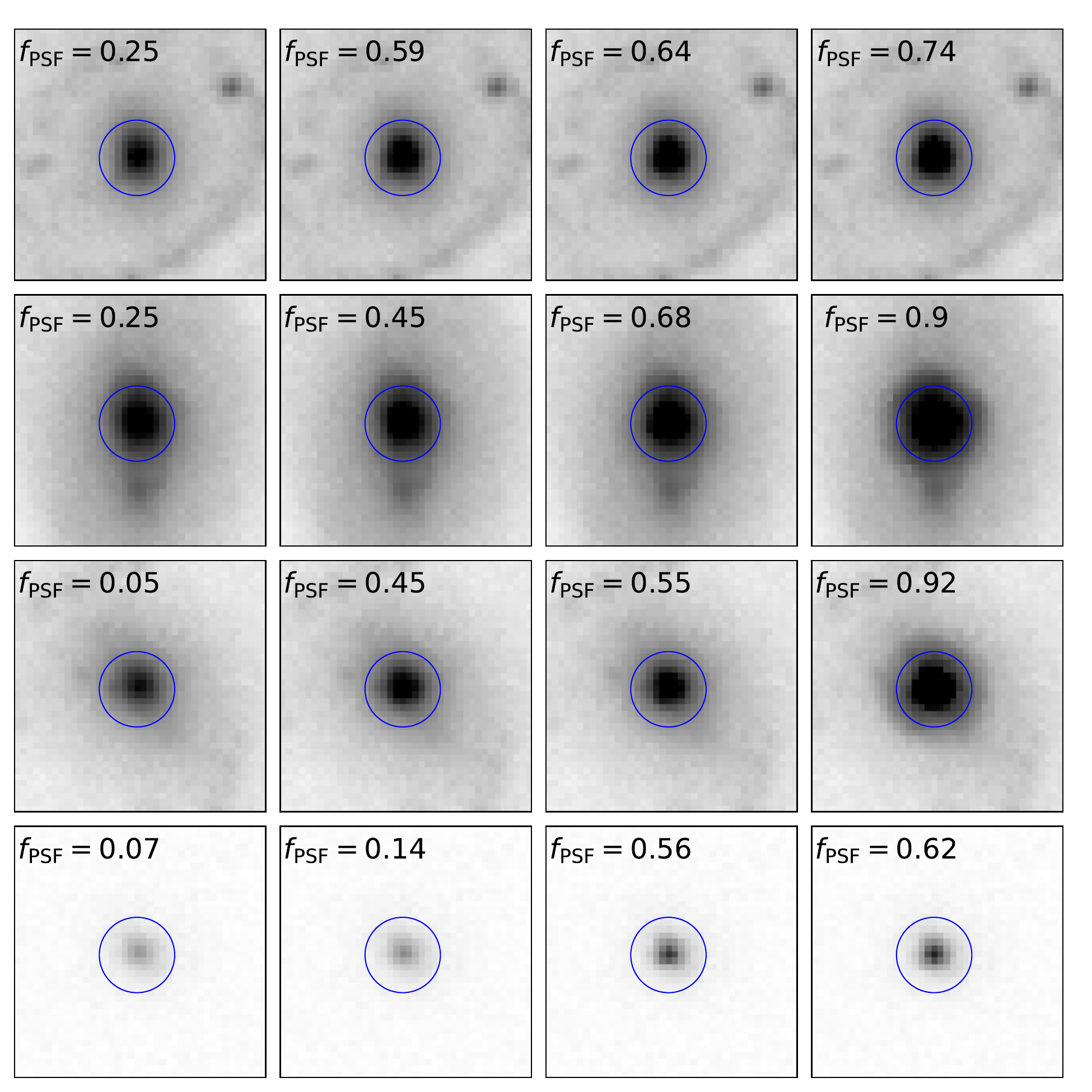}
    \caption{Example mock \Euclid VIS images with varying levels of PSF contribution ($f_{\rm PSF}$) to the total flux. The images have been generated to mimic \Euclid observations and include realistic \Euclid noise and background. Each row corresponds to a different galaxy with increasing PSF contribution fractions from left to right, \textbf{at different magnitudes}. Images correspond to a physical size of around 25\,kpc and are displayed with an inverse arcsinh scaling. \textbf{The blue circles show an aperture of \ang{;;0.5} radius}.}
    \label{fig:injected_AGN}
\end{figure}

\textbf{While IllustrisTNG includes SMBH feedback in their physical models, the simulated images produced do not include the light emission from} the possible presence of AGN, therefore, we need to add the contribution of the AGN by injecting a PSF component at varying contribution levels. The PSF \textbf{contribution} fraction (i.e., contribution to the total light) can be defined as 
\begin{equation}
    \label{eq:fAGN}
    f_{\rm PSF} = \frac{F_{\rm PSF}}{F_{\rm host} + F_{\rm PSF}}\;,
\end{equation}
where $F_{\rm PSF}$ is the aperture flux of the PSF component and $F_{\rm host}$ is the aperture flux of the host galaxy. We want to create a diverse training sample with different values of $f_{\rm PSF}$. To do so, we injected a central point source at different levels into the host galaxy image. The observed \Euclid PSF \textbf{effective models \citep{Cropper16}} were used as the central point source.  
%The PSF fractions were chosen to be . 
For each galaxy, five different images were created with five different $f_{\rm PSF}$ values, chosen randomly in the range [0, 1). The  PSF injected images were created as follows:
\begin{enumerate}
  \item a random PSF was selected;
  \item the flux of the PSF and the host galaxy was measured within an aperture of \mbox{\ang{;;0.5}} radius for all galaxies, using the \verb|aperture_photometry| function of the \verb|photutils| package \citep{Bradley2024_10967176};
  \item the PSF image was scaled to satisfy Eq.\,(\ref{eq:fAGN}), for a chosen $f_{\rm PSF}$;
  \item the scaled PSF image was added to the host galaxy image, \textbf{at the centre of the galaxy}. 
\end{enumerate}
%First, a random PSF was selected. Secondly, the flux of the PSF and the host galaxy was measured within a 0.5\arcsec aperture using the \verb|aperture_photometry| function of the \verb|photutils| package \citep{Bradley2024_10967176}. Third, the PSF image was scaled to satisfy Eq. \ref{eq:fAGN}. Finally, the scaled PSF image was added to the host galaxy image.  
We chose this aperture to capture the majority of the galaxy flux for most sources. This is particularly true at the highest redshifts, where most galaxies above the stellar mass limit are smaller than our aperture \citep{vanderWel2014}. However, at lower redshifts, an increasing number of galaxies exceed the aperture size, potentially leading to a slight overestimation of $f_{\rm PSF}$. Conversely, selecting a larger aperture to accommodate these extended low-redshift galaxies could introduce flux contamination from nearby sources, biasing $f_{\rm PSF}$ toward lower values. Finding an optimal aperture across a wide redshift range is challenging, and we leave a more detailed exploration of aperture selection to future work. This procedure resulted in a final sample of 750\,000 mock galaxies with different levels of $f_{\rm PSF}$. Example images of these simulated galaxy images with varying levels of $f_{\rm PSF}$ can be seen in Fig.\,\ref{fig:injected_AGN}. It is clear that as we increase the relative contribution of the PSF, the galaxy image becomes increasingly more dominated by an unresolved point source.

\subsection{\label{subsc:method.model}Deep learning model and training}

\texttt{Zoobot} \citep{Walmsley2023} is a Python package used to measure detailed morphologies of galaxies with DL. \texttt{Zoobot} includes different DL architectures pre-trained on millions of labelled galaxies, derived from visual classifications of the Galaxy Zoo project \citep{Lintott2008} on real images of galaxies selected from surveys such as the Sloan Digital Sky Survey (SDSS), Hyper Suprime-Cam (HSC) and Hubble \citep{Willett2013, Willett2017, Simmons2017, Walmsley2022a, Walmsley2022b, Omori2023}. The models can be adapted to new tasks and new galaxy surveys without needing a large amount of labelled data, since they rely on the learned representations. \textbf{This process, known as `transfer learning' \citep{Lu2015}, allows a previously trained machine-learning model to be applied to a new problem. Instead of retraining all parameters from scratch, the existing model architecture and learned weights from prior training can be reused, making adaptation more efficient.}

For this work, \textbf{to train a deep-learning model to predict the PSF contribution fraction, $f_{\rm PSF}$, from a galaxy image}, we follow the same procedure as in \citet{Margalef2024b}. We use a ConvNeXt \citep{Liu2022} model, in particular, the ConvNeXt-Base architecture, \textbf{which consists of 36 convolutional blocks that are designed to resemble transformer blocks, while maintaining the efficiency of CNNs} pre-trained on the Galaxy Zoo data set of over 820\,000 images and 100 million volunteer votes to morphological questions. ConvNext architectures incorporate enhancements inspired by transformer \textbf{models \citep{dosovitskiy2021a}} into traditional convolutional networks, resulting in improved performance and efficiency for vision-based tasks. We adapted the model to perform a regression task \textbf{by replacing the original model head (top layer) with a single dense layer with one neuron (corresponding to the predicted output of the network), using a sigmoid activation function for the final layer} (to restrict the output between 0 and 1), and a mean-square-error loss function to train the network. \textbf{First, we load the pre-trained parameters of the architecture. Then, we retrain the last four blocks and the linear head, while keeping the rest of the network's parameters frozen to the optimal values found for the pre-trained data from Galaxy Zoo.} Our sample of mock galaxies was split into train, validation, and test sets, containing $80\%$, $10\%$, and $10\%$ of the total sample, respectively. The split is done in a way that the five iterations of a single galaxy (the mock images of the same galaxy with \textbf{five} different levels of the PSF component injected) are only contained in one of the splits. The train and validation data sets are used during training and to optimise the model's hyperparameters, while the test data set is only used for evaluating the best-performing model presented here. The model was trained on a v100 GPU and took 24 hours to complete.

\section{\label{sc:Results} Results}

In this section, we first analyse the overall performance of our DL model in estimating $f_{\rm PSF}$, using common metrics \textbf{for regression tasks} such as the root mean squared error (RMSE), the relative absolute error (RAE) and outlier fraction. Then we present an analysis on how our method compares with other AGN-selection techniques, and how the overlaps with other selections change with respect to AGN properties such as its luminosity and relative dominance compared to the host galaxy. Finally, we examine the dependence of our DL-identified AGN on host galaxy stellar mass and location in the star-formation main sequence (SFMS) diagram.

\subsection{\label{subsc:results.performance}Zoobot model performance}

\begin{figure}
    \centering
    \includegraphics[width=0.48
\textwidth]{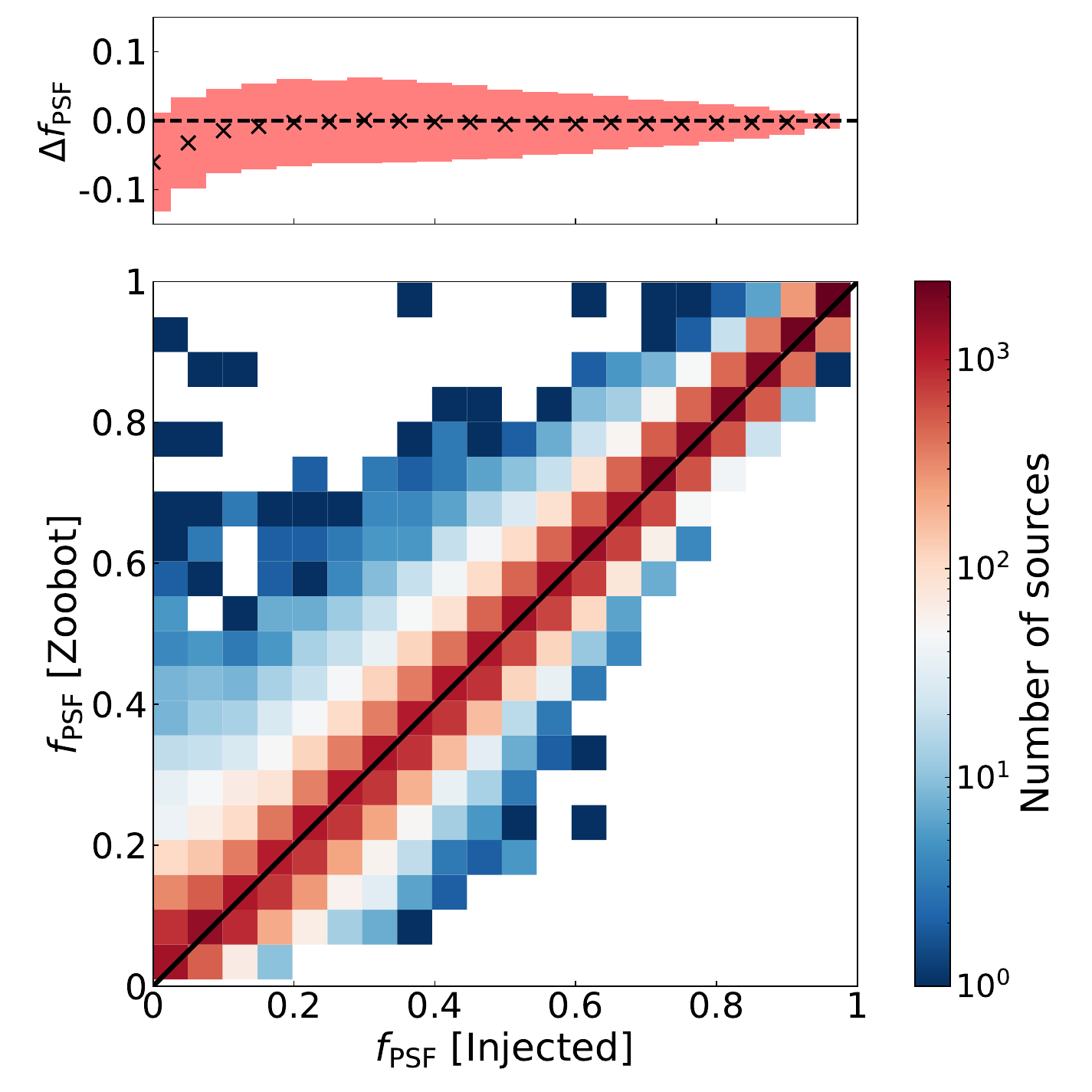}
    \caption{Comparison between the injected PSF \textbf{contribution} fraction and the predicted \textbf{contribution} fraction from \texttt{Zoobot} on the test set across the whole redshift range ($0.5<z<2$). 
    %The comparison shows a mean difference between the two quantities ($\Delta f_{\rm PSF} = f_{\rm PSF}$ [Injected] $- f_{\rm PSF}$ [Zoobot]) of $-0.0078$ and an overall RMSE = 0.052. 
    The diagonal line is the 1:1 relation. The top plot shows the mean difference and the standard deviation as a function of the injected PSF contribution fraction. The colour bar indicates the number of sources in each 2D bin.}
    \label{fig.pred_zoobot}
\end{figure}

\begin{figure*}
    \centering  \includegraphics[width=0.4\textwidth]{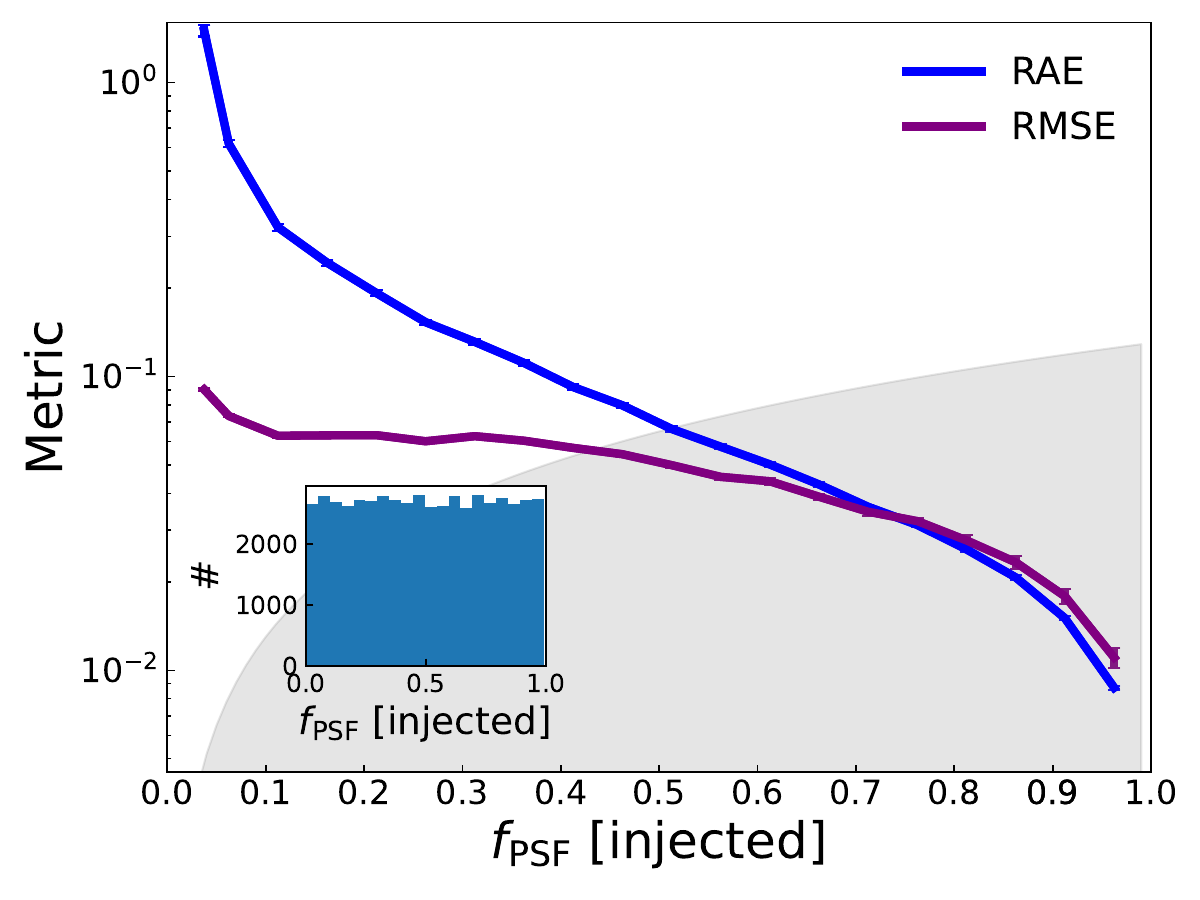}
\includegraphics[width=0.4\textwidth]{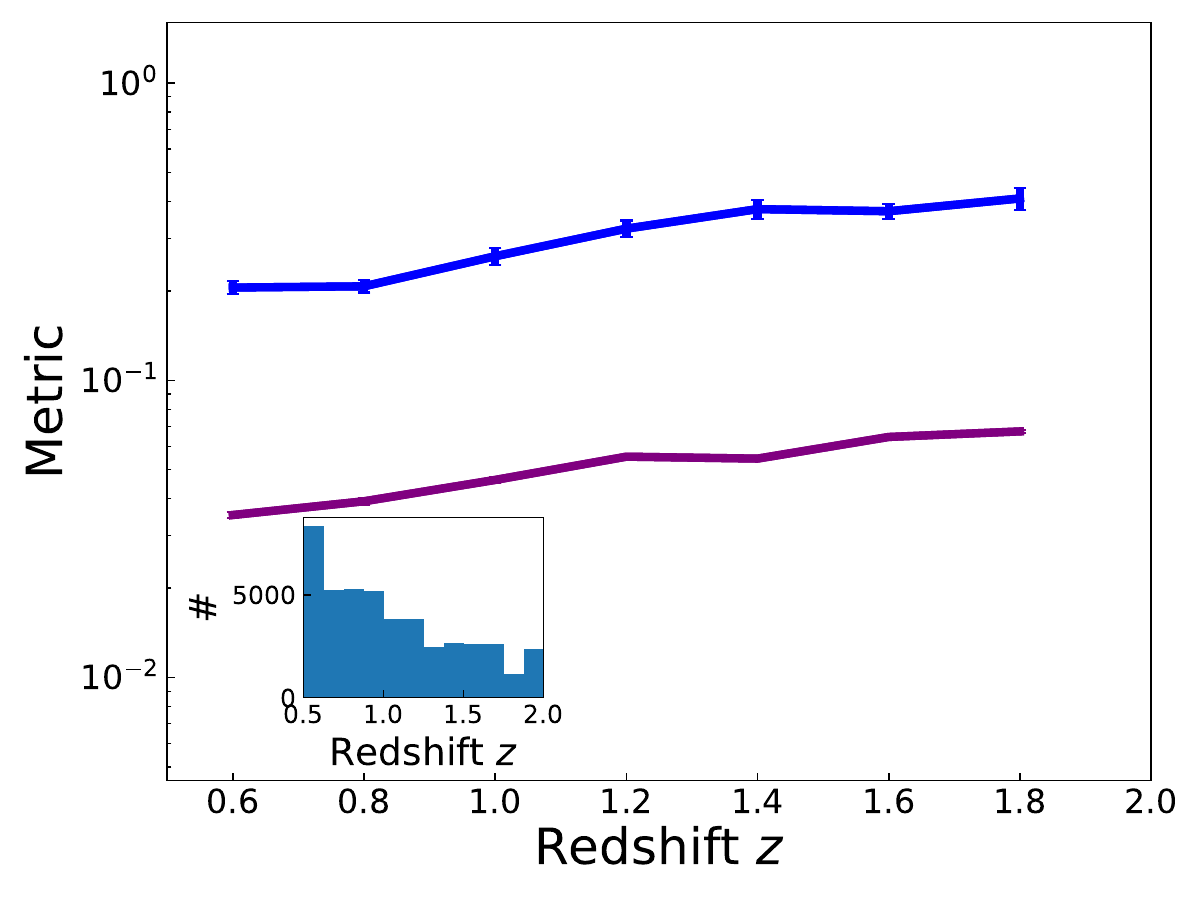}
\includegraphics[width=0.4\textwidth]{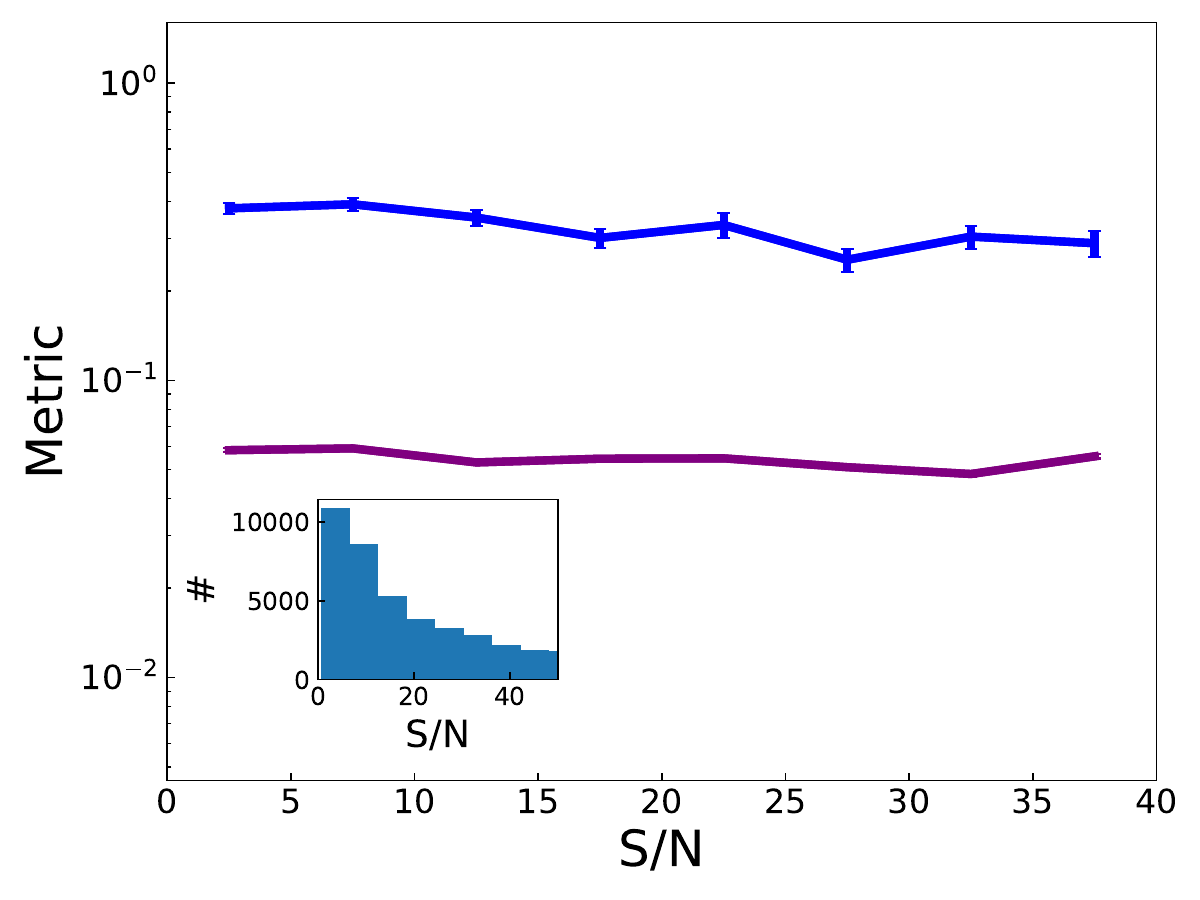}    \includegraphics[width=0.4\textwidth]{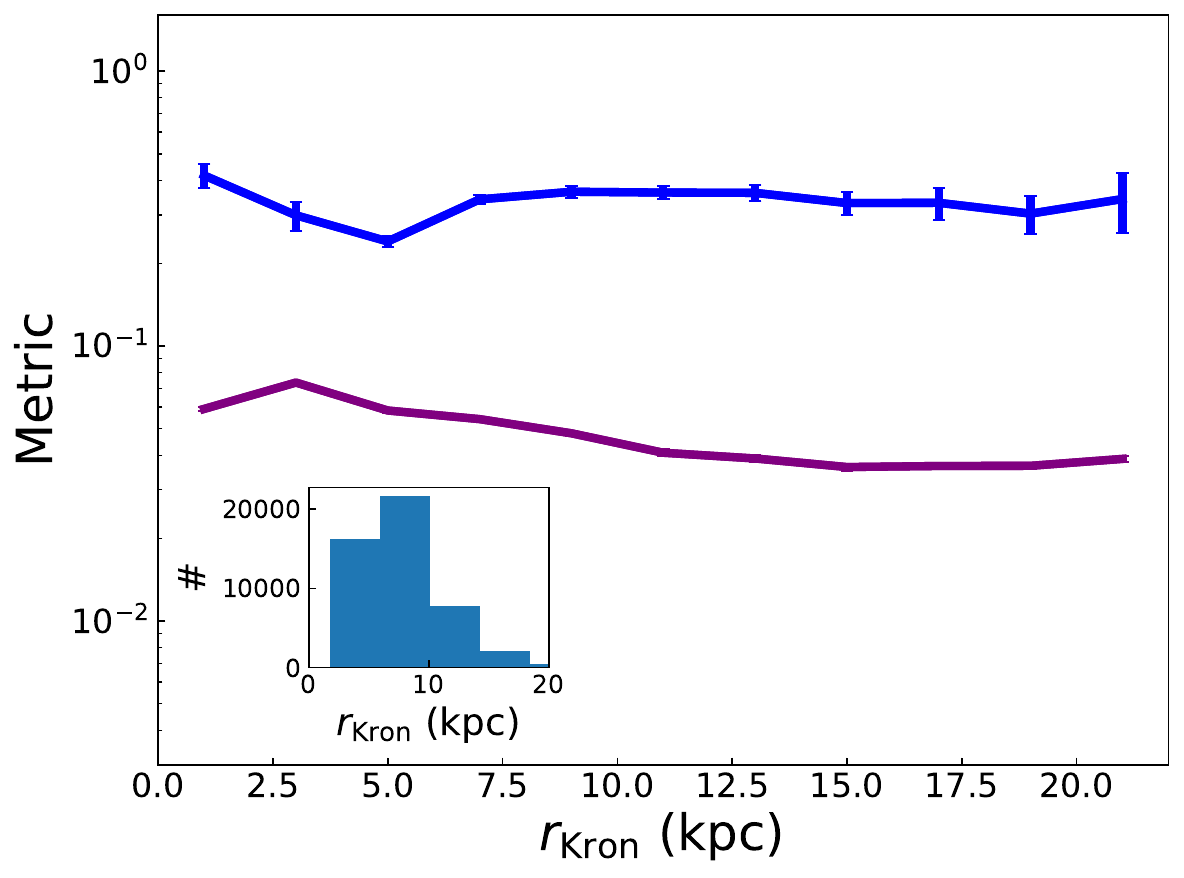}
    \caption{RMSE and average RAE as a function of the injected PSF \textbf{contribution} fraction (top left panel), redshift (top right), S/N (bottom left), and size (bottom right).  The error bars show the 95\% interval from bootstrapping. The insets show the distribution of each quantity. The grey area of the top left panel corresponds to the level of the intrinsic fractional variation (standard deviation divided by the mean), considering the spatial and temporal variations in the observed \Euclid VIS PSF.}
    \label{fig.rmse_z}
\end{figure*}

\begin{figure}
    \centering
    \includegraphics[width=0.5\textwidth]{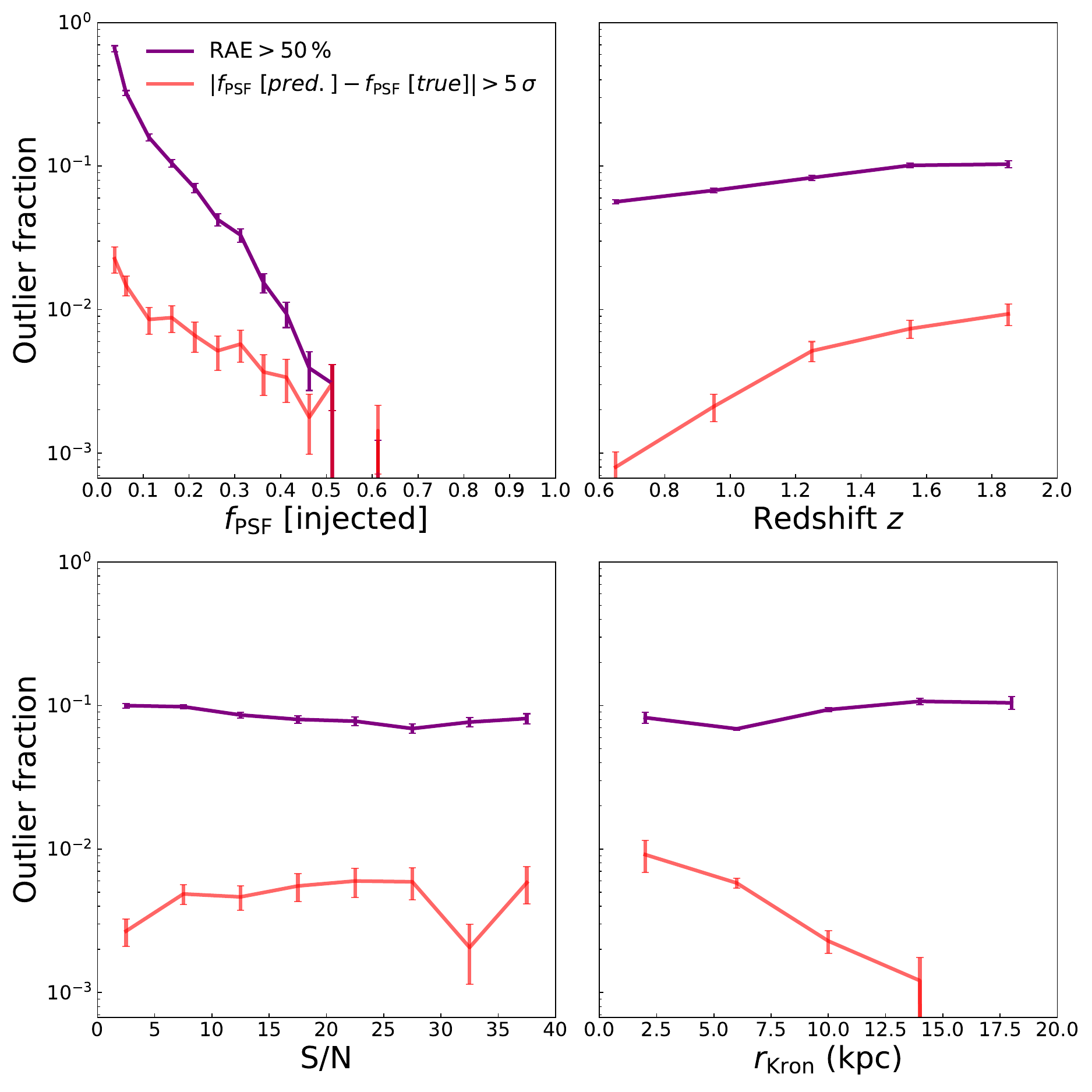}

    \caption{
    Similar to Fig.\,\ref{fig.rmse_z}, but for outlier fraction, calculated as the fraction of galaxies with ${\rm RAE} >  50\%$ or $|f_{\rm PSF} [\rm predicted]-f_{\rm PSF} [\rm true]| > 5\,\sigma$.
    %Outlier fraction (calculated as the fraction of galaxies with RAE $> 10\%, 20\%$, or 30\%, shown in purple, green and yellow, respectively) as a function of the injected AGN fraction (top left panel), redshift (top right), S/N (bottom left) and size (bottom right). 
    }
    \label{fig.outlier_fraction}
\end{figure}

\begin{figure}
    \centering
    \includegraphics[width=1\linewidth]{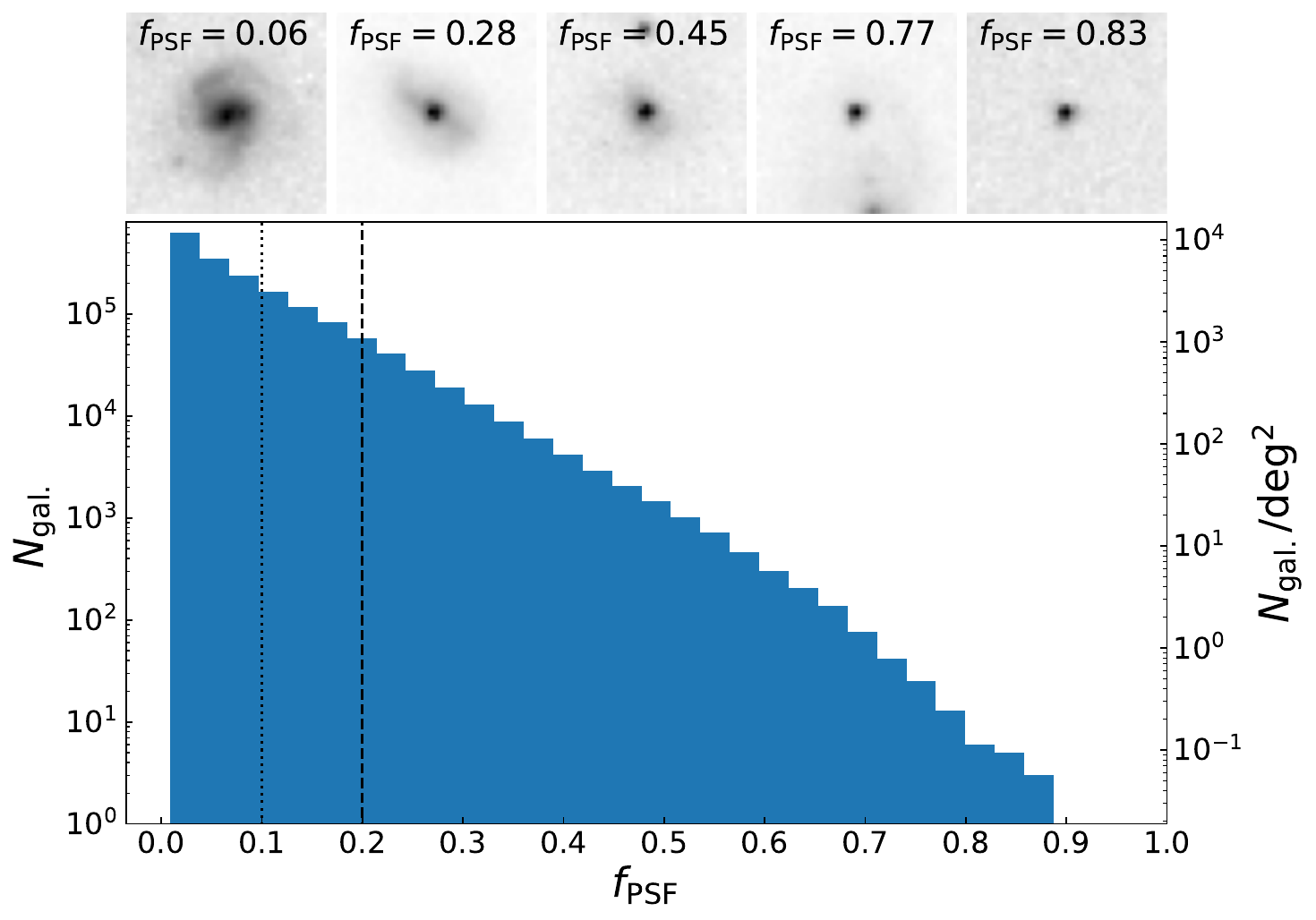}
    \caption{Distribution of $f_{\rm PSF}$ for the whole \Euclid stellar-mass-selected sample of galaxies with $0.5<z<2$ and $\logten(M_{*}/M_{\odot})>9.8 $. The vertical dashed lines represent the two adopted thresholds above which we classify galaxies as AGN according to the DL model, \textbf{for the purpose of comparing with other AGN-selection methods.}}
    \label{fig.hist_AGN_frac}
\end{figure}

\begin{figure*}
    \centering
    \includegraphics[width=1\textwidth]{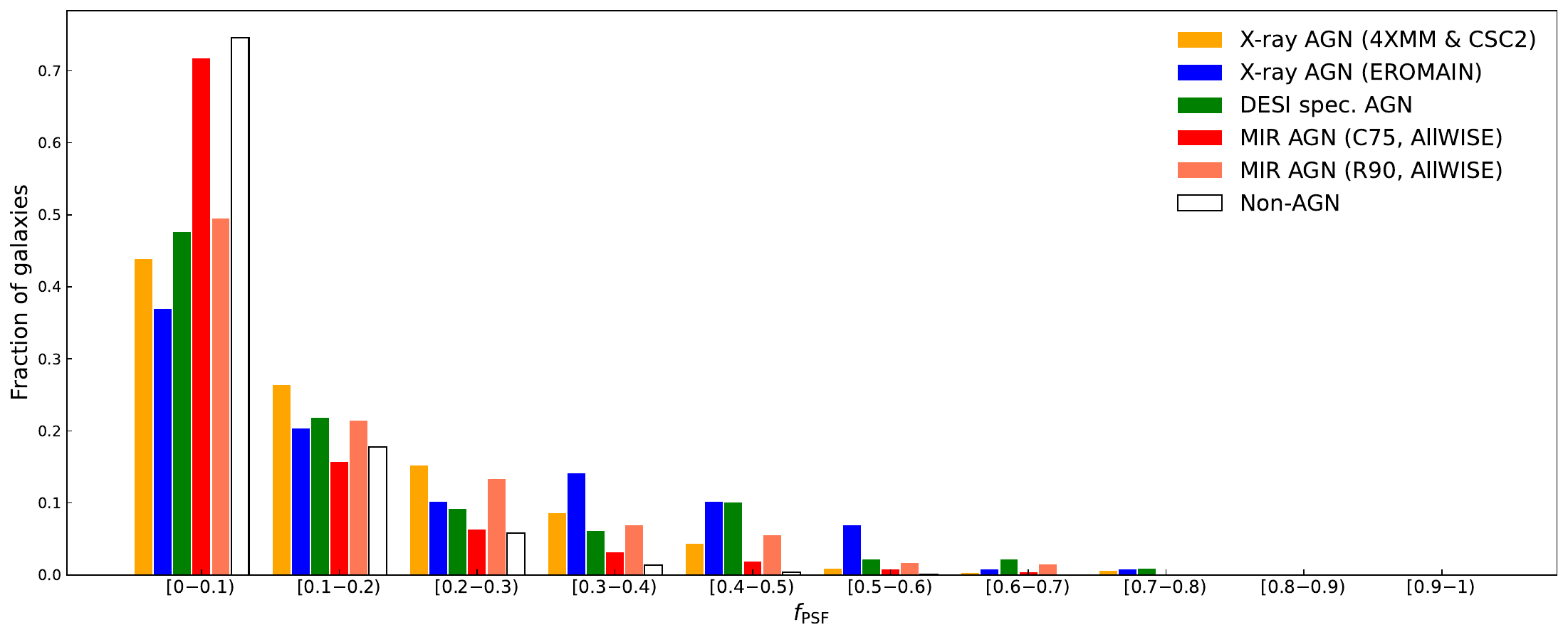}
    \caption{Fraction of galaxies in bins of \textbf{the predicted PSF contribution fraction in the AGN samples from different selections (X-ray detection, MIR colours, and optical spectroscopy) and the `non-AGN' sample } (see text in Sec.\,\ref{subsc:data.agn_selection}).}
    \label{fig.comparison}
\end{figure*}

We analyse the performance of the trained DL model on the test data constructed from the TNG simulations, \textbf{evaluating its ability to predict the PSF contribution fraction ($f_{\rm PSF}$}. The bottom panel of Fig.\,\ref{fig.pred_zoobot} shows the predicted $f_{\rm PSF}$ versus the injected $f_{\rm PSF}$ (i.e., the true value), colour-coded by the number density of objects. It is evident that the vast majority of the objects lie close to the 1:1 line across the whole range of the injected $f_{\rm PSF}$, demonstrating that the model is able to recover the true $f_{\rm PSF}$ with high accuracy and precision. \textbf{For galaxies with $f_{\rm  PSF} [\rm injected] < 0.05$, where the mean difference between the predicted and true $f_{\rm PSF}$ is larger, only around 5\% of them have $f_{\rm  PSF} [\rm predicted] > 0.2$ and 0.4\% have $f_{\rm  PSF} [\rm predicted] > 0.5$. Furthermore, the galaxies with the largest differences tend to be in the highest redshift bins and appear to be compact sources. As shown in \citet{Margalef2024b}, when sources are very compact, and possibly unresolved, this method will not give accurate results in terms of the predicted $f_{\rm PSF}$}. The top panel of Fig.\,\ref{fig.pred_zoobot} shows the mean difference between the real and predicted values ($\Delta f_{\rm PSF} = f_{\rm PSF}$ [injected] $- f_{\rm PSF}$ [Zoobot]) and its dispersion as a function of the injected $f_{\rm PSF}$. The mean bias across the whole test set is $\langle{\Delta f_{\rm PSF}}\rangle=-0.0078$. When the intrinsic PSF contribution is very low (i.e., $f_{\rm PSF}\ \rm [injected]<10\%$), the difference $\Delta f_{\rm PSF}$ starts to increase (to $-0.06$ in the lowest $f_{\rm PSF}$ [injected] bin), with a slight overestimation in the predicted $f_{\rm PSF}$. The dispersion also increases with decreasing $f_{\rm PSF}$ [injected], from 0.02 to 0.14.

%We see a tight correlation between predicted and real values and very little bias for $f_{\rm PSF}>0.1$. 
%Bellow this PSF fraction our model tends to slightly overestimate $f_{\rm PSF}$. 

We further analyse the RMSE, RAE, and outlier fraction as a function of different properties, including the injected \textbf{(i.e., true)} $f_{\rm PSF}$, redshift, S/N, and size. The RMSE can be derived as follows,
\begin{equation}
\textrm{RMSE} = \sqrt{\frac{1}{n}\sum_{i=1} ^{n} (f^i_{\rm PSF} [\textrm{injected}] - f^i_{\rm PSF} [\textrm{predicted}])^2}\;,
\end{equation}
which measures the average difference between the predicted values and the actual injected values. 
The RAE is the ratio between the absolute error divided by the real value,
\begin{equation}
\textrm{RAE} = \frac{|f_{\rm PSF}[\textrm{injected}] - f_{\rm PSF} [\textrm{predicted}] |}{f_{\rm PSF} [\textrm{injected}]}\;. 
\end{equation}
Note that the RAE is not well defined when the real $f_{\rm PSF}$ is equal to zero, therefore we do not calculate RAE in that case. Based on the RAE, we can also define the outlier fraction in the \texttt{Zoobot} predictions, as the fraction of galaxies that have RAE higher than a given threshold ($t_{\rm outlier}$). That is the fraction of galaxies that satisfy
\begin{equation}
{\rm RAE} >t_{\rm outlier}\;, 
\end{equation}
with the adopted thresholds being $50\%$ in this study \textbf{(i.e., the prediction error is at least $50\%$)}. 
We use \texttt{Sextractor} to determine the physical size of the simulated galaxy, as measured by the
Kron radius ($r_{\rm w}$), in kpc. To calculate the S/N, we measure the flux within an aperture of \mbox{\ang{;;0.5}} centred on the source and divide by the flux corresponding to the background noise in an aperture of the same size \textbf{in an empty region of the sky near the source.}  
 
Our trained DL model has an overall \textbf{mean value} of $\text{RMSE}=0.052$ and ${\rm RAE} = 0.30$. \textbf{In Fig.\,\ref{fig.rmse_z},  we show the RMSE and RAE of the \texttt{Zoobot} predictions as a function of the injected PSF contribution fraction, redshift, S/N and Kron radius (calculated in the $I_{\rm E}$ filter). The error bars in Fig.\,\ref{fig.rmse_z} represent the 95\% confidence interval obtained through bootstrapping.} We can see that the RMSE decreases with increasing contribution from the PSF, which is expected because a more dominant PSF can help us estimate its contribution more precisely. In fact, when  $f_{\rm PSF}$ [injected] $\gtrsim40\%$, the precision of $f_{\rm PSF} [\textrm{predicted}]$ is higher than the intrinsic variation (fractional change of about $13\%$ as shown in Fig.\,\ref{fig.psf}) in the observed \Euclid PSF (indicated by the grey shaded region). The RAE increases rapidly with decreasing  $f_{\rm PSF}$ [injected], which is expected \textbf{due to RAE being sensitive to small values of $f_{\rm PSF} [\rm injected]$}. Both the RMSE and RAE increase slowly with increasing redshift \textbf{(with a minimum value of RMSE of 0.035 at the lowest redshift bin and a maximum of 0.67 at the highest redshift bin), and remain mostly constant with S/N, probably due to the training sample having enough galaxies with low S/N}. The RMSE increases with decreasing galaxy size \textbf{(with a minimum value of 0.36 for larger galaxies and a maximum of 0.74 for the smallest ones)}, which can be explained by the fact that it is more difficult to estimate $f_{\rm PSF}$ precisely in more compact galaxies. 
 %The RMSE and the RAE increase at higher redshifts and lower $f_{\rm PSF}$. 
 %There is only a small increase in RMSE for lower galaxy sizes (in kpc) and no change in RMSE as a function of S/N. 

The overall outlier fraction is $8 \pm 1\%$, based on outliers defined as those with RAE $>50\%$. \textbf{This definition of outlier fraction is most sensitive to smaller values of the true $f_{\rm PSF}$ and can miss large absolute errors; that is why we} also adopt another definition for selecting outliers \textbf{based on the residuals, and } using the overall RMSE value of 0.052. We define as outliers those galaxies for which the difference between the predicted and true $f_{\rm PSF}$  is more than $5\,\sigma$ (i.e., $>0.26$). Based on this alternative definition, we find an overall outlier fraction of $0.43\pm0.03\%$. In Fig.\,\ref{fig.outlier_fraction} we show how the two different outlier fractions change as a function of $f_{\rm PSF}$ [injected], redshift, S/N, and galaxy size. 
\textbf{The residual-based outlier fraction generally increases with decreasing $f_{\rm PSF}$ [injected], increasing $z$, and galaxy size, while remaining relatively constant with S/N. The outlier fraction based on RAE increases more drastically with decreasing $f_{\rm PSF}$, as expected, and only slightly with increasing $z$, while remaining constant with S/N and galaxy size}. 
 %\textcolor{violet}{Add discussions on Fig. 3, 4 and 5.}

\subsection{\label{subsc:results.comparison}Comparison with other AGN selections}

We apply the trained DL model to our stellar-mass-selected sample of real \Euclid galaxies described in Sect.\,\ref{subsc:data.agn_selection}. Figure\,\ref{fig.hist_AGN_frac} shows the cumulative distribution of the estimated $f_{\rm PSF}$ for the whole sample across the EDFs. \textbf{Traditional methods typically adopt a binary AGN versus non-AGN classification, but it has been shown that, for massive galaxies, the AGN fraction depends on the sensitivity of the survey \citep{Sabater2019}. With our approach, by estimating the AGN contribution fraction, we can move beyond this simplistic binary classification. Galaxies can, instead, be classified as AGN candidates based on a specific threshold of the fractional PSF contribution. While these AGN candidates have a measurable contribution from the PSF, they need to be confirmed as AGN, since other compact central sources, such as stellar clusters or starburst regions, particularly at high redshift, may remain unresolved and contribute to the detected PSF contribution fraction in our method. We can select the most appropriate threshold for a given science case, whether we aim to focus on more dominant AGN or include galaxies with lower AGN contributions.} \textbf{First, we apply a threshold of $f_{\rm PSF} > 0.2$ to classify AGN candidates. This threshold, chosen as a conservative cut (approximately 4$\,\sigma$), is based on the overall RMSE of the model. %, as for lower injected $f_PSF$ the prediction becomes more uncertain. 
Using this criterion, our model identifies} \textbf{$48\,840\pm 78$} galaxies being classified as AGN over the entire area of the EDFs by our model, which represents \textbf{$7.8\pm0.1\%$ of our whole stellar-mass-limited sample and corresponds to $774\pm 2$ deg$^{-2}$. To estimate the number of AGN candidates, we performed Monte Carlo realisations of the $f_{\rm PSF}$ values by sampling from a Gaussian distribution, where the width of the distribution was set to the RMSE value. The mean of these realisations represents the value of the classified AGN candidates, while the standard deviation of the realisations determines the error.} We also adopt a less conservative cut at $f_{\rm PSF}> 0.1$, motivated by the fact that the mean difference between the predicted $f_{\rm PSF}$ and the true fraction is close to zero in this regime, as demonstrated in Fig.\,\ref{fig.pred_zoobot}. Adopting this cut, we find a total of \textbf{$161\,503\pm 140$ AGN in the EDFs, representing to $25.9\pm0.1\%$ of the whole stellar-mass-limited sample and corresponding to $2559\pm 3$ deg$^{-2}$}. \textbf{This highlights the power of our method in identifying an unprecedentedly large sample of AGN-hosting galaxies, spanning a wide range in the relative dominance of the central point source, by selecting more AGN candidates with a measurable AGN component than other methods we compare against, whose numbers are shown in Table \ref{tab.agn_counts}.}
Furthermore, there are \textbf{$1123\pm 11$} galaxies with $f_{\rm PSF}> 0.5$, that is to say, galaxies in which the AGN overshine the host galaxy. This number corresponds to \textbf{$18\pm 1$} of such AGN per deg$^2$.

\begin{figure}
\includegraphics[width=0.48
\textwidth]{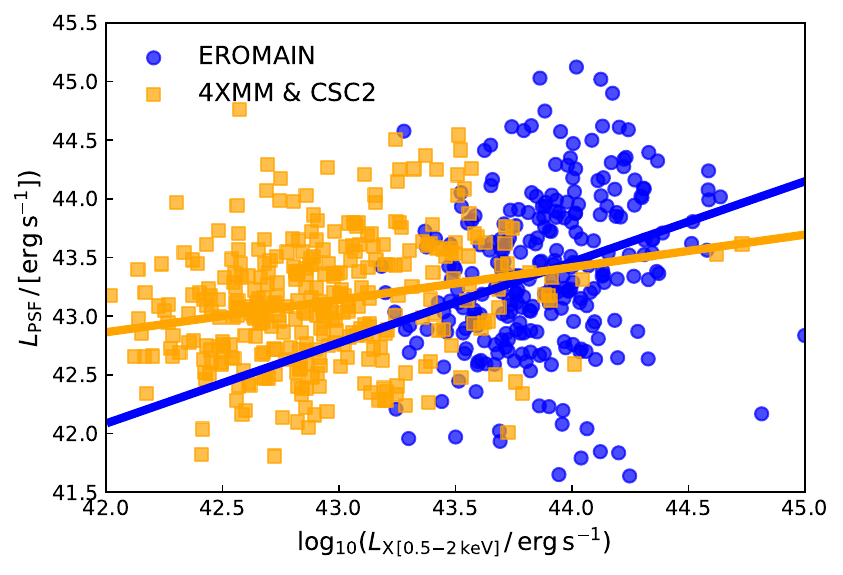}
\includegraphics[width=0.48
\textwidth]{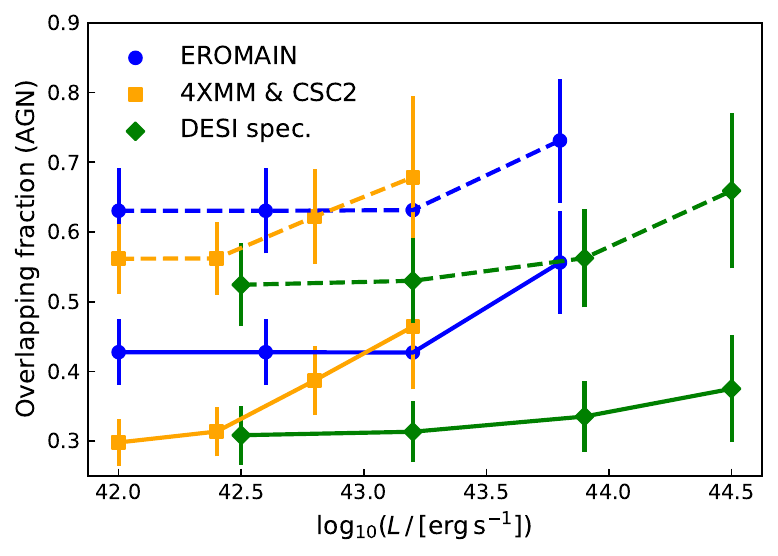}
%\includegraphics[width=0.48
%\textwidth]{Figures/frac_agns_lbol.pdf}
\caption{X-ray and bolometric luminosity relations. The top panel show the AGN luminosity $L_{\rm PSF}$. The bottom panel shows the overlapping fraction of AGN as a function of the adopted cut on the X-ray luminosity (for the X-ray sample) or the bolometric luminosity (for the DESI spectroscopic sample). The solid lines show the overlapping fraction of AGN defined by $f_{\rm PSF}> 0.2$, while the dashed lines show the fraction if we select AGN as defined by $f_{\rm PSF} > 0.1$.}
    \label{fig.AGNfrac_vs_lum}
\end{figure}

To compare our AGN based on the estimated $f_{\rm PSF}$ with AGN samples selected using other selection criteria in Sect.\,\ref{subsc:data.agn_selection}, we summarise in Table\,\ref{tab.agn_comparison} the percentage of AGN in each selection that are also selected as AGN by our DL model. Using the cut at $f_{\rm PSF}> 0.2$, 30\% of the X-ray AGN from the combined 4XMM and CSC2 surveys and 43\% of the X-ray AGN from the EROMAIN survey are also selected as AGN based on the estimated $f_{\rm PSF}$. The larger overlap with the EROMAIN AGN sample is possibly due to the fact that these X-ray AGN are more luminous than the ones from XMM and {\it Chandra} (as shown in Fig.\,\ref{fig.xray_sources}). With respect to the MIR-selected AGN, 29\% (13\%) of the AGN selected by the R90 (C75) diagnostic are also selected by our method. The smaller overlap with the C75-selected MIR AGN is consistent with the fact that this selection has a higher contamination rate compared to the R90 selection. Finally 31\% of DESI spectroscopic AGN are also identified as AGN according to our selection based on PSF fraction. \textbf{However, when considering only the QSO subclass within the DESI spectroscopic AGN, 74\% meet our selection criteria}. If we use a less conservative cut at $f_{\rm PSF}> 0.1$, then the overlapping fractions increase significantly for all three AGN selections (i.e., X-ray detection, MIR colour, and optical spectroscopy), with 28\% overlap with the C75 selected MIR AGN and 63\% overlap with the EROMAIN X-ray AGN at the two extreme ends\textbf{ (the overlap with the QSO sample increases to 87\%)}.

%The excess of these MIR AGN in galaxies with $f_{\rm PSF} > 0.2$ is also seen in Fig \ref{fig.comparison}. 
%This means that our DL-model is selecting as AGN galaxies that are confidently selected as MIR. On the other hand, only 13\% of MIR AGN, from the c75 diagnostic are also classified as such by our model, which may be a sign of the high contamination in that sample. 
%Finally only 13\% of DESI spectroscopic AGN are also identifed as AGN according to our selection based on PSF fraction. However, the fraction of DESI spectroscopic AGN that our model classifies as AGN increases with its bolometric luminosity (see Fig.\,\ref{fig.AGNfrac_vs_lum}).

In Fig.\,\ref{fig.comparison} we show the normalised distributions of the predicted $f_{\rm PSF}$ in the different AGN samples. \textbf{The X-ray-selected AGN are the most common AGN population among galaxies with higher PSF contribution fractions ($f_{\rm PSF}>0.3$)}, indicating better correspondence with optically dominant AGN \citep[which would be naturally linked to less dust-obscured AGN; see][]{Q1-SP003}. For comparison, we also plot the distribution of the predicted $f_{\rm PSF}$ in what we call the `non-AGN' sample. These `non-AGN' galaxies are, by construction, not identified as X-ray or MIR AGN in a small sky area (${\rm RA} = [51.7,\,53.7]$, ${\rm Dec} = [-29,\,-28]$) for which we have both X-ray and MIR coverage. Unfortunately DESI does not overlap with the X-ray surveys in the EDFs. We can see that while reassuringly a large fraction  (over 70\%) of the `non-AGN' have $f_{\rm PSF}<0.1$, a small fraction (around 8\%) of them do have predicted $f_{\rm PSF}>0.2$, indicating that they can in fact be AGN that are missed by the X-ray and MIR selections. In future work, we will investigate whether these AGN are picked up by other selection techniques, for example, using deeper IRAC MIR data, radio data or spectroscopic data. \textbf{ Additionally, we explore the most AGN-dominated galaxies in our sample that lack counterparts in the comparison methods. In Fig.\,\ref{fig.mass_z_dist}, we present the properties of galaxies with $f_{\rm PSF} > 0.7$ (purple histograms) and find that they tend to be less massive and fainter than AGN selected by other methods. This suggests that our approach is particularly effective at detecting AGN activity in lower-mass and fainter galaxies, where traditional selection techniques may be less sensitive. Despite their lower overall luminosity, these AGN can still exhibit a very strong central contribution in the VIS filter, as reflected by their high PSF contribution fraction. This result opens a new parameter space for studying AGN in lower-mass galaxies, providing valuable insights into SMBH growth in this regime.} 

For the X-ray-selected AGN, clearly the more luminous ones detected in EROMAIN have systematically higher $f_{\rm PSF}$ values. Similarly, for the MIR-selected AGN, the distribution corresponding to the R90 selection, which includes more secure and possibly brighter AGN than the C75 selection, is systematically skewed towards higher $f_{\rm PSF}$ values. \textbf{However, a large fraction of the X-ray-selected, MIR, and DESI spectroscopic AGN exhibit significantly lower $f_{\rm PSF}$ values, as seen in Fig.\,\ref{fig.comparison}, indicating that they are more likely to be obscured by dust. These galaxies show little to no contribution from the central point-source component in the $I_{\rm E}$ images, which explains the low predicted $f_{\rm PSF}$ values despite their classification as AGN from their respective selections. This can be seen clearly in Figs.\,\ref{fig.examples_xray_no_dl}, \ref{fig.examples_mir_no_dl}, and \ref{fig.examples_desi_no_dl}, where random examples of each AGN type with $f_{\rm PSF}<0.1$ are shown. Many of these galaxies display spiral, clumpy, or edge-on morphologies, which are expected to have higher dust content.}

%This suggests that our DL model preferentially selects as AGN the most luminous X-ray galaxies. 

\textbf{We can use the PSF contribution fraction to estimate the AGN luminosity, defined as the luminosity of the PSF component $L_{\rm PSF}$ in the \Euclid $I_{\rm E}$ filter and calculated as
\begin{equation}
    \label{eq.Lpsf}
    L_{\rm PSF}= f_{\rm PSF}\, L_{\rm total}\;, 
\end{equation}
where $L_{\rm total}$ is the total luminosity of a galaxy in the $I_{\rm E}$ filter, \textbf{estimated from the total flux derived in \cite{Q1-TP004}}. We show the relation between $L_{\rm PSF}$ and the X-ray luminosity in the top panel of Fig.\,\ref{fig.AGNfrac_vs_lum}.}
The bottom panel in Fig.\,\ref{fig.AGNfrac_vs_lum} shows the fraction of AGN identified using our method based on $f_{\rm PSF}$ as a function of the adopted cut on the X-ray luminosity $L_{\rm X}$, \textbf{for the X-ray sample, or the AGN bolometric luminosity $L_{\rm bol}$  for AGN selected in the X-ray or via DESI optical spectroscopy.} We see that the fraction of AGN selected by our model increases with increasing $L_{\rm X}$. This is expected because $L_{\rm X}$ generally correlates with the luminosity from the PSF component in the VIS images, albeit with significant scatter. For the DESI spectroscopic AGN, estimates of $L_{\rm bol}$ have been estimated through SED fitting \citep{Siudek2024}. To convert from $L_{\rm X}$ to $L_{\rm bol}$ we follow the luminosity-dependent correction presented in \citet{shenBolometricQuasarLuminosity2020}:
\begin{equation}
    \frac{L_{\rm bol}}{L_{\rm X\,[0.5\--2\,\rm{keV}]}} = c_1 \left( \frac{L_{\rm bol}}{10^{10}L_{\odot}} \right)^{k_1} + c_2 \left( \frac{L_{\rm bol}}{10^{10}L_{\odot}} \right)^{k_2}\;,
\end{equation}
where $c_1=5.712$, $k_1 = -0.026$, $c_2 = 12.60$, and $k_2=0.278$. Again we see a generally increasing fraction of AGN identified based on $f_{\rm PSF}$ with increasing $L_{\rm bol}$.

%In Fig.\,\ref{fig.comparison} we show the fraction of AGN and non-AGN according to each selection in bins of $f_{\rm PSF}$. Galaxies with higher PSF fractions are more likely X-ray AGN, particularly those detected in EROMAIN, and therefore the most luminous. This suggests that our DL model preferentially selects as AGN the most luminous X-ray galaxies. This is also observed in Fig.\,\ref{fig.AGNfrac_vs_lum}, where we see that the fraction of AGN selected by our model increases with X-ray luminosity. 

\begin{table}[]
    \centering
    \caption{Percentage of AGN from each selection method that we also identify as AGN using a cut on the PSF contribution fraction.}
    \begin{tabular}{lcc}
        \hline\hline
        \noalign{\vskip 1pt}

         & \multicolumn{2}{c}{Percentage (\%)} \\
        AGN-selection method & $f_{\rm PSF}>0.2$ & $f_{\rm PSF}>0.1$ \\
        \hline
        \noalign{\vskip 1pt}
        X-ray (All) & $35 \pm 3$ & $60 \pm 4$\\
        X-ray (4XMM \& CSC2) & $30 \pm 3$ & $56 \pm 5$\\
        X-ray (EROMAIN) & $43 \pm 5$ & $63 \pm 6$ \\
        DESI Spectroscopic & $31 \pm 4$ & $52 \pm 6$ \\
        MIR colours (C75, AllWISE) & $13 \pm 4$ & $28 \pm 1$ \\
        MIR colours (R90, AllWISE) & $29 \pm 2$ & $51 \pm 3$ \\
        \hline
    \end{tabular}
    \label{tab.agn_comparison}
\end{table}

\subsection{\label{subsc:results.sf}Dependence on stellar mass and SFR}

\begin{figure*}
    \centering
    \includegraphics[width=0.8\textwidth]{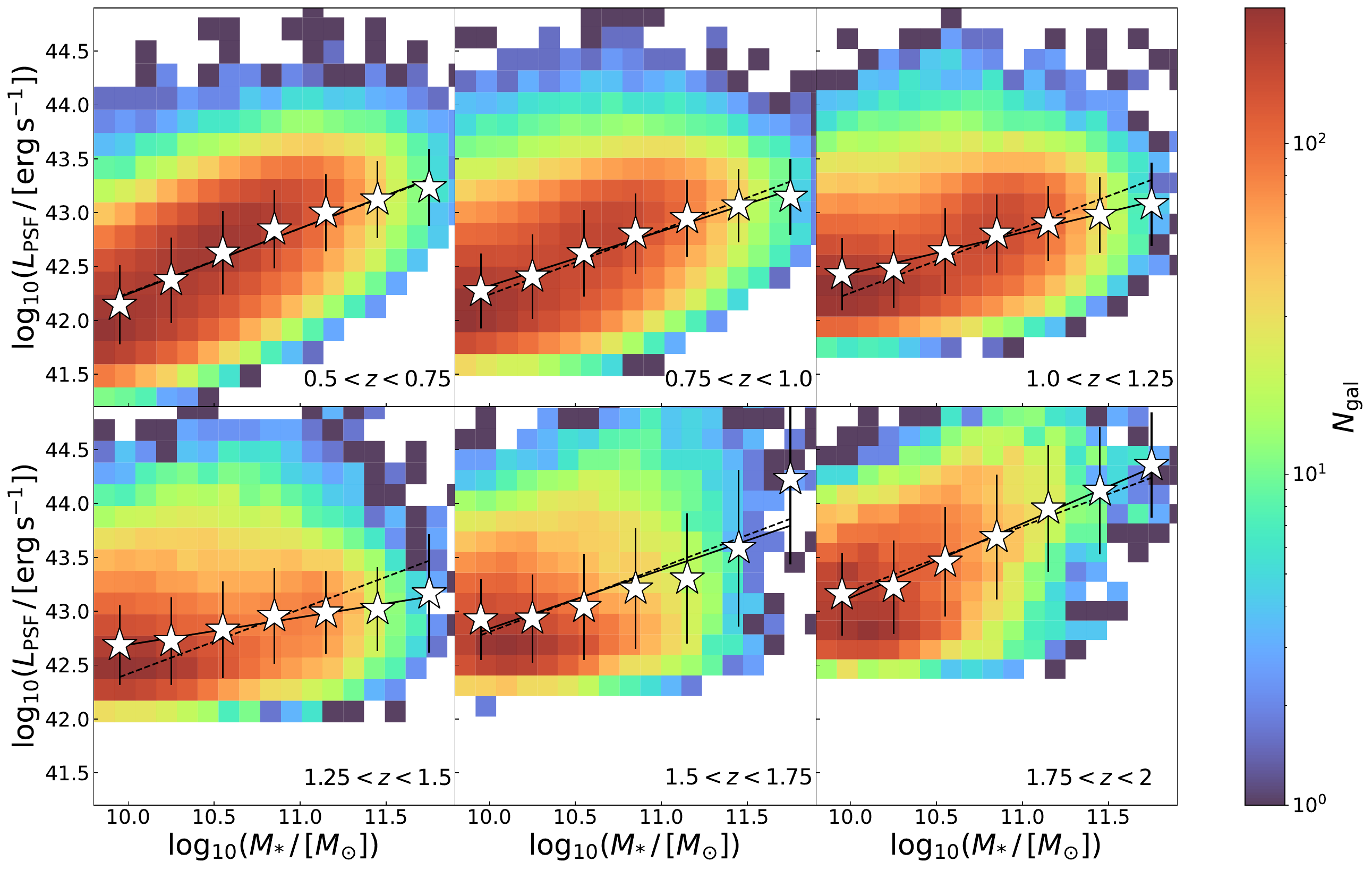}
    \caption{2D histogram of the AGN luminosity (luminosity \textbf{in the $I_{\rm E}$ filter} multiplied by $f_{\rm PSF}$) versus stellar mass, in different redshift bins. The white stars show the mean value and the error bars show the dispersion of the data. The colours of the 2D histogram show the number of points in each bin.}
    \label{fig.AGNlum_vs_mass_z}
\end{figure*}

Many previous studies have shown that the X-ray luminosity or the SMBH accretion rate increases with increasing host galaxy stellar mass across a wide range of redshift \citep[e.g.,][]{Mullaney2012, Rodighiero2015, Aird2018, Yang2018MNRAS, Carraro2020}. 
%Similar correlations can be seen between X-ray luminosity and stellar mass across a wide range of redshift \citep{ }. 
Therefore, we first analyse whether stellar mass plays a major role in determining how luminous the AGN is (as a proxy for the growth rate of the SMBH) in a galaxy, and whether there is any significant redshift evolution.

\begin{figure*}
    \centering
    \includegraphics[width=0.49\textwidth]{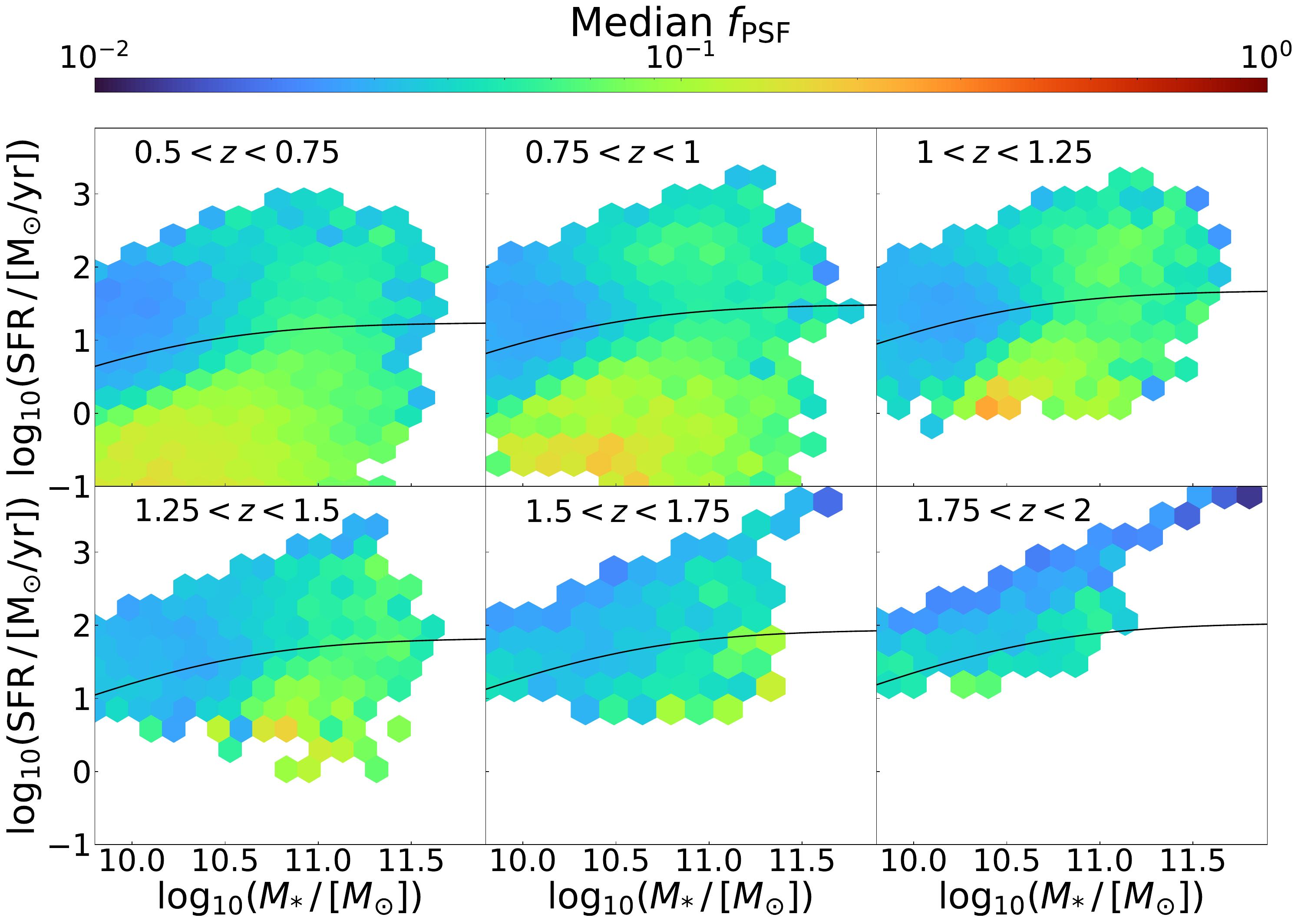}
    \includegraphics[width=0.49\textwidth]{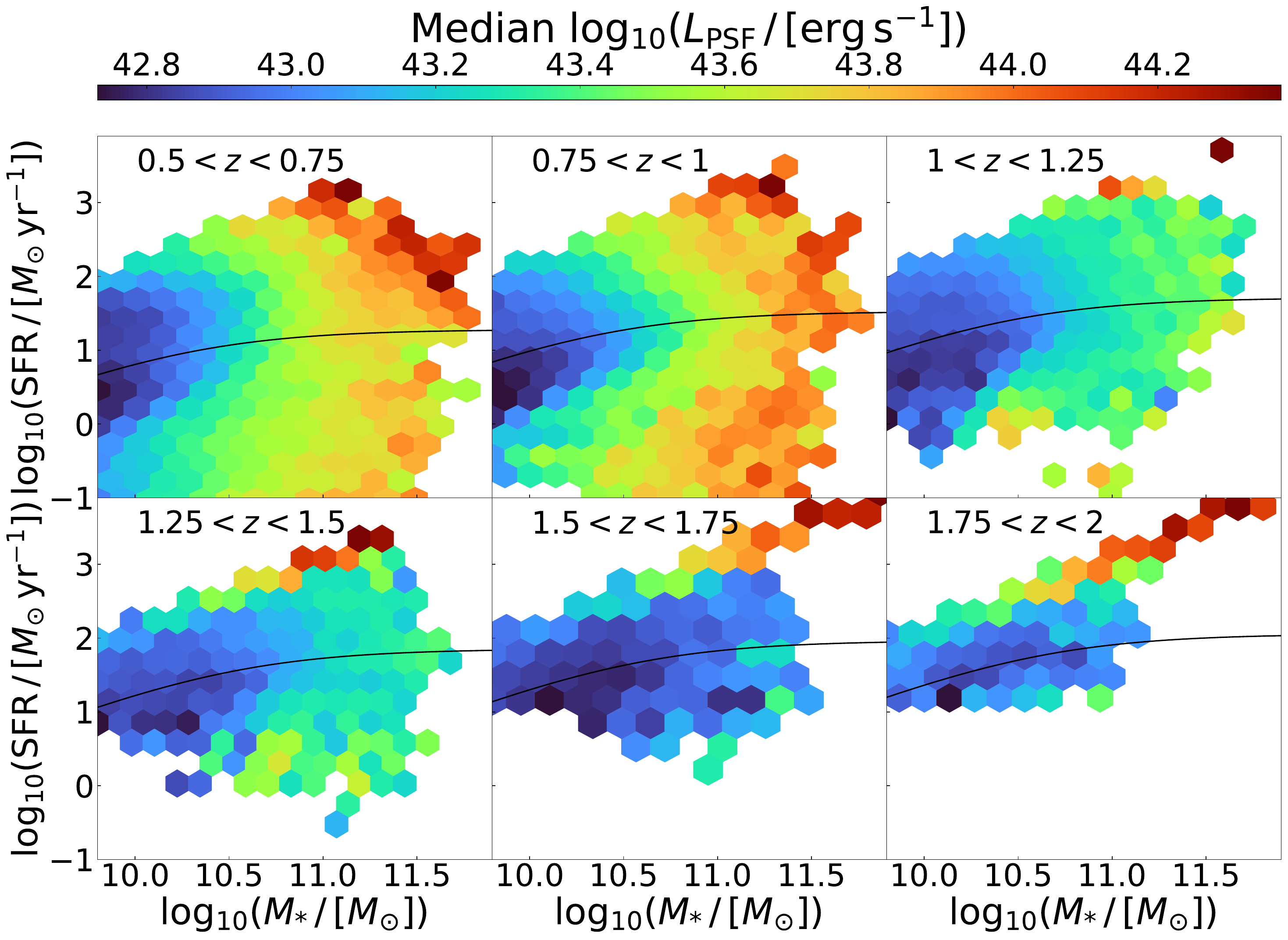}
    \caption{2D histogram of SFR versus stellar mass in different redshift bins. Each 2D bin \textbf{is colour-coded by the median value of $f_{\rm PSF}$ (left) or the median value of $\log_{10}(L_{\rm PSF})$ (right) in that bin}. The black lines show the \citet{Popesso2023} SFMS at the median redshift of the bin. Only bins with at least 10 galaxies are plotted.}
    \label{fig.sfr1}
\end{figure*}

\begin{figure}
    \centering
    \includegraphics[width=0.49\textwidth]{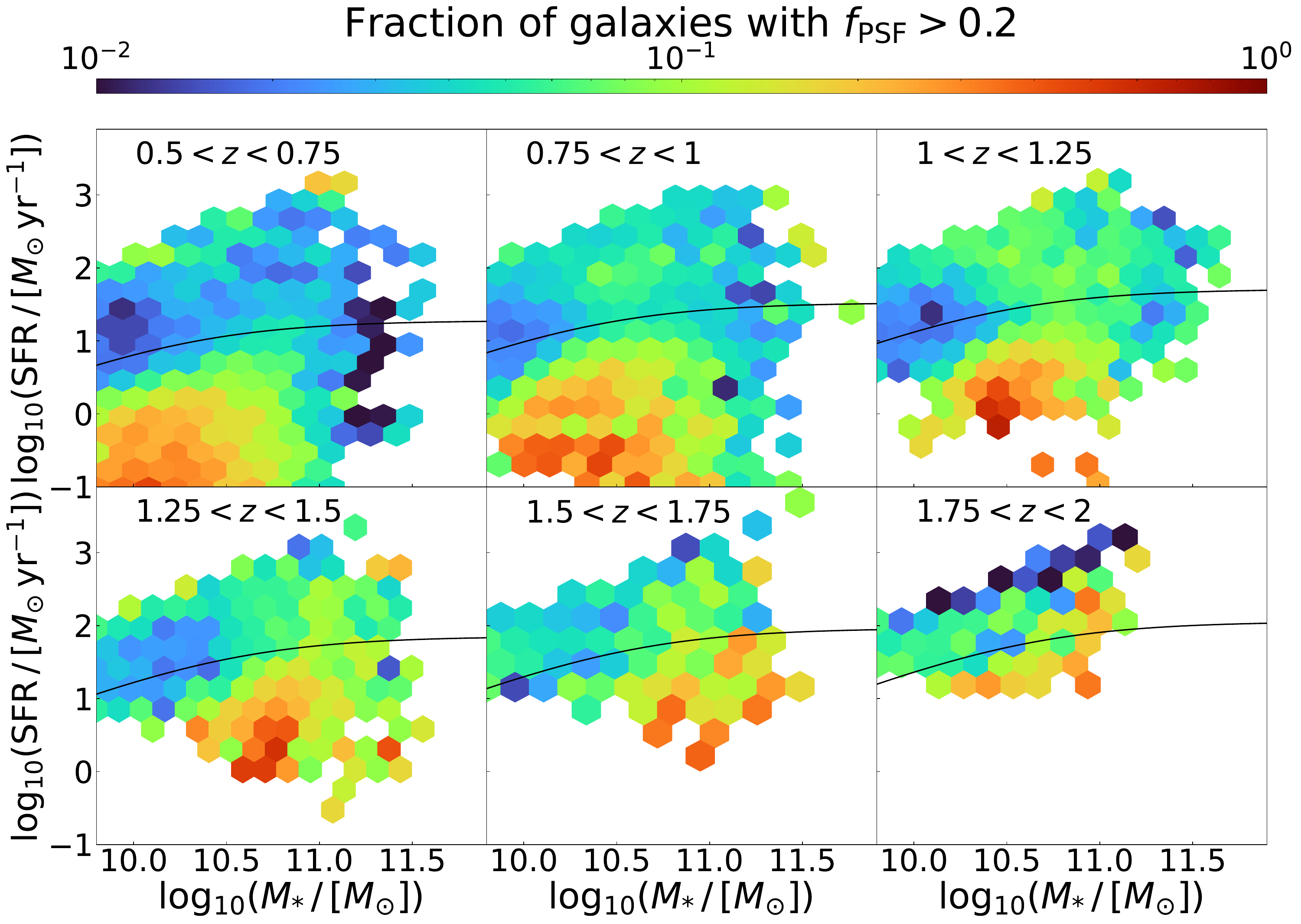}
    \includegraphics[width=0.49\textwidth]{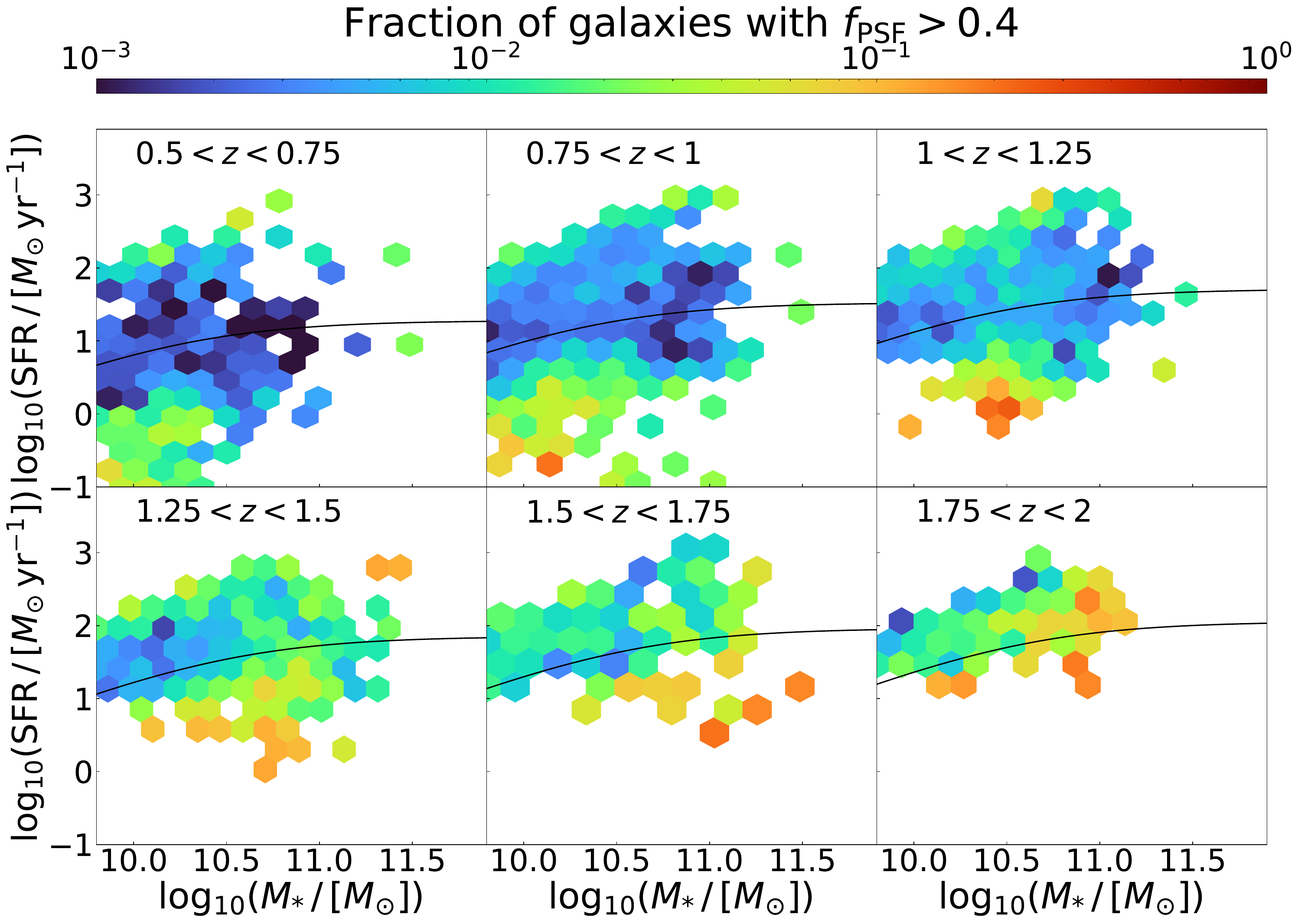}
    \includegraphics[width=0.49\textwidth]{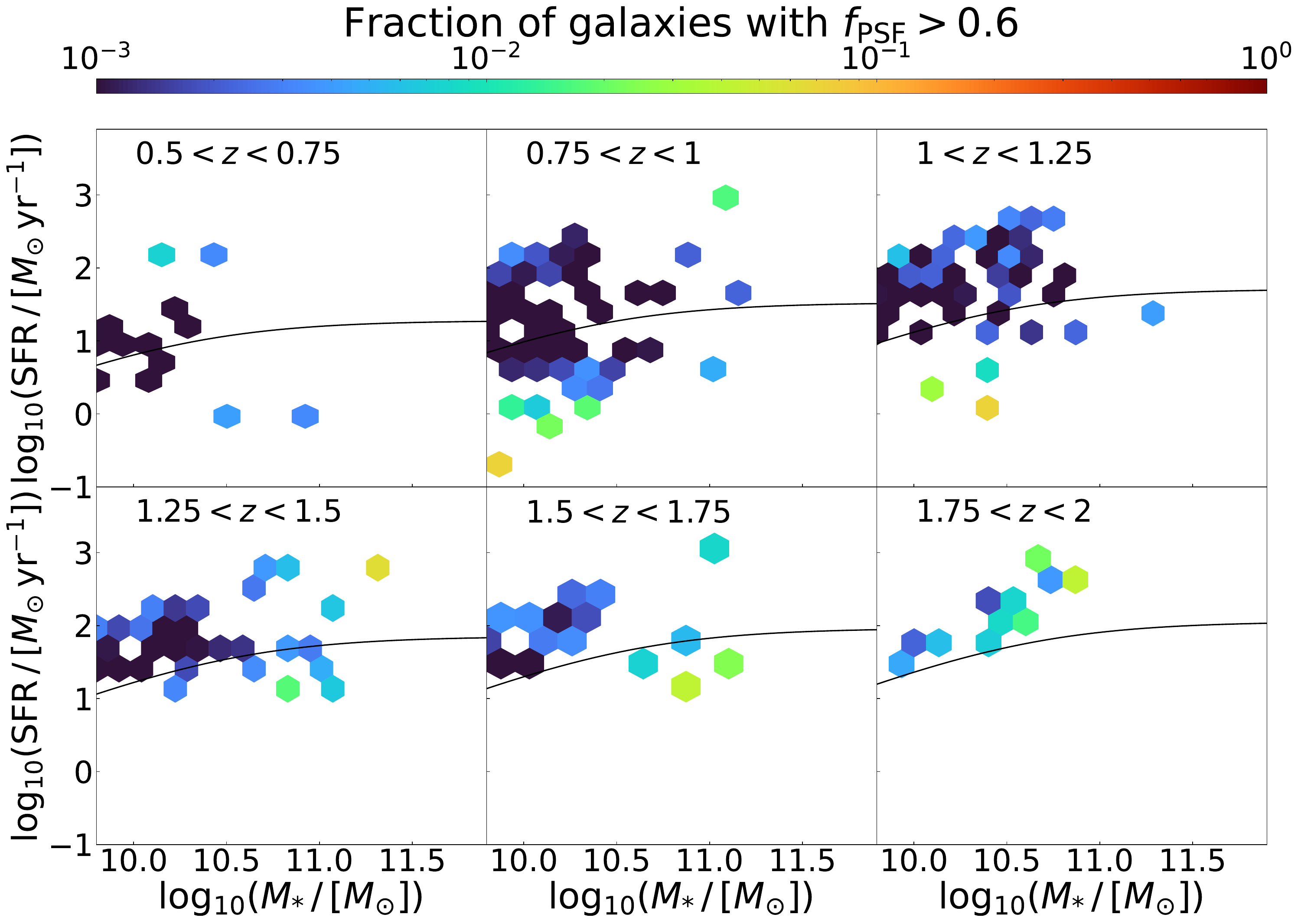}
    \caption{
    Similar to Fig.\,\ref{fig.sfr1}, but colour-coded by the fraction of galaxies with $f_{\rm PSF}>0.2$ \textbf{(top), $f_{\rm PSF}>0.4$ (centre) and $f_{\rm PSF}>0.6$ (bottom)}. %2D histogram of SFR vs. mass at different redshift bins. Each 2D bin in colour coded by the fraction of galaxies with $f_{\rm PSF}>0.2$ within that bin. The black lines show the \citep{Popesso2023} SF main sequence at the median redshift of the bin.
    }
    \label{fig.sfr2}
\end{figure}

%\begin{figure*}
%    \centering
%\includegraphics[width=0.8\textwidth]{Figures/ms_maks_n_tot_zall2.png}
 %   \caption{
 %   Similar to Fig.\,\ref{fig.sfr1}, but colour-coded by the number of galaxies with $f_{\rm PSF}>0.2$.
    %2D histogram of SFR vs. mass at different redshift bins. Each 2D bin in colour coded by the number of galaxies with  $f_{\rm PSF}>0.2$ within that bin. The black lines show the \citep{Popesso2023} SF main sequence at the median redshift of the bin.
  %  }
  %  \label{fig.sfr3}
%\end{figure*}

In Fig.\,\ref{fig.AGNlum_vs_mass_z} we show the luminosity of the AGN, which corresponds to the luminosity of the PSF component $L_{\rm PSF}$ in the \Euclid $I_{\rm E}$ filter, calculated from Eq.\,(\ref{eq.Lpsf}) as a function of stellar mass and redshift. The panels show the 2D histogram of the AGN luminosity and stellar mass in six redshift bins. At all redshifts we observe a broad positive correlation between  $L_{\rm PSF}$ and stellar mass, supporting previous claims that SMBHs generally grow faster in more massive systems. This could be due to a larger supply of gas in more massive galaxies and/or a more efficient way of transporting the gas to the central region (e.g., galaxy mergers and the presence of compact cores, which are more prevalent in galaxies with larger stellar masses). We parameterise this correlation in each redshift bin by fitting a log-log linear function ($\logten L_{\rm PSF}\,[{\rm erg\,s}^{-1}]=a\logten M_{*} [M_{\odot}] + b$) and report the best-fit parameters and their uncertainties in Table\,\ref{tab.fit}. Based on the best-fit values, we also derive another set of fits by fixing the slope $a$ at 0.61. \textbf{This choice is motivated by the fact that, at higher redshifts, detecting fainter AGN becomes progressively more challenging, as shown in Fig.\,\ref{fig.AGNlum_vs_mass_z}, where the lower boundary of the distribution shifts upward with increasing redshift. This effect could potentially lead to an artificial flattening of the slope. By fixing it to the value derived from the more complete lower-redshift bin, we ensure that any observed evolution is reflected in the normalisation rather than in a potentially biased slope.}This positive correlation between AGN luminosity and galaxy stellar mass bears similarity to the well-studied SFMS \citep[e.g.,][]{Brinchmann2004, Elbaz2007, Speagle2014, Pearson2018, Popesso2023}, suggesting that a common supply of gas could be used to fuel both the assembly of the SMBH and the host galaxy.  In addition, we also observe a mild redshift evolution in the $L_{\rm PSF}$ versus stellar mass relation, indicating SMBHs grow faster in host galaxies at the same mass at higher redshifts. This behaviour is also qualitatively similar to the observed redshift evolution of the SFMS, which could be partly explained by an increasing molecular gas fraction at higher redshifts \citep{Scoville2017, Liu2019, Tacconi2020, Wang2022}. Several studies have also shown differences in the correlation between SMBH accretion rates and host galaxy stellar mass in different galaxy types \textbf{\citep{Carraro2020, Aird2022}}. For example, star-forming galaxies are shown to have steeper slopes than quiescent galaxies. Given that the fraction of quiescent galaxies also evolves with redshift, we defer a proper characterisation of the evolution in the $L_{\rm PSF}$ versus stellar mass relation to future work when we can reliably separate different galaxy types.

%The Wang et al. 2022 reference is this https://ui.adsabs.harvard.edu/abs/2022A%26A...660A.142W/abstract

%The Liu et al. 2019 reference is this https://ui.adsabs.harvard.edu/abs/2019ApJ...887..235L/abstract

%This correlation is an indication that the most massive galaxies tend to host more luminous AGN (in the visible light), and that AGN are a relevant process particularly in high stellar mass galaxies. 

\begin{table}[]
    \centering
    \caption{Best-fit parameters for the $\logten(L_{\rm PSF}/[\rm erg\,s^{-1}])$ versus $\logten( M_*/[M_{\odot}])$ linear relation, in different redshift bins.}
    \begin{tabular}{lccc}
        \hline\hline
        \noalign{\vskip 1pt}
        Redshift range & $b$ & $a$ & $b'(a=0.61)$\\
        \hline
        \noalign{\vskip 1pt}
        $0.5<z<0.75$ & $36.1\pm0.5$ & $0.61\pm0.04$ & $36.2\pm0.2$\\
        $0.75<z<1$   & $37.2\pm0.3$ & $0.51\pm0.03$ & $36.2\pm0.3$\\
        $1<z<0.25$   & $38.6\pm0.2$ & $0.38\pm0.02$ & $36.3\pm0.6$\\
        $1.25<z<1.5$ & $40.1\pm0.2$ & $0.25\pm0.02$ & $36.4\pm0.8$\\
        $1.5<z<1.75$ & $37.4\pm1.2$ & $0.55\pm0.12$ & $36.8\pm0.6$ \\
        $1.75<z<1$   & $36.2\pm0.4$ & $0.69\pm0.04$ & $37.2\pm0.3$ \\
        \hline
    \end{tabular}
    \label{tab.fit}
\end{table}

%In Fig.\,\ref{fig.AGNf_vs_mass_z} we show the fraction of AGN as selected by our model (with $f_{\rm PSF}> 0.2$) as a function of stellar mass, and for three redshift bins. We observe that there is a peak in the fraction of AGN at M=11 for $1<z<2$, and that peak shifts to lower masses (M=10.7) at lower redshifts. In general, for a given stellar mass, the fraction of AGN increases as redshift increases.

Lastly, we explore the connection between AGN identified based on the contribution of the PSF component and their location in the SFR versus stellar mass plane. Figure\,\ref{fig.sfr1} shows the 2D histogram of SFR versus stellar mass, in different redshift bins, colour-coded by the median $f_{\rm PSF}$ (left panel) and the median $\logten L_{\rm PSF}$. Only bins with at least 10 galaxies are plotted. The \cite{Popesso2023} SFMS is also over-plotted in all redshift bins. We can see that, in terms of the relative \textbf{contribution fraction} of the AGN (as characterised by $f_{\rm PSF}$), quiescent galaxies (i.e., galaxies \textbf{significantly offset below the SFMS}) or galaxies at the high-mass end (across the full range in the specific SFR, i.e., SFR divided by stellar mass) tend to be more dominated by their AGN. In terms of the absolute power of the AGN (as characterised by $L_{\rm PSF}$), the starburst galaxies (i.e., galaxies above the SFMS) or very massive galaxies tend to host the most powerful AGN. The observation that starburst galaxies can host very powerful AGN might indicate that rapid build-up of the SMBH and the host galaxy can occur concomitantly. \textbf{Galaxy mergers provide a possible pathway for this coevolution by funnelling gas into the central regions, triggering intense star formation while also fuelling the SMBH’s growth.} 
In Fig.\,\ref{fig.sfr2}, we plot the 2D histogram of SFR versus stellar mass, colour-coded by the fraction of galaxies with $f_{\rm PSF}>0.2$ \textbf{(top panel), $f_{\rm PSF}>0.4$ (middle panel) and $f_{\rm PSF}>0.6$ (bottom panel)}. 
%The most luminous AGN tend to reside in the most massive galaxies, along the SFMS or above (starburst regime) particular at $z>1$. 
We can see that while a larger fraction of the quiescent galaxies host AGN (as defined by $f_{\rm PSF}>0.2$ in this work) compared to the star-forming galaxy population, this is not the case at the highest redshift bins (where the fraction of quiescent galaxies is very small). \textbf{This could indicate that, while at high redshift, star-forming galaxies and their central SMBHs grow together, at lower redshifts this co-evolution weakens, and AGN instead play a role in the quenching of galaxies by suppressing star formation. We observe the same trend with different thresholds of $f_{\rm PSF}$, as shown by the different panels in Fig.\,\ref{fig.sfr2}. A cautionary note to consider is that stellar masses and SFRs are currently estimated without accounting for the potential contribution of the AGN component in a galaxy. Future work should focus on deriving physical properties that incorporate the AGN contribution for more accurate results.} 
%In terms of absolute numbers, most galaxies with $f_{\rm PSF} > 0.2$ can be found along the SFMS or above. We also show another version of Fig.\,\ref{fig.sfr2} in the appendix by focusing galaxies with a dominant AGN, i.e., $f_{\rm PSF} > 0.5$.

\section{\label{sc:Conc} Conclusions}

We have presented a DL-based image decomposition method to quantify the AGN contribution, which is calculated as the contribution of the point-source component ($f_{\rm PSF}$) in galaxy imaging data. We trained the DL model with a large sample of mock galaxy images, produced from the TNG simulations to mimic the \Euclid VIS observations and with artificially injected AGN, in the form of varying $f_{\rm PSF}$. 
%The DL model is then trained to calculated the level of the injected PSF. 
%Our model perform extremely well, with a RMSE of 0.052 in the test set. 
We applied the trained model to estimate $f_{\rm PSF}$ in a stellar-mass-limited sample of galaxies selected from the \Euclid Q1 data. Our main findings are the following.

\begin{itemize}
    \item The DL model trained on the mock data is able to recover the intrinsic contribution of the PSF, with high precision and accuracy. The mean difference between the true and predicted $f_{\rm PSF}$ is $-0.0078$. The overall \textbf{mean RMSE} and RAE are $0.052$ and $0.30$, respectively. The outlier fraction defined as RAE $>50\%$ (difference $>5\,\sigma$) is $8.0\pm0.1\%$ ($0.43\pm0.03\%$). In addition, when the intrinsic $f_{\rm PSF}$ is $>40\%$, the precision of our method exceeds the level of the intrinsic variation in the observed \Euclid VIS PSF.
    \item Based on the estimated AGN contribution, \textbf{$7.8 \pm 0.1 \%$} galaxies can be classified as AGN in the \Euclid sample across the EDFs, if we impose a condition of $f_{\rm PSF} > 0.2$. By adopting a less conservative threshold of $f_{\rm PSF} > 0.1$, we can identify a total of \textbf{$25.9 \pm 0.1 \%$} AGN. Because our DL-based method can select AGN even if the AGN component is not the main contributor to the luminosity of the host galaxy,  
    %We find $48\,085$ AGN in the \Euclid Q1 data based on this criteria. 
    this technique gives many more AGN compared to the other AGN-selection methods explored in this study. In addition, we can go beyond a simple binary AGN or non-AGN classification by quantifying the contribution of the AGN.
    \item We compare our AGN sample selected using cuts on $f_{\rm PSF}$ with other commonly used AGN selections, based on X-ray detections, MIR colours, and optical spectroscopy. We find that 13--43\% of the AGN (depending on the specific selection technique) are also selected as AGN by our criterion ($f_{\rm PSF} > 0.2$). The overlap increases to 28--63\% when we select our AGN using a less conservative criterion $f_{\rm PSF} > 0.1$. In addition, we find that the overlap increases with increasing X-ray luminosity \textbf{(for the X-ray AGN) }or bolometric luminosity of the AGN \textbf{(for the DESI spectroscopic AGN)}.
    \item Galaxies with larger stellar masses tend to host more luminous AGN (i.e., a more luminous point source), indicating faster growth of the SMBH in more massive systems. The correlation also seems to evolve mildly with redshift, with AGN becoming more luminous at higher redshifts. 
    \textbf{\item We find that quiescent galaxies are more likely to host AGN (as determined by our DL method) compared to star-forming galaxies, particularly at lower redshifts, with a stronger dominance of the AGN in terms of its contribution to the total observed light. This suggests that the presence of AGN is closely linked to the quenching process in galaxy evolution.}
    \textbf{\item The most massive and starbursting galaxies host the most luminous AGN, suggesting that these galaxies undergo a phase of intense SMBH growth alongside starburst activity. Additionally, the higher number of galaxies with $f_{\rm PSF}>0.2$ above and along the SFMS suggests that many star-forming galaxies and starbursts undergo a crucial AGN phase in their evolutionary path, highlighting the interplay between galaxy formation, starburst activity, and black hole growth.}

    %On the other hand, the most massive galaxies as well as the starburst galaxies host the most luminous AGN. In terms of absolute numbers, there are more galaxies with $f_{\rm PSF}>0.2$ above and along the SFMS, indicating that many star-forming galaxies and starbursts go through an AGN phase.
    %the mean PSF fraction tends to be higher in the region occupied by quiescent galaxies. The fraction of galaxies with PSF contribution $>0.2$ also is higher in the quiescent population. Quiescent galaxies are more dominated by galaxies hosting AGN. 

\end{itemize}

In future work, we will extend this DL-based approach to higher redshifts and the \Euclid NISP bands, \textbf{for which the model can easily be adapted and can output $f_{\rm PSF}$ predictions of thousands of galaxies in a few seconds, making it an ideal method for future data releases from \Euclid.} With the future data releases covering significantly larger areas, we will also extend the comparison of our AGN sample with other AGN-selection techniques, such as radio-selected AGN, \Euclid type I and type II AGN, and variability-selected AGN. \textbf{This will also allow us to better investigate galaxies with high $f_{\rm PSF}$ that are not classified as AGN by any of the methods presented here. By examining whether these galaxies are detected through alternative selection techniques or remain uniquely identified by our approach, we can explore the nature of this potential new population of AGN candidates and assess their role in galaxy evolution. A complementary approach would be to follow up a subset of these sources at higher resolution to determine whether some fraction of the PSF contribution originates from compact galactic cores rather than AGN. Such an investigation would provide further insight into the nature of these objects.} In the current work, our estimates of the photo-$z$s and galaxy physical properties such as stellar mass and SFR are not optimised for galaxies with a dominant AGN. \textbf{For future analysis,} using our method, we can remove the contribution of the AGN in the photometric bands and then carry out SED fitting using the decomposed photometric measurements to obtain more reliable photo-$z$ and physical property estimates. Consequently, we can properly study the co-evolution of the growth of the SMBHs and their host galaxies in different galaxy populations (i.e., star-forming galaxies along the SFMS, starburst galaxies and quiescent galaxies) and how the relative pace of the two assembly histories evolve with cosmic time.
%. Derivation of AGN bolometric luminosity and SMBH accretion rates. With the statistical power of \Euclid, study the relative growth of the SMBH and host galaxies, via ratio of BHAR to SFR as a function of stellar mass and host it evolves with cosmic time.
%
% Add the acknowledgement using the achnowledgements environment.
% Do not use \acknowledgement{....} as this affects the formatting
% of the references.
%

\begin{acknowledgements}
%\AckERO  
\AckEC  
\AckQone
\AckDatalabs
Based on data from UNIONS, a scientific collaboration using three Hawaii-based telescopes: CFHT, Pan-STARRS, and Subaru (\url{www.skysurvey.cc}\,).

Based on data from the Dark Energy Camera (DECam) on the Blanco 4-m Telescope at CTIO in Chile (\url{https://www.darkenergysurvey.org}\,).

Based on data from the ESA mission {\it Gaia}, whose data are being processed by the Gaia Data Processing and Analysis Consortium (\url{https://www.cosmos.esa.int/gaia}\,).

This publication is part of the project ``Clash of the Titans: deciphering the enigmatic role of cosmic collisions'' (with project number VI.Vidi.193.113 of the research programme Vidi, which is (partly) financed by the Dutch Research Council (NWO). 
This research was supported by the International Space Science Institute (ISSI) in Bern, through ISSI International Team project \#23-573 "Active Galactic Nuclei in Next Generation Surveys".
We thank the Center for Information Technology of the University of Groningen for their support and for providing access to the Hábrók high-performance computing cluster. We thank SURF (\url{www.surf.nl}) for the support in using the National Supercomputer Snellius. \end{acknowledgements}

%
% Here comes the reference list, generated via bibtex from
% your bibfile my.bib and Euclid.bib. Please make sure that
% the same paper is not referenced twice, one from your my.bib
% file, and once from Euclid.bib.
%

\bibliography{Euclid, Euclid_Q1_AGN, Q1}

%
% Now you can add appendices.
% In this example, the appendices are in one column mode.
% If that is not requires, comment out \onecolumn
% Note that appendices in A\&A come {\it after\/} the references.

\begin{appendix}
  \onecolumn %If you don't want single column for the Appendix, please
             %comment this out
  
\section{AGN samples - additional information}

\begin{figure*}[h]
    \centering
    \includegraphics[width=0.47\linewidth]{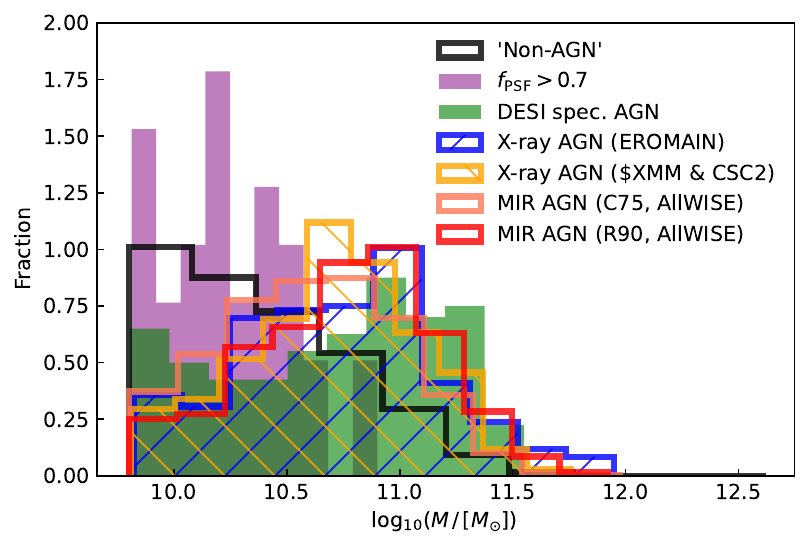}
    \includegraphics[width=0.47\linewidth]{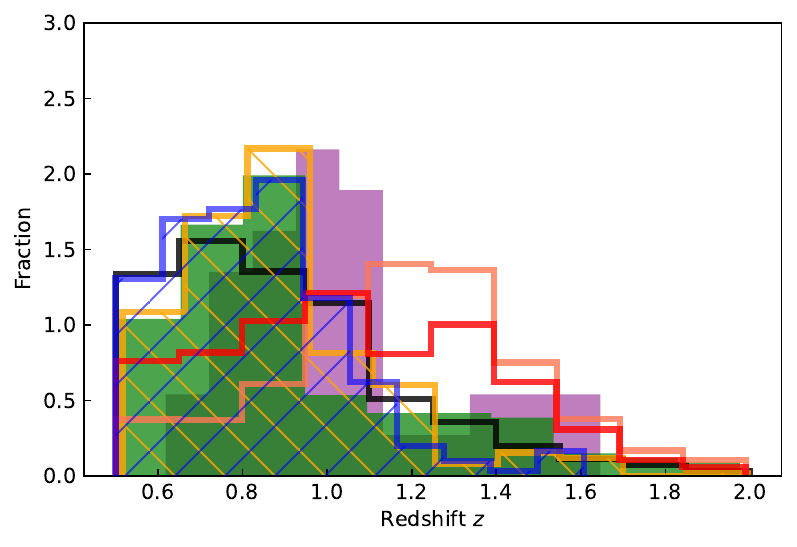}
    \includegraphics[width=0.47\linewidth]{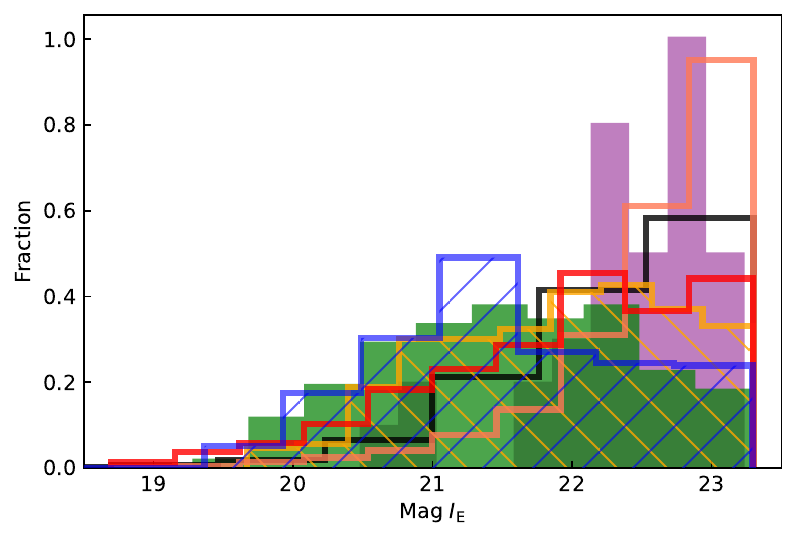}
    \includegraphics[width=0.47\linewidth]{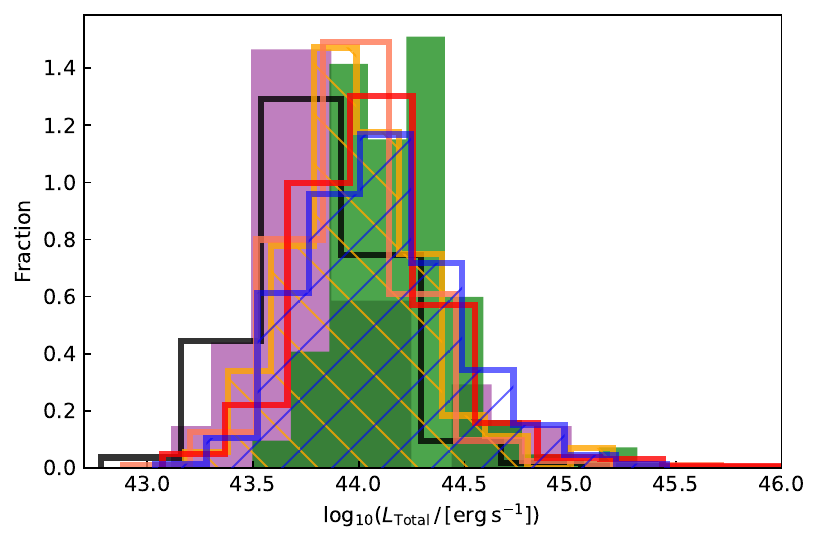}
    \caption{Normalised distribution of stellar mass (top left), redshift (top right), magnitude (bottom left), and total luminosity (bottom right) for the different types of AGN.}
    \label{fig.mass_z_dist}
\end{figure*}

We show in Fig.\,\ref{fig.mass_z_dist} the distribution of stellar mass, redshift, magnitude in the $I_{\rm E}$ filter and total luminosity of the galaxy in the $I_{\rm E}$ filter (estimated from the total flux derived in \citealt{Q1-TP004}) for the sample of AGN from the various selections (from X-ray, MIR colours, and DESI spectroscopy). \textbf{The black histograms represent the samples we refer to as `non-AGN' (explained in Sec.\,\ref{subsc:data.agn_selection}). The purple histograms depict the sample of galaxies with $f_{\rm PSF} >0.7$ that are not classified as AGN by any other method.}

\begin{figure*}[h]
    \centering
    \includegraphics[width=1\linewidth]{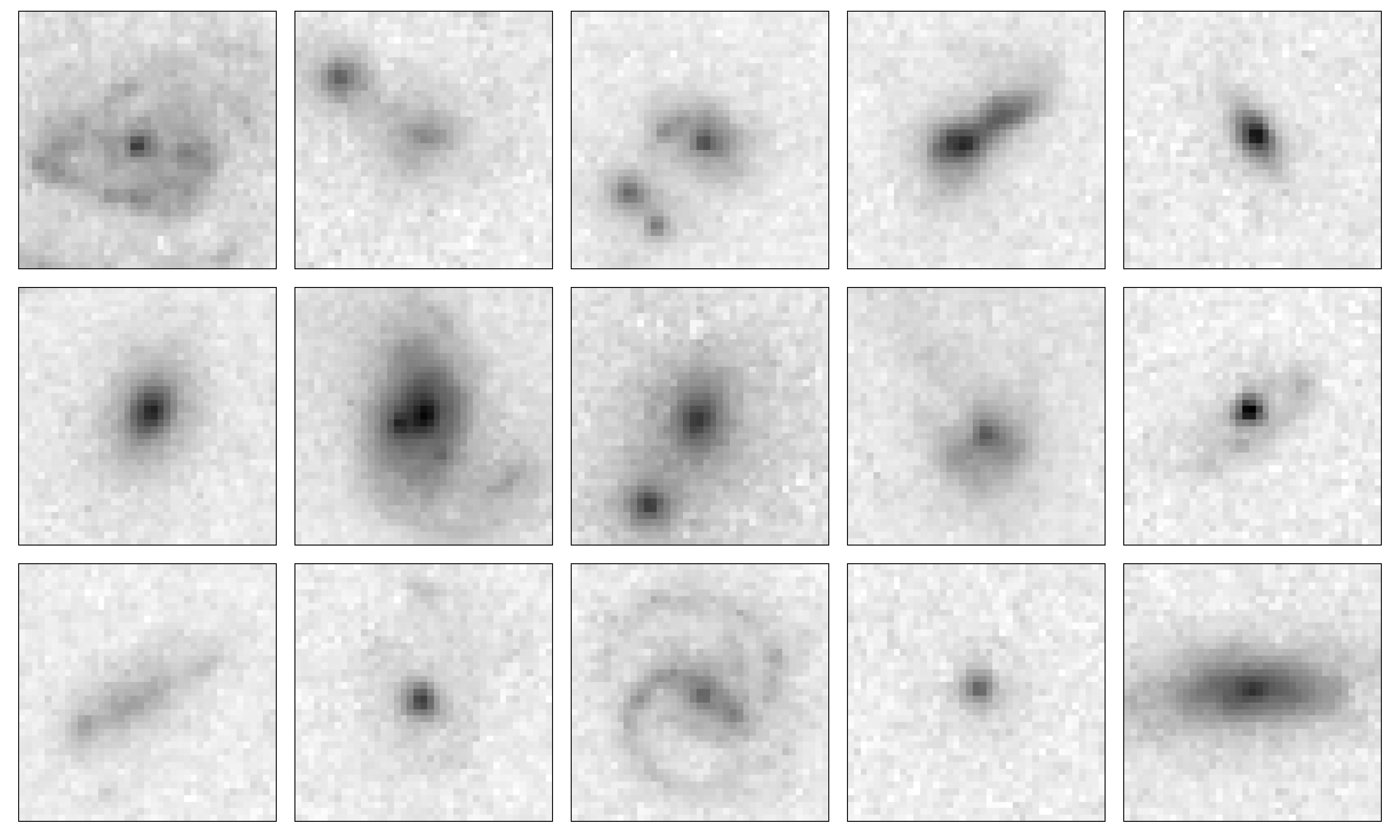}
    \caption{Example of X-ray AGN with $f_{\rm PSF}<0.1$. These images correspond to a physical size of around 25 kpc and
are displayed with an inverse arcsinh scaling.}\label{fig.examples_xray_no_dl}
\end{figure*}

\begin{figure*}[h]
    \centering
    \includegraphics[width=1\linewidth]{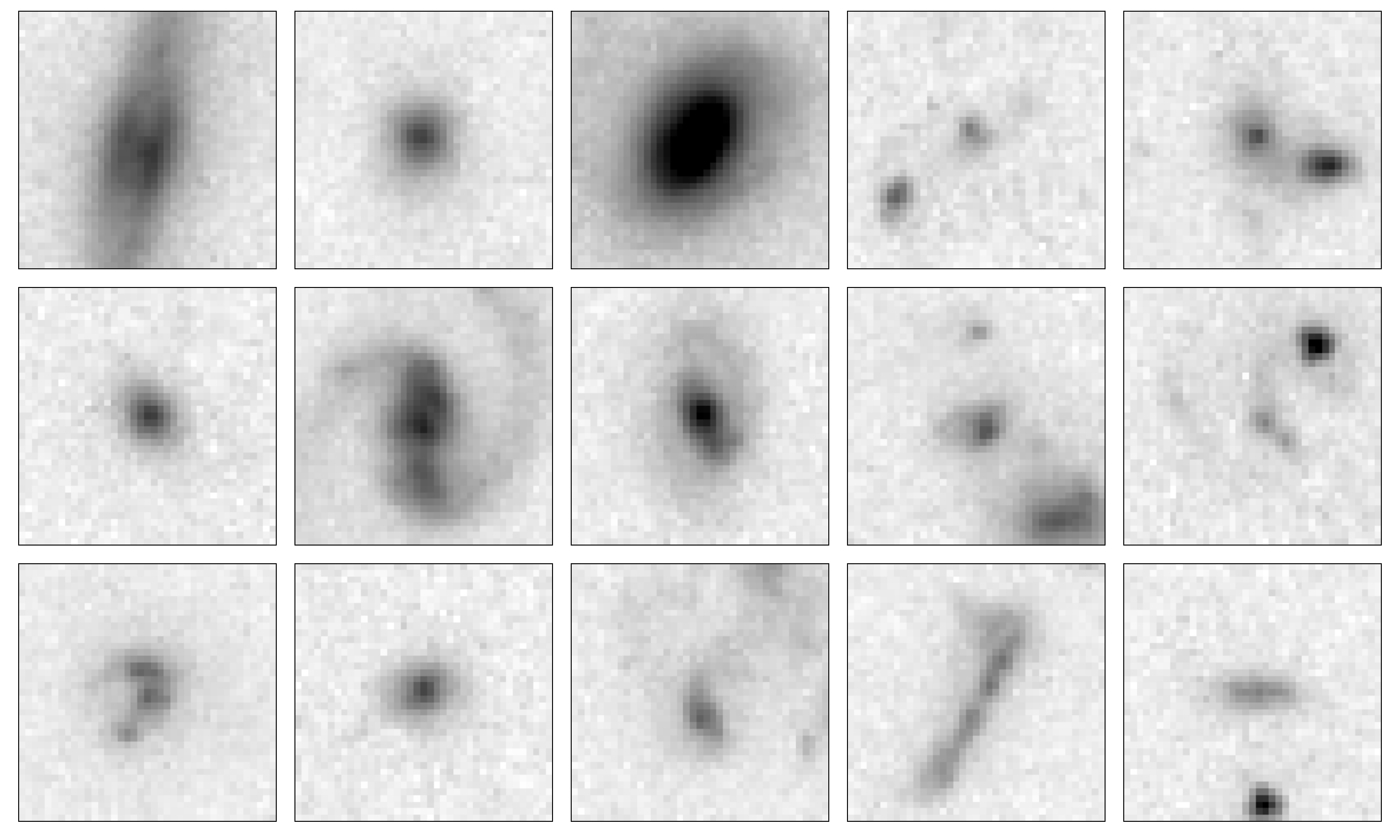}
    \caption{Example of MIR AGN with $f_{\rm PSF}<0.1$. These images correspond to a physical size of around 25 kpc and
are displayed with an inverse arcsinh scaling.}\label{fig.examples_mir_no_dl}
\end{figure*}

\begin{figure*}[h]
    \centering
    \includegraphics[width=1\linewidth]{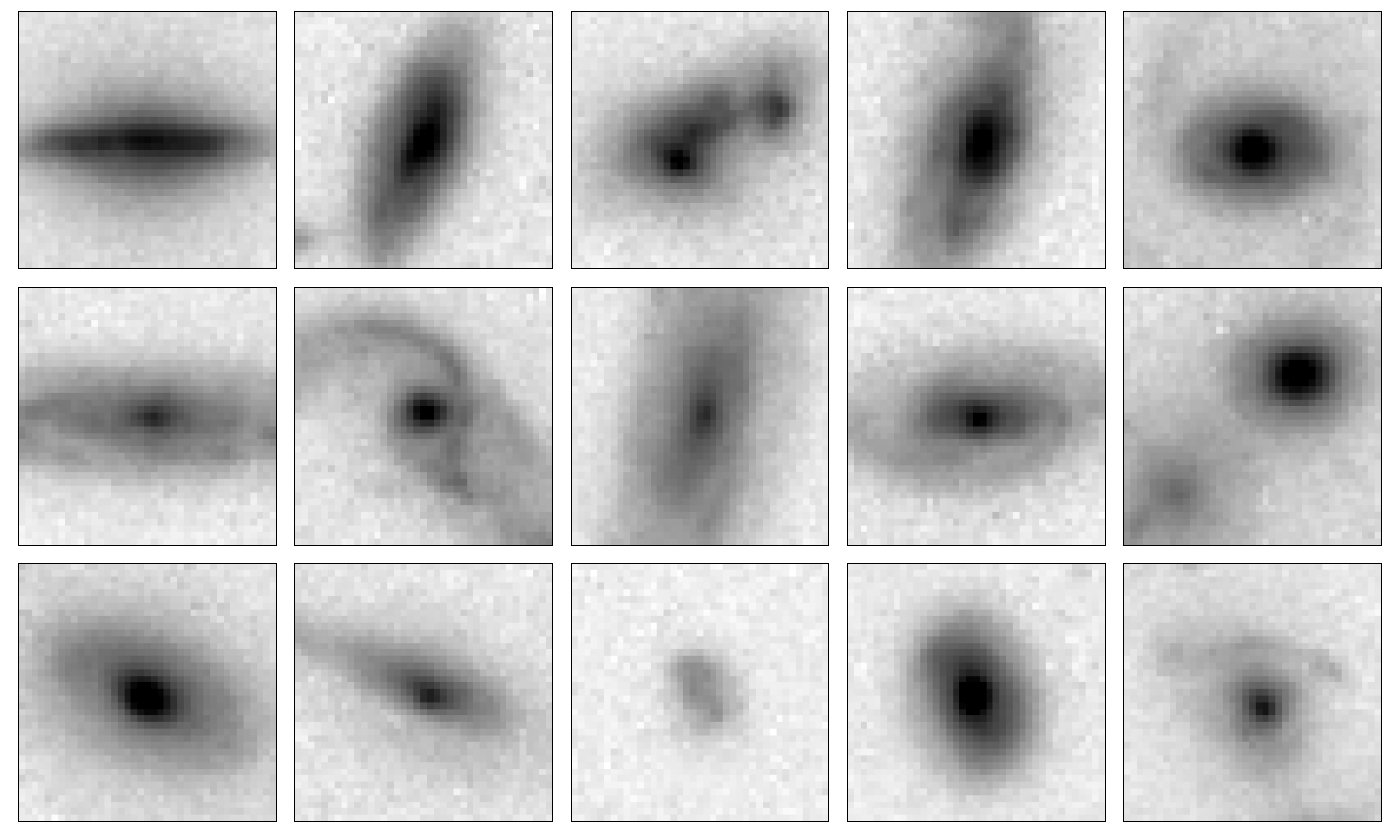}
    \caption{Example of DESI spectroscopic AGN with $f_{\rm PSF}<0.1$. These images correspond to a physical size of around 25 kpc and
are displayed with an inverse arcsinh scaling.}\label{fig.examples_desi_no_dl}
\label{LastPage}
\end{figure*}

Figures \ref{fig.examples_xray_no_dl}, \ref{fig.examples_mir_no_dl}, and \ref{fig.examples_desi_no_dl} show random examples of galaxies selected as X-ray, MIR, and DESI spectroscopic AGN, respectively, for which our DL model predicts low values of PSF contribution fraction ($f_{\rm PSF}<0.1$).

\end{appendix}

\end{document}

%% file: authors.tex
%%%% Version Sunday 16th of March 2025 09:19:27 AM UT												
%%%% Please do not edit the author list -- contact ECEB Bureau for changes
\newcommand{\orcid}[1]{} %% if already defined in aa.cls: comment, or use renewcommand			   
\author{Euclid Collaboration: B.~Margalef-Bentabol\orcid{0000-0001-8702-7019}\thanks{\email{B.Margalef.Bentabol@sron.nl}}\inst{\ref{aff1}}
\and L.~Wang\orcid{0000-0002-6736-9158}\inst{\ref{aff1},\ref{aff2}}
\and A.~La~Marca\orcid{0000-0002-7217-5120}\inst{\ref{aff1},\ref{aff2}}
\and V.~Rodriguez-Gomez\orcid{0000-0002-9495-0079}\inst{\ref{aff3}}
\and A.~Humphrey\orcid{0000-0002-0510-2351}\inst{\ref{aff4},\ref{aff5}}
\and S.~Fotopoulou\orcid{0000-0002-9686-254X}\inst{\ref{aff6}}
\and F.~Ricci\orcid{0000-0001-5742-5980}\inst{\ref{aff7},\ref{aff8}}
\and Y.~Toba\orcid{0000-0002-3531-7863}\inst{\ref{aff9},\ref{aff10},\ref{aff11}}
\and G.~Stevens\orcid{0000-0002-8885-4443}\inst{\ref{aff6}}
\and M.~Mezcua\orcid{0000-0003-4440-259X}\inst{\ref{aff12},\ref{aff13}}
\and W.~Roster\orcid{0000-0002-9149-6528}\inst{\ref{aff14}}
\and J.~H.~Knapen\orcid{0000-0003-1643-0024}\inst{\ref{aff15},\ref{aff16}}
\and M.~Salvato\orcid{0000-0001-7116-9303}\inst{\ref{aff14}}
\and M.~Siudek\orcid{0000-0002-2949-2155}\inst{\ref{aff17},\ref{aff12}}
\and F.~Shankar\orcid{0000-0001-8973-5051}\inst{\ref{aff18}}
\and T.~Matamoro~Zatarain\orcid{0009-0007-2976-293X}\inst{\ref{aff6}}
\and L.~Spinoglio\orcid{0000-0001-8840-1551}\inst{\ref{aff19}}
\and P.~Dayal\orcid{0000-0001-8460-1564}\inst{\ref{aff2}}
\and J.~Petley\orcid{0000-0002-4496-0754}\inst{\ref{aff20}}
\and R.~Kondapally\orcid{0000-0001-6127-8151}\inst{\ref{aff21}}
\and N.~Aghanim\orcid{0000-0002-6688-8992}\inst{\ref{aff22}}
\and B.~Altieri\orcid{0000-0003-3936-0284}\inst{\ref{aff23}}
\and A.~Amara\inst{\ref{aff24}}
\and S.~Andreon\orcid{0000-0002-2041-8784}\inst{\ref{aff25}}
\and N.~Auricchio\orcid{0000-0003-4444-8651}\inst{\ref{aff26}}
\and H.~Aussel\orcid{0000-0002-1371-5705}\inst{\ref{aff27}}
\and C.~Baccigalupi\orcid{0000-0002-8211-1630}\inst{\ref{aff28},\ref{aff29},\ref{aff30},\ref{aff31}}
\and M.~Baldi\orcid{0000-0003-4145-1943}\inst{\ref{aff32},\ref{aff26},\ref{aff33}}
\and A.~Balestra\orcid{0000-0002-6967-261X}\inst{\ref{aff34}}
\and S.~Bardelli\orcid{0000-0002-8900-0298}\inst{\ref{aff26}}
\and P.~Battaglia\orcid{0000-0002-7337-5909}\inst{\ref{aff26}}
\and A.~Biviano\orcid{0000-0002-0857-0732}\inst{\ref{aff29},\ref{aff28}}
\and A.~Bonchi\orcid{0000-0002-2667-5482}\inst{\ref{aff35}}
\and D.~Bonino\orcid{0000-0002-3336-9977}\inst{\ref{aff36}}
\and E.~Branchini\orcid{0000-0002-0808-6908}\inst{\ref{aff37},\ref{aff38},\ref{aff25}}
\and M.~Brescia\orcid{0000-0001-9506-5680}\inst{\ref{aff39},\ref{aff40}}
\and J.~Brinchmann\orcid{0000-0003-4359-8797}\inst{\ref{aff4},\ref{aff41}}
\and S.~Camera\orcid{0000-0003-3399-3574}\inst{\ref{aff42},\ref{aff43},\ref{aff36}}
\and G.~Ca\~nas-Herrera\orcid{0000-0003-2796-2149}\inst{\ref{aff44},\ref{aff45},\ref{aff20}}
\and V.~Capobianco\orcid{0000-0002-3309-7692}\inst{\ref{aff36}}
\and C.~Carbone\orcid{0000-0003-0125-3563}\inst{\ref{aff46}}
\and J.~Carretero\orcid{0000-0002-3130-0204}\inst{\ref{aff47},\ref{aff48}}
\and S.~Casas\orcid{0000-0002-4751-5138}\inst{\ref{aff49}}
\and M.~Castellano\orcid{0000-0001-9875-8263}\inst{\ref{aff8}}
\and G.~Castignani\orcid{0000-0001-6831-0687}\inst{\ref{aff26}}
\and S.~Cavuoti\orcid{0000-0002-3787-4196}\inst{\ref{aff40},\ref{aff50}}
\and K.~C.~Chambers\orcid{0000-0001-6965-7789}\inst{\ref{aff51}}
\and A.~Cimatti\inst{\ref{aff52}}
\and C.~Colodro-Conde\inst{\ref{aff15}}
\and G.~Congedo\orcid{0000-0003-2508-0046}\inst{\ref{aff21}}
\and C.~J.~Conselice\orcid{0000-0003-1949-7638}\inst{\ref{aff53}}
\and L.~Conversi\orcid{0000-0002-6710-8476}\inst{\ref{aff54},\ref{aff23}}
\and Y.~Copin\orcid{0000-0002-5317-7518}\inst{\ref{aff55}}
\and A.~Costille\inst{\ref{aff56}}
\and F.~Courbin\orcid{0000-0003-0758-6510}\inst{\ref{aff57},\ref{aff58}}
\and H.~M.~Courtois\orcid{0000-0003-0509-1776}\inst{\ref{aff59}}
\and M.~Cropper\orcid{0000-0003-4571-9468}\inst{\ref{aff60}}
\and A.~Da~Silva\orcid{0000-0002-6385-1609}\inst{\ref{aff61},\ref{aff62}}
\and H.~Degaudenzi\orcid{0000-0002-5887-6799}\inst{\ref{aff63}}
\and G.~De~Lucia\orcid{0000-0002-6220-9104}\inst{\ref{aff29}}
\and A.~M.~Di~Giorgio\orcid{0000-0002-4767-2360}\inst{\ref{aff19}}
\and C.~Dolding\orcid{0009-0003-7199-6108}\inst{\ref{aff60}}
\and H.~Dole\orcid{0000-0002-9767-3839}\inst{\ref{aff22}}
\and F.~Dubath\orcid{0000-0002-6533-2810}\inst{\ref{aff63}}
\and C.~A.~J.~Duncan\orcid{0009-0003-3573-0791}\inst{\ref{aff53}}
\and X.~Dupac\inst{\ref{aff23}}
\and A.~Ealet\orcid{0000-0003-3070-014X}\inst{\ref{aff55}}
\and S.~Escoffier\orcid{0000-0002-2847-7498}\inst{\ref{aff64}}
\and M.~Farina\orcid{0000-0002-3089-7846}\inst{\ref{aff19}}
\and R.~Farinelli\inst{\ref{aff26}}
\and F.~Faustini\orcid{0000-0001-6274-5145}\inst{\ref{aff35},\ref{aff8}}
\and S.~Ferriol\inst{\ref{aff55}}
\and F.~Finelli\orcid{0000-0002-6694-3269}\inst{\ref{aff26},\ref{aff65}}
\and M.~Frailis\orcid{0000-0002-7400-2135}\inst{\ref{aff29}}
\and E.~Franceschi\orcid{0000-0002-0585-6591}\inst{\ref{aff26}}
\and S.~Galeotta\orcid{0000-0002-3748-5115}\inst{\ref{aff29}}
\and K.~George\orcid{0000-0002-1734-8455}\inst{\ref{aff66}}
\and B.~Gillis\orcid{0000-0002-4478-1270}\inst{\ref{aff21}}
\and C.~Giocoli\orcid{0000-0002-9590-7961}\inst{\ref{aff26},\ref{aff33}}
\and P.~G\'omez-Alvarez\orcid{0000-0002-8594-5358}\inst{\ref{aff67},\ref{aff23}}
\and J.~Gracia-Carpio\inst{\ref{aff14}}
\and B.~R.~Granett\orcid{0000-0003-2694-9284}\inst{\ref{aff25}}
\and A.~Grazian\orcid{0000-0002-5688-0663}\inst{\ref{aff34}}
\and F.~Grupp\inst{\ref{aff14},\ref{aff66}}
\and L.~Guzzo\orcid{0000-0001-8264-5192}\inst{\ref{aff68},\ref{aff25},\ref{aff69}}
\and S.~Gwyn\orcid{0000-0001-8221-8406}\inst{\ref{aff70}}
\and S.~V.~H.~Haugan\orcid{0000-0001-9648-7260}\inst{\ref{aff71}}
\and W.~Holmes\inst{\ref{aff72}}
\and I.~M.~Hook\orcid{0000-0002-2960-978X}\inst{\ref{aff73}}
\and F.~Hormuth\inst{\ref{aff74}}
\and A.~Hornstrup\orcid{0000-0002-3363-0936}\inst{\ref{aff75},\ref{aff76}}
\and P.~Hudelot\inst{\ref{aff77}}
\and K.~Jahnke\orcid{0000-0003-3804-2137}\inst{\ref{aff78}}
\and M.~Jhabvala\inst{\ref{aff79}}
\and E.~Keih\"anen\orcid{0000-0003-1804-7715}\inst{\ref{aff80}}
\and S.~Kermiche\orcid{0000-0002-0302-5735}\inst{\ref{aff64}}
\and A.~Kiessling\orcid{0000-0002-2590-1273}\inst{\ref{aff72}}
\and B.~Kubik\orcid{0009-0006-5823-4880}\inst{\ref{aff55}}
\and M.~K\"ummel\orcid{0000-0003-2791-2117}\inst{\ref{aff66}}
\and M.~Kunz\orcid{0000-0002-3052-7394}\inst{\ref{aff81}}
\and H.~Kurki-Suonio\orcid{0000-0002-4618-3063}\inst{\ref{aff82},\ref{aff83}}
\and Q.~Le~Boulc'h\inst{\ref{aff84}}
\and A.~M.~C.~Le~Brun\orcid{0000-0002-0936-4594}\inst{\ref{aff85}}
\and D.~Le~Mignant\orcid{0000-0002-5339-5515}\inst{\ref{aff56}}
\and P.~Liebing\inst{\ref{aff60}}
\and S.~Ligori\orcid{0000-0003-4172-4606}\inst{\ref{aff36}}
\and P.~B.~Lilje\orcid{0000-0003-4324-7794}\inst{\ref{aff71}}
\and V.~Lindholm\orcid{0000-0003-2317-5471}\inst{\ref{aff82},\ref{aff83}}
\and I.~Lloro\orcid{0000-0001-5966-1434}\inst{\ref{aff86}}
\and G.~Mainetti\orcid{0000-0003-2384-2377}\inst{\ref{aff84}}
\and D.~Maino\inst{\ref{aff68},\ref{aff46},\ref{aff69}}
\and E.~Maiorano\orcid{0000-0003-2593-4355}\inst{\ref{aff26}}
\and O.~Mansutti\orcid{0000-0001-5758-4658}\inst{\ref{aff29}}
\and S.~Marcin\inst{\ref{aff87}}
\and O.~Marggraf\orcid{0000-0001-7242-3852}\inst{\ref{aff88}}
\and M.~Martinelli\orcid{0000-0002-6943-7732}\inst{\ref{aff8},\ref{aff89}}
\and N.~Martinet\orcid{0000-0003-2786-7790}\inst{\ref{aff56}}
\and F.~Marulli\orcid{0000-0002-8850-0303}\inst{\ref{aff90},\ref{aff26},\ref{aff33}}
\and R.~Massey\orcid{0000-0002-6085-3780}\inst{\ref{aff91}}
\and S.~Maurogordato\inst{\ref{aff92}}
\and E.~Medinaceli\orcid{0000-0002-4040-7783}\inst{\ref{aff26}}
\and S.~Mei\orcid{0000-0002-2849-559X}\inst{\ref{aff93},\ref{aff94}}
\and M.~Melchior\inst{\ref{aff95}}
\and Y.~Mellier\inst{\ref{aff96},\ref{aff77}}
\and M.~Meneghetti\orcid{0000-0003-1225-7084}\inst{\ref{aff26},\ref{aff33}}
\and E.~Merlin\orcid{0000-0001-6870-8900}\inst{\ref{aff8}}
\and G.~Meylan\inst{\ref{aff97}}
\and A.~Mora\orcid{0000-0002-1922-8529}\inst{\ref{aff98}}
\and M.~Moresco\orcid{0000-0002-7616-7136}\inst{\ref{aff90},\ref{aff26}}
\and L.~Moscardini\orcid{0000-0002-3473-6716}\inst{\ref{aff90},\ref{aff26},\ref{aff33}}
\and R.~Nakajima\orcid{0009-0009-1213-7040}\inst{\ref{aff88}}
\and C.~Neissner\orcid{0000-0001-8524-4968}\inst{\ref{aff99},\ref{aff48}}
\and S.-M.~Niemi\inst{\ref{aff44}}
\and J.~W.~Nightingale\orcid{0000-0002-8987-7401}\inst{\ref{aff100}}
\and C.~Padilla\orcid{0000-0001-7951-0166}\inst{\ref{aff99}}
\and S.~Paltani\orcid{0000-0002-8108-9179}\inst{\ref{aff63}}
\and F.~Pasian\orcid{0000-0002-4869-3227}\inst{\ref{aff29}}
\and K.~Pedersen\inst{\ref{aff101}}
\and W.~J.~Percival\orcid{0000-0002-0644-5727}\inst{\ref{aff102},\ref{aff103},\ref{aff104}}
\and V.~Pettorino\inst{\ref{aff44}}
\and S.~Pires\orcid{0000-0002-0249-2104}\inst{\ref{aff27}}
\and G.~Polenta\orcid{0000-0003-4067-9196}\inst{\ref{aff35}}
\and M.~Poncet\inst{\ref{aff105}}
\and L.~A.~Popa\inst{\ref{aff106}}
\and L.~Pozzetti\orcid{0000-0001-7085-0412}\inst{\ref{aff26}}
\and F.~Raison\orcid{0000-0002-7819-6918}\inst{\ref{aff14}}
\and R.~Rebolo\orcid{0000-0003-3767-7085}\inst{\ref{aff15},\ref{aff107},\ref{aff16}}
\and A.~Renzi\orcid{0000-0001-9856-1970}\inst{\ref{aff108},\ref{aff109}}
\and J.~Rhodes\orcid{0000-0002-4485-8549}\inst{\ref{aff72}}
\and G.~Riccio\inst{\ref{aff40}}
\and E.~Romelli\orcid{0000-0003-3069-9222}\inst{\ref{aff29}}
\and M.~Roncarelli\orcid{0000-0001-9587-7822}\inst{\ref{aff26}}
\and H.~J.~A.~Rottgering\orcid{0000-0001-8887-2257}\inst{\ref{aff20}}
\and B.~Rusholme\orcid{0000-0001-7648-4142}\inst{\ref{aff110}}
\and R.~Saglia\orcid{0000-0003-0378-7032}\inst{\ref{aff66},\ref{aff14}}
\and Z.~Sakr\orcid{0000-0002-4823-3757}\inst{\ref{aff111},\ref{aff112},\ref{aff113}}
\and D.~Sapone\orcid{0000-0001-7089-4503}\inst{\ref{aff114}}
\and B.~Sartoris\orcid{0000-0003-1337-5269}\inst{\ref{aff66},\ref{aff29}}
\and J.~A.~Schewtschenko\orcid{0000-0002-4913-6393}\inst{\ref{aff21}}
\and P.~Schneider\orcid{0000-0001-8561-2679}\inst{\ref{aff88}}
\and T.~Schrabback\orcid{0000-0002-6987-7834}\inst{\ref{aff115}}
\and A.~Secroun\orcid{0000-0003-0505-3710}\inst{\ref{aff64}}
\and G.~Seidel\orcid{0000-0003-2907-353X}\inst{\ref{aff78}}
\and M.~Seiffert\orcid{0000-0002-7536-9393}\inst{\ref{aff72}}
\and S.~Serrano\orcid{0000-0002-0211-2861}\inst{\ref{aff13},\ref{aff116},\ref{aff12}}
\and P.~Simon\inst{\ref{aff88}}
\and C.~Sirignano\orcid{0000-0002-0995-7146}\inst{\ref{aff108},\ref{aff109}}
\and G.~Sirri\orcid{0000-0003-2626-2853}\inst{\ref{aff33}}
\and L.~Stanco\orcid{0000-0002-9706-5104}\inst{\ref{aff109}}
\and J.~Steinwagner\orcid{0000-0001-7443-1047}\inst{\ref{aff14}}
\and P.~Tallada-Cresp\'{i}\orcid{0000-0002-1336-8328}\inst{\ref{aff47},\ref{aff48}}
\and A.~N.~Taylor\inst{\ref{aff21}}
\and I.~Tereno\inst{\ref{aff61},\ref{aff117}}
\and S.~Toft\orcid{0000-0003-3631-7176}\inst{\ref{aff118},\ref{aff119}}
\and R.~Toledo-Moreo\orcid{0000-0002-2997-4859}\inst{\ref{aff120}}
\and F.~Torradeflot\orcid{0000-0003-1160-1517}\inst{\ref{aff48},\ref{aff47}}
\and I.~Tutusaus\orcid{0000-0002-3199-0399}\inst{\ref{aff112}}
\and L.~Valenziano\orcid{0000-0002-1170-0104}\inst{\ref{aff26},\ref{aff65}}
\and J.~Valiviita\orcid{0000-0001-6225-3693}\inst{\ref{aff82},\ref{aff83}}
\and T.~Vassallo\orcid{0000-0001-6512-6358}\inst{\ref{aff66},\ref{aff29}}
\and G.~Verdoes~Kleijn\orcid{0000-0001-5803-2580}\inst{\ref{aff2}}
\and A.~Veropalumbo\orcid{0000-0003-2387-1194}\inst{\ref{aff25},\ref{aff38},\ref{aff37}}
\and Y.~Wang\orcid{0000-0002-4749-2984}\inst{\ref{aff121}}
\and J.~Weller\orcid{0000-0002-8282-2010}\inst{\ref{aff66},\ref{aff14}}
\and A.~Zacchei\orcid{0000-0003-0396-1192}\inst{\ref{aff29},\ref{aff28}}
\and G.~Zamorani\orcid{0000-0002-2318-301X}\inst{\ref{aff26}}
\and F.~M.~Zerbi\inst{\ref{aff25}}
\and I.~A.~Zinchenko\orcid{0000-0002-2944-2449}\inst{\ref{aff66}}
\and E.~Zucca\orcid{0000-0002-5845-8132}\inst{\ref{aff26}}
\and V.~Allevato\orcid{0000-0001-7232-5152}\inst{\ref{aff40}}
\and M.~Ballardini\orcid{0000-0003-4481-3559}\inst{\ref{aff122},\ref{aff123},\ref{aff26}}
\and M.~Bolzonella\orcid{0000-0003-3278-4607}\inst{\ref{aff26}}
\and E.~Bozzo\orcid{0000-0002-8201-1525}\inst{\ref{aff63}}
\and C.~Burigana\orcid{0000-0002-3005-5796}\inst{\ref{aff124},\ref{aff65}}
\and R.~Cabanac\orcid{0000-0001-6679-2600}\inst{\ref{aff112}}
\and A.~Cappi\inst{\ref{aff26},\ref{aff92}}
\and D.~Di~Ferdinando\inst{\ref{aff33}}
\and J.~A.~Escartin~Vigo\inst{\ref{aff14}}
\and L.~Gabarra\orcid{0000-0002-8486-8856}\inst{\ref{aff125}}
\and M.~Huertas-Company\orcid{0000-0002-1416-8483}\inst{\ref{aff15},\ref{aff17},\ref{aff126},\ref{aff127}}
\and J.~Mart\'{i}n-Fleitas\orcid{0000-0002-8594-569X}\inst{\ref{aff98}}
\and S.~Matthew\orcid{0000-0001-8448-1697}\inst{\ref{aff21}}
\and N.~Mauri\orcid{0000-0001-8196-1548}\inst{\ref{aff52},\ref{aff33}}
\and R.~B.~Metcalf\orcid{0000-0003-3167-2574}\inst{\ref{aff90},\ref{aff26}}
\and A.~Pezzotta\orcid{0000-0003-0726-2268}\inst{\ref{aff128},\ref{aff14}}
\and M.~P\"ontinen\orcid{0000-0001-5442-2530}\inst{\ref{aff82}}
\and C.~Porciani\orcid{0000-0002-7797-2508}\inst{\ref{aff88}}
\and I.~Risso\orcid{0000-0003-2525-7761}\inst{\ref{aff129}}
\and V.~Scottez\inst{\ref{aff96},\ref{aff130}}
\and M.~Sereno\orcid{0000-0003-0302-0325}\inst{\ref{aff26},\ref{aff33}}
\and M.~Tenti\orcid{0000-0002-4254-5901}\inst{\ref{aff33}}
\and M.~Viel\orcid{0000-0002-2642-5707}\inst{\ref{aff28},\ref{aff29},\ref{aff31},\ref{aff30},\ref{aff131}}
\and M.~Wiesmann\orcid{0009-0000-8199-5860}\inst{\ref{aff71}}
\and Y.~Akrami\orcid{0000-0002-2407-7956}\inst{\ref{aff132},\ref{aff133}}
\and I.~T.~Andika\orcid{0000-0001-6102-9526}\inst{\ref{aff134},\ref{aff135}}
\and S.~Anselmi\orcid{0000-0002-3579-9583}\inst{\ref{aff109},\ref{aff108},\ref{aff136}}
\and M.~Archidiacono\orcid{0000-0003-4952-9012}\inst{\ref{aff68},\ref{aff69}}
\and F.~Atrio-Barandela\orcid{0000-0002-2130-2513}\inst{\ref{aff137}}
\and C.~Benoist\inst{\ref{aff92}}
\and K.~Benson\inst{\ref{aff60}}
\and D.~Bertacca\orcid{0000-0002-2490-7139}\inst{\ref{aff108},\ref{aff34},\ref{aff109}}
\and M.~Bethermin\orcid{0000-0002-3915-2015}\inst{\ref{aff138}}
\and L.~Bisigello\orcid{0000-0003-0492-4924}\inst{\ref{aff34}}
\and A.~Blanchard\orcid{0000-0001-8555-9003}\inst{\ref{aff112}}
\and L.~Blot\orcid{0000-0002-9622-7167}\inst{\ref{aff139},\ref{aff136}}
\and H.~B\"ohringer\orcid{0000-0001-8241-4204}\inst{\ref{aff14},\ref{aff140},\ref{aff141}}
\and S.~Borgani\orcid{0000-0001-6151-6439}\inst{\ref{aff142},\ref{aff28},\ref{aff29},\ref{aff30},\ref{aff131}}
\and M.~L.~Brown\orcid{0000-0002-0370-8077}\inst{\ref{aff53}}
\and S.~Bruton\orcid{0000-0002-6503-5218}\inst{\ref{aff143}}
\and A.~Calabro\orcid{0000-0003-2536-1614}\inst{\ref{aff8}}
\and B.~Camacho~Quevedo\orcid{0000-0002-8789-4232}\inst{\ref{aff13},\ref{aff12}}
\and F.~Caro\inst{\ref{aff8}}
\and C.~S.~Carvalho\inst{\ref{aff117}}
\and T.~Castro\orcid{0000-0002-6292-3228}\inst{\ref{aff29},\ref{aff30},\ref{aff28},\ref{aff131}}
\and F.~Cogato\orcid{0000-0003-4632-6113}\inst{\ref{aff90},\ref{aff26}}
\and A.~R.~Cooray\orcid{0000-0002-3892-0190}\inst{\ref{aff144}}
\and O.~Cucciati\orcid{0000-0002-9336-7551}\inst{\ref{aff26}}
\and S.~Davini\orcid{0000-0003-3269-1718}\inst{\ref{aff38}}
\and F.~De~Paolis\orcid{0000-0001-6460-7563}\inst{\ref{aff145},\ref{aff146},\ref{aff147}}
\and G.~Desprez\orcid{0000-0001-8325-1742}\inst{\ref{aff2}}
\and A.~D\'iaz-S\'anchez\orcid{0000-0003-0748-4768}\inst{\ref{aff148}}
\and J.~J.~Diaz\inst{\ref{aff17}}
\and S.~Di~Domizio\orcid{0000-0003-2863-5895}\inst{\ref{aff37},\ref{aff38}}
\and J.~M.~Diego\orcid{0000-0001-9065-3926}\inst{\ref{aff149}}
\and P.-A.~Duc\orcid{0000-0003-3343-6284}\inst{\ref{aff138}}
\and A.~Enia\orcid{0000-0002-0200-2857}\inst{\ref{aff32},\ref{aff26}}
\and Y.~Fang\inst{\ref{aff66}}
\and A.~G.~Ferrari\orcid{0009-0005-5266-4110}\inst{\ref{aff33}}
\and P.~G.~Ferreira\orcid{0000-0002-3021-2851}\inst{\ref{aff125}}
\and A.~Finoguenov\orcid{0000-0002-4606-5403}\inst{\ref{aff82}}
\and A.~Fontana\orcid{0000-0003-3820-2823}\inst{\ref{aff8}}
\and A.~Franco\orcid{0000-0002-4761-366X}\inst{\ref{aff146},\ref{aff145},\ref{aff147}}
\and K.~Ganga\orcid{0000-0001-8159-8208}\inst{\ref{aff93}}
\and J.~Garc\'ia-Bellido\orcid{0000-0002-9370-8360}\inst{\ref{aff132}}
\and T.~Gasparetto\orcid{0000-0002-7913-4866}\inst{\ref{aff29}}
\and V.~Gautard\inst{\ref{aff150}}
\and R.~Gavazzi\orcid{0000-0002-5540-6935}\inst{\ref{aff56},\ref{aff77}}
\and E.~Gaztanaga\orcid{0000-0001-9632-0815}\inst{\ref{aff12},\ref{aff13},\ref{aff151}}
\and F.~Giacomini\orcid{0000-0002-3129-2814}\inst{\ref{aff33}}
\and F.~Gianotti\orcid{0000-0003-4666-119X}\inst{\ref{aff26}}
\and G.~Gozaliasl\orcid{0000-0002-0236-919X}\inst{\ref{aff152},\ref{aff82}}
\and M.~Guidi\orcid{0000-0001-9408-1101}\inst{\ref{aff32},\ref{aff26}}
\and C.~M.~Gutierrez\orcid{0000-0001-7854-783X}\inst{\ref{aff153}}
\and A.~Hall\orcid{0000-0002-3139-8651}\inst{\ref{aff21}}
\and W.~G.~Hartley\inst{\ref{aff63}}
\and S.~Hemmati\orcid{0000-0003-2226-5395}\inst{\ref{aff110}}
\and C.~Hern\'andez-Monteagudo\orcid{0000-0001-5471-9166}\inst{\ref{aff16},\ref{aff15}}
\and H.~Hildebrandt\orcid{0000-0002-9814-3338}\inst{\ref{aff154}}
\and J.~Hjorth\orcid{0000-0002-4571-2306}\inst{\ref{aff101}}
\and J.~J.~E.~Kajava\orcid{0000-0002-3010-8333}\inst{\ref{aff155},\ref{aff156}}
\and Y.~Kang\orcid{0009-0000-8588-7250}\inst{\ref{aff63}}
\and V.~Kansal\orcid{0000-0002-4008-6078}\inst{\ref{aff157},\ref{aff158}}
\and D.~Karagiannis\orcid{0000-0002-4927-0816}\inst{\ref{aff122},\ref{aff159}}
\and K.~Kiiveri\inst{\ref{aff80}}
\and C.~C.~Kirkpatrick\inst{\ref{aff80}}
\and S.~Kruk\orcid{0000-0001-8010-8879}\inst{\ref{aff23}}
\and J.~Le~Graet\orcid{0000-0001-6523-7971}\inst{\ref{aff64}}
\and L.~Legrand\orcid{0000-0003-0610-5252}\inst{\ref{aff160},\ref{aff161}}
\and M.~Lembo\orcid{0000-0002-5271-5070}\inst{\ref{aff122},\ref{aff123}}
\and F.~Lepori\orcid{0009-0000-5061-7138}\inst{\ref{aff162}}
\and G.~Leroy\orcid{0009-0004-2523-4425}\inst{\ref{aff163},\ref{aff91}}
\and G.~F.~Lesci\orcid{0000-0002-4607-2830}\inst{\ref{aff90},\ref{aff26}}
\and J.~Lesgourgues\orcid{0000-0001-7627-353X}\inst{\ref{aff49}}
\and L.~Leuzzi\orcid{0009-0006-4479-7017}\inst{\ref{aff90},\ref{aff26}}
\and T.~I.~Liaudat\orcid{0000-0002-9104-314X}\inst{\ref{aff164}}
\and S.~J.~Liu\orcid{0000-0001-7680-2139}\inst{\ref{aff19}}
\and A.~Loureiro\orcid{0000-0002-4371-0876}\inst{\ref{aff165},\ref{aff166}}
\and J.~Macias-Perez\orcid{0000-0002-5385-2763}\inst{\ref{aff167}}
\and G.~Maggio\orcid{0000-0003-4020-4836}\inst{\ref{aff29}}
\and M.~Magliocchetti\orcid{0000-0001-9158-4838}\inst{\ref{aff19}}
\and F.~Mannucci\orcid{0000-0002-4803-2381}\inst{\ref{aff168}}
\and R.~Maoli\orcid{0000-0002-6065-3025}\inst{\ref{aff169},\ref{aff8}}
\and C.~J.~A.~P.~Martins\orcid{0000-0002-4886-9261}\inst{\ref{aff170},\ref{aff4}}
\and L.~Maurin\orcid{0000-0002-8406-0857}\inst{\ref{aff22}}
\and M.~Miluzio\inst{\ref{aff23},\ref{aff171}}
\and P.~Monaco\orcid{0000-0003-2083-7564}\inst{\ref{aff142},\ref{aff29},\ref{aff30},\ref{aff28}}
\and C.~Moretti\orcid{0000-0003-3314-8936}\inst{\ref{aff31},\ref{aff131},\ref{aff29},\ref{aff28},\ref{aff30}}
\and G.~Morgante\inst{\ref{aff26}}
\and C.~Murray\inst{\ref{aff93}}
\and K.~Naidoo\orcid{0000-0002-9182-1802}\inst{\ref{aff151}}
\and A.~Navarro-Alsina\orcid{0000-0002-3173-2592}\inst{\ref{aff88}}
\and S.~Nesseris\orcid{0000-0002-0567-0324}\inst{\ref{aff132}}
\and F.~Passalacqua\orcid{0000-0002-8606-4093}\inst{\ref{aff108},\ref{aff109}}
\and K.~Paterson\orcid{0000-0001-8340-3486}\inst{\ref{aff78}}
\and L.~Patrizii\inst{\ref{aff33}}
\and A.~Pisani\orcid{0000-0002-6146-4437}\inst{\ref{aff64},\ref{aff172}}
\and D.~Potter\orcid{0000-0002-0757-5195}\inst{\ref{aff162}}
\and S.~Quai\orcid{0000-0002-0449-8163}\inst{\ref{aff90},\ref{aff26}}
\and M.~Radovich\orcid{0000-0002-3585-866X}\inst{\ref{aff34}}
\and P.-F.~Rocci\inst{\ref{aff22}}
\and G.~Rodighiero\orcid{0000-0002-9415-2296}\inst{\ref{aff108},\ref{aff34}}
\and S.~Sacquegna\orcid{0000-0002-8433-6630}\inst{\ref{aff145},\ref{aff146},\ref{aff147}}
\and M.~Sahl\'en\orcid{0000-0003-0973-4804}\inst{\ref{aff173}}
\and D.~B.~Sanders\orcid{0000-0002-1233-9998}\inst{\ref{aff51}}
\and E.~Sarpa\orcid{0000-0002-1256-655X}\inst{\ref{aff31},\ref{aff131},\ref{aff30}}
\and C.~Scarlata\orcid{0000-0002-9136-8876}\inst{\ref{aff174}}
\and A.~Schneider\orcid{0000-0001-7055-8104}\inst{\ref{aff162}}
\and D.~Sciotti\orcid{0009-0008-4519-2620}\inst{\ref{aff8},\ref{aff89}}
\and E.~Sellentin\inst{\ref{aff175},\ref{aff20}}
\and L.~C.~Smith\orcid{0000-0002-3259-2771}\inst{\ref{aff176}}
\and K.~Tanidis\orcid{0000-0001-9843-5130}\inst{\ref{aff125}}
\and G.~Testera\inst{\ref{aff38}}
\and R.~Teyssier\orcid{0000-0001-7689-0933}\inst{\ref{aff172}}
\and S.~Tosi\orcid{0000-0002-7275-9193}\inst{\ref{aff37},\ref{aff129}}
\and A.~Troja\orcid{0000-0003-0239-4595}\inst{\ref{aff108},\ref{aff109}}
\and M.~Tucci\inst{\ref{aff63}}
\and C.~Valieri\inst{\ref{aff33}}
\and A.~Venhola\orcid{0000-0001-6071-4564}\inst{\ref{aff177}}
\and D.~Vergani\orcid{0000-0003-0898-2216}\inst{\ref{aff26}}
\and G.~Verza\orcid{0000-0002-1886-8348}\inst{\ref{aff178}}
\and P.~Vielzeuf\orcid{0000-0003-2035-9339}\inst{\ref{aff64}}
\and N.~A.~Walton\orcid{0000-0003-3983-8778}\inst{\ref{aff176}}
\and D.~Scott\orcid{0000-0002-6878-9840}\inst{\ref{aff179}}}
										   
%%%% please do not edit the affiliation list -- contact ECEB Bureau for changes
\institute{SRON Netherlands Institute for Space Research, Landleven 12, 9747 AD, Groningen, The Netherlands\label{aff1}
\and
Kapteyn Astronomical Institute, University of Groningen, PO Box 800, 9700 AV Groningen, The Netherlands\label{aff2}
\and
Instituto de Radioastronom\'ia y Astrof\'isica, Universidad Nacional Aut\'onoma de M\'exico, A.P. 72-3, 58089 Morelia, Mexico\label{aff3}
\and
Instituto de Astrof\'isica e Ci\^encias do Espa\c{c}o, Universidade do Porto, CAUP, Rua das Estrelas, PT4150-762 Porto, Portugal\label{aff4}
\and
DTx -- Digital Transformation CoLAB, Building 1, Azur\'em Campus, University of Minho, 4800-058 Guimar\~aes, Portugal\label{aff5}
\and
School of Physics, HH Wills Physics Laboratory, University of Bristol, Tyndall Avenue, Bristol, BS8 1TL, UK\label{aff6}
\and
Department of Mathematics and Physics, Roma Tre University, Via della Vasca Navale 84, 00146 Rome, Italy\label{aff7}
\and
INAF-Osservatorio Astronomico di Roma, Via Frascati 33, 00078 Monteporzio Catone, Italy\label{aff8}
\and
Department of Physical Sciences, Ritsumeikan University, Kusatsu, Shiga 525-8577, Japan\label{aff9}
\and
National Astronomical Observatory of Japan, 2-21-1 Osawa, Mitaka, Tokyo 181-8588, Japan\label{aff10}
\and
Academia Sinica Institute of Astronomy and Astrophysics (ASIAA), 11F of ASMAB, No.~1, Section 4, Roosevelt Road, Taipei 10617, Taiwan\label{aff11}
\and
Institute of Space Sciences (ICE, CSIC), Campus UAB, Carrer de Can Magrans, s/n, 08193 Barcelona, Spain\label{aff12}
\and
Institut d'Estudis Espacials de Catalunya (IEEC),  Edifici RDIT, Campus UPC, 08860 Castelldefels, Barcelona, Spain\label{aff13}
\and
Max Planck Institute for Extraterrestrial Physics, Giessenbachstr. 1, 85748 Garching, Germany\label{aff14}
\and
Instituto de Astrof\'{\i}sica de Canarias, V\'{\i}a L\'actea, 38205 La Laguna, Tenerife, Spain\label{aff15}
\and
Universidad de La Laguna, Departamento de Astrof\'{\i}sica, 38206 La Laguna, Tenerife, Spain\label{aff16}
\and
Instituto de Astrof\'isica de Canarias (IAC); Departamento de Astrof\'isica, Universidad de La Laguna (ULL), 38200, La Laguna, Tenerife, Spain\label{aff17}
\and
School of Physics \& Astronomy, University of Southampton, Highfield Campus, Southampton SO17 1BJ, UK\label{aff18}
\and
INAF-Istituto di Astrofisica e Planetologia Spaziali, via del Fosso del Cavaliere, 100, 00100 Roma, Italy\label{aff19}
\and
Leiden Observatory, Leiden University, Einsteinweg 55, 2333 CC Leiden, The Netherlands\label{aff20}
\and
Institute for Astronomy, University of Edinburgh, Royal Observatory, Blackford Hill, Edinburgh EH9 3HJ, UK\label{aff21}
\and
Universit\'e Paris-Saclay, CNRS, Institut d'astrophysique spatiale, 91405, Orsay, France\label{aff22}
\and
ESAC/ESA, Camino Bajo del Castillo, s/n., Urb. Villafranca del Castillo, 28692 Villanueva de la Ca\~nada, Madrid, Spain\label{aff23}
\and
School of Mathematics and Physics, University of Surrey, Guildford, Surrey, GU2 7XH, UK\label{aff24}
\and
INAF-Osservatorio Astronomico di Brera, Via Brera 28, 20122 Milano, Italy\label{aff25}
\and
INAF-Osservatorio di Astrofisica e Scienza dello Spazio di Bologna, Via Piero Gobetti 93/3, 40129 Bologna, Italy\label{aff26}
\and
Universit\'e Paris-Saclay, Universit\'e Paris Cit\'e, CEA, CNRS, AIM, 91191, Gif-sur-Yvette, France\label{aff27}
\and
IFPU, Institute for Fundamental Physics of the Universe, via Beirut 2, 34151 Trieste, Italy\label{aff28}
\and
INAF-Osservatorio Astronomico di Trieste, Via G. B. Tiepolo 11, 34143 Trieste, Italy\label{aff29}
\and
INFN, Sezione di Trieste, Via Valerio 2, 34127 Trieste TS, Italy\label{aff30}
\and
SISSA, International School for Advanced Studies, Via Bonomea 265, 34136 Trieste TS, Italy\label{aff31}
\and
Dipartimento di Fisica e Astronomia, Universit\`a di Bologna, Via Gobetti 93/2, 40129 Bologna, Italy\label{aff32}
\and
INFN-Sezione di Bologna, Viale Berti Pichat 6/2, 40127 Bologna, Italy\label{aff33}
\and
INAF-Osservatorio Astronomico di Padova, Via dell'Osservatorio 5, 35122 Padova, Italy\label{aff34}
\and
Space Science Data Center, Italian Space Agency, via del Politecnico snc, 00133 Roma, Italy\label{aff35}
\and
INAF-Osservatorio Astrofisico di Torino, Via Osservatorio 20, 10025 Pino Torinese (TO), Italy\label{aff36}
\and
Dipartimento di Fisica, Universit\`a di Genova, Via Dodecaneso 33, 16146, Genova, Italy\label{aff37}
\and
INFN-Sezione di Genova, Via Dodecaneso 33, 16146, Genova, Italy\label{aff38}
\and
Department of Physics "E. Pancini", University Federico II, Via Cinthia 6, 80126, Napoli, Italy\label{aff39}
\and
INAF-Osservatorio Astronomico di Capodimonte, Via Moiariello 16, 80131 Napoli, Italy\label{aff40}
\and
Faculdade de Ci\^encias da Universidade do Porto, Rua do Campo de Alegre, 4150-007 Porto, Portugal\label{aff41}
\and
Dipartimento di Fisica, Universit\`a degli Studi di Torino, Via P. Giuria 1, 10125 Torino, Italy\label{aff42}
\and
INFN-Sezione di Torino, Via P. Giuria 1, 10125 Torino, Italy\label{aff43}
\and
European Space Agency/ESTEC, Keplerlaan 1, 2201 AZ Noordwijk, The Netherlands\label{aff44}
\and
Institute Lorentz, Leiden University, Niels Bohrweg 2, 2333 CA Leiden, The Netherlands\label{aff45}
\and
INAF-IASF Milano, Via Alfonso Corti 12, 20133 Milano, Italy\label{aff46}
\and
Centro de Investigaciones Energ\'eticas, Medioambientales y Tecnol\'ogicas (CIEMAT), Avenida Complutense 40, 28040 Madrid, Spain\label{aff47}
\and
Port d'Informaci\'{o} Cient\'{i}fica, Campus UAB, C. Albareda s/n, 08193 Bellaterra (Barcelona), Spain\label{aff48}
\and
Institute for Theoretical Particle Physics and Cosmology (TTK), RWTH Aachen University, 52056 Aachen, Germany\label{aff49}
\and
INFN section of Naples, Via Cinthia 6, 80126, Napoli, Italy\label{aff50}
\and
Institute for Astronomy, University of Hawaii, 2680 Woodlawn Drive, Honolulu, HI 96822, USA\label{aff51}
\and
Dipartimento di Fisica e Astronomia "Augusto Righi" - Alma Mater Studiorum Universit\`a di Bologna, Viale Berti Pichat 6/2, 40127 Bologna, Italy\label{aff52}
\and
Jodrell Bank Centre for Astrophysics, Department of Physics and Astronomy, University of Manchester, Oxford Road, Manchester M13 9PL, UK\label{aff53}
\and
European Space Agency/ESRIN, Largo Galileo Galilei 1, 00044 Frascati, Roma, Italy\label{aff54}
\and
Universit\'e Claude Bernard Lyon 1, CNRS/IN2P3, IP2I Lyon, UMR 5822, Villeurbanne, F-69100, France\label{aff55}
\and
Aix-Marseille Universit\'e, CNRS, CNES, LAM, Marseille, France\label{aff56}
\and
Institut de Ci\`{e}ncies del Cosmos (ICCUB), Universitat de Barcelona (IEEC-UB), Mart\'{i} i Franqu\`{e}s 1, 08028 Barcelona, Spain\label{aff57}
\and
Instituci\'o Catalana de Recerca i Estudis Avan\c{c}ats (ICREA), Passeig de Llu\'{\i}s Companys 23, 08010 Barcelona, Spain\label{aff58}
\and
UCB Lyon 1, CNRS/IN2P3, IUF, IP2I Lyon, 4 rue Enrico Fermi, 69622 Villeurbanne, France\label{aff59}
\and
Mullard Space Science Laboratory, University College London, Holmbury St Mary, Dorking, Surrey RH5 6NT, UK\label{aff60}
\and
Departamento de F\'isica, Faculdade de Ci\^encias, Universidade de Lisboa, Edif\'icio C8, Campo Grande, PT1749-016 Lisboa, Portugal\label{aff61}
\and
Instituto de Astrof\'isica e Ci\^encias do Espa\c{c}o, Faculdade de Ci\^encias, Universidade de Lisboa, Campo Grande, 1749-016 Lisboa, Portugal\label{aff62}
\and
Department of Astronomy, University of Geneva, ch. d'Ecogia 16, 1290 Versoix, Switzerland\label{aff63}
\and
Aix-Marseille Universit\'e, CNRS/IN2P3, CPPM, Marseille, France\label{aff64}
\and
INFN-Bologna, Via Irnerio 46, 40126 Bologna, Italy\label{aff65}
\and
Universit\"ats-Sternwarte M\"unchen, Fakult\"at f\"ur Physik, Ludwig-Maximilians-Universit\"at M\"unchen, Scheinerstrasse 1, 81679 M\"unchen, Germany\label{aff66}
\and
FRACTAL S.L.N.E., calle Tulip\'an 2, Portal 13 1A, 28231, Las Rozas de Madrid, Spain\label{aff67}
\and
Dipartimento di Fisica "Aldo Pontremoli", Universit\`a degli Studi di Milano, Via Celoria 16, 20133 Milano, Italy\label{aff68}
\and
INFN-Sezione di Milano, Via Celoria 16, 20133 Milano, Italy\label{aff69}
\and
NRC Herzberg, 5071 West Saanich Rd, Victoria, BC V9E 2E7, Canada\label{aff70}
\and
Institute of Theoretical Astrophysics, University of Oslo, P.O. Box 1029 Blindern, 0315 Oslo, Norway\label{aff71}
\and
Jet Propulsion Laboratory, California Institute of Technology, 4800 Oak Grove Drive, Pasadena, CA, 91109, USA\label{aff72}
\and
Department of Physics, Lancaster University, Lancaster, LA1 4YB, UK\label{aff73}
\and
Felix Hormuth Engineering, Goethestr. 17, 69181 Leimen, Germany\label{aff74}
\and
Technical University of Denmark, Elektrovej 327, 2800 Kgs. Lyngby, Denmark\label{aff75}
\and
Cosmic Dawn Center (DAWN), Denmark\label{aff76}
\and
Institut d'Astrophysique de Paris, UMR 7095, CNRS, and Sorbonne Universit\'e, 98 bis boulevard Arago, 75014 Paris, France\label{aff77}
\and
Max-Planck-Institut f\"ur Astronomie, K\"onigstuhl 17, 69117 Heidelberg, Germany\label{aff78}
\and
NASA Goddard Space Flight Center, Greenbelt, MD 20771, USA\label{aff79}
\and
Department of Physics and Helsinki Institute of Physics, Gustaf H\"allstr\"omin katu 2, 00014 University of Helsinki, Finland\label{aff80}
\and
Universit\'e de Gen\`eve, D\'epartement de Physique Th\'eorique and Centre for Astroparticle Physics, 24 quai Ernest-Ansermet, CH-1211 Gen\`eve 4, Switzerland\label{aff81}
\and
Department of Physics, P.O. Box 64, 00014 University of Helsinki, Finland\label{aff82}
\and
Helsinki Institute of Physics, Gustaf H{\"a}llstr{\"o}min katu 2, University of Helsinki, Helsinki, Finland\label{aff83}
\and
Centre de Calcul de l'IN2P3/CNRS, 21 avenue Pierre de Coubertin 69627 Villeurbanne Cedex, France\label{aff84}
\and
Laboratoire d'etude de l'Univers et des phenomenes eXtremes, Observatoire de Paris, Universit\'e PSL, Sorbonne Universit\'e, CNRS, 92190 Meudon, France\label{aff85}
\and
SKA Observatory, Jodrell Bank, Lower Withington, Macclesfield, Cheshire SK11 9FT, UK\label{aff86}
\and
University of Applied Sciences and Arts of Northwestern Switzerland, School of Computer Science, 5210 Windisch, Switzerland\label{aff87}
\and
Universit\"at Bonn, Argelander-Institut f\"ur Astronomie, Auf dem H\"ugel 71, 53121 Bonn, Germany\label{aff88}
\and
INFN-Sezione di Roma, Piazzale Aldo Moro, 2 - c/o Dipartimento di Fisica, Edificio G. Marconi, 00185 Roma, Italy\label{aff89}
\and
Dipartimento di Fisica e Astronomia "Augusto Righi" - Alma Mater Studiorum Universit\`a di Bologna, via Piero Gobetti 93/2, 40129 Bologna, Italy\label{aff90}
\and
Department of Physics, Institute for Computational Cosmology, Durham University, South Road, Durham, DH1 3LE, UK\label{aff91}
\and
Universit\'e C\^{o}te d'Azur, Observatoire de la C\^{o}te d'Azur, CNRS, Laboratoire Lagrange, Bd de l'Observatoire, CS 34229, 06304 Nice cedex 4, France\label{aff92}
\and
Universit\'e Paris Cit\'e, CNRS, Astroparticule et Cosmologie, 75013 Paris, France\label{aff93}
\and
CNRS-UCB International Research Laboratory, Centre Pierre Binetruy, IRL2007, CPB-IN2P3, Berkeley, USA\label{aff94}
\and
University of Applied Sciences and Arts of Northwestern Switzerland, School of Engineering, 5210 Windisch, Switzerland\label{aff95}
\and
Institut d'Astrophysique de Paris, 98bis Boulevard Arago, 75014, Paris, France\label{aff96}
\and
Institute of Physics, Laboratory of Astrophysics, Ecole Polytechnique F\'ed\'erale de Lausanne (EPFL), Observatoire de Sauverny, 1290 Versoix, Switzerland\label{aff97}
\and
Aurora Technology for European Space Agency (ESA), Camino bajo del Castillo, s/n, Urbanizacion Villafranca del Castillo, Villanueva de la Ca\~nada, 28692 Madrid, Spain\label{aff98}
\and
Institut de F\'{i}sica d'Altes Energies (IFAE), The Barcelona Institute of Science and Technology, Campus UAB, 08193 Bellaterra (Barcelona), Spain\label{aff99}
\and
School of Mathematics, Statistics and Physics, Newcastle University, Herschel Building, Newcastle-upon-Tyne, NE1 7RU, UK\label{aff100}
\and
DARK, Niels Bohr Institute, University of Copenhagen, Jagtvej 155, 2200 Copenhagen, Denmark\label{aff101}
\and
Waterloo Centre for Astrophysics, University of Waterloo, Waterloo, Ontario N2L 3G1, Canada\label{aff102}
\and
Department of Physics and Astronomy, University of Waterloo, Waterloo, Ontario N2L 3G1, Canada\label{aff103}
\and
Perimeter Institute for Theoretical Physics, Waterloo, Ontario N2L 2Y5, Canada\label{aff104}
\and
Centre National d'Etudes Spatiales -- Centre spatial de Toulouse, 18 avenue Edouard Belin, 31401 Toulouse Cedex 9, France\label{aff105}
\and
Institute of Space Science, Str. Atomistilor, nr. 409 M\u{a}gurele, Ilfov, 077125, Romania\label{aff106}
\and
Consejo Superior de Investigaciones Cientificas, Calle Serrano 117, 28006 Madrid, Spain\label{aff107}
\and
Dipartimento di Fisica e Astronomia "G. Galilei", Universit\`a di Padova, Via Marzolo 8, 35131 Padova, Italy\label{aff108}
\and
INFN-Padova, Via Marzolo 8, 35131 Padova, Italy\label{aff109}
\and
Caltech/IPAC, 1200 E. California Blvd., Pasadena, CA 91125, USA\label{aff110}
\and
Institut f\"ur Theoretische Physik, University of Heidelberg, Philosophenweg 16, 69120 Heidelberg, Germany\label{aff111}
\and
Institut de Recherche en Astrophysique et Plan\'etologie (IRAP), Universit\'e de Toulouse, CNRS, UPS, CNES, 14 Av. Edouard Belin, 31400 Toulouse, France\label{aff112}
\and
Universit\'e St Joseph; Faculty of Sciences, Beirut, Lebanon\label{aff113}
\and
Departamento de F\'isica, FCFM, Universidad de Chile, Blanco Encalada 2008, Santiago, Chile\label{aff114}
\and
Universit\"at Innsbruck, Institut f\"ur Astro- und Teilchenphysik, Technikerstr. 25/8, 6020 Innsbruck, Austria\label{aff115}
\and
Satlantis, University Science Park, Sede Bld 48940, Leioa-Bilbao, Spain\label{aff116}
\and
Instituto de Astrof\'isica e Ci\^encias do Espa\c{c}o, Faculdade de Ci\^encias, Universidade de Lisboa, Tapada da Ajuda, 1349-018 Lisboa, Portugal\label{aff117}
\and
Cosmic Dawn Center (DAWN)\label{aff118}
\and
Niels Bohr Institute, University of Copenhagen, Jagtvej 128, 2200 Copenhagen, Denmark\label{aff119}
\and
Universidad Polit\'ecnica de Cartagena, Departamento de Electr\'onica y Tecnolog\'ia de Computadoras,  Plaza del Hospital 1, 30202 Cartagena, Spain\label{aff120}
\and
Infrared Processing and Analysis Center, California Institute of Technology, Pasadena, CA 91125, USA\label{aff121}
\and
Dipartimento di Fisica e Scienze della Terra, Universit\`a degli Studi di Ferrara, Via Giuseppe Saragat 1, 44122 Ferrara, Italy\label{aff122}
\and
Istituto Nazionale di Fisica Nucleare, Sezione di Ferrara, Via Giuseppe Saragat 1, 44122 Ferrara, Italy\label{aff123}
\and
INAF, Istituto di Radioastronomia, Via Piero Gobetti 101, 40129 Bologna, Italy\label{aff124}
\and
Department of Physics, Oxford University, Keble Road, Oxford OX1 3RH, UK\label{aff125}
\and
Universit\'e PSL, Observatoire de Paris, Sorbonne Universit\'e, CNRS, LERMA, 75014, Paris, France\label{aff126}
\and
Universit\'e Paris-Cit\'e, 5 Rue Thomas Mann, 75013, Paris, France\label{aff127}
\and
INAF - Osservatorio Astronomico di Brera, via Emilio Bianchi 46, 23807 Merate, Italy\label{aff128}
\and
INAF-Osservatorio Astronomico di Brera, Via Brera 28, 20122 Milano, Italy, and INFN-Sezione di Genova, Via Dodecaneso 33, 16146, Genova, Italy\label{aff129}
\and
ICL, Junia, Universit\'e Catholique de Lille, LITL, 59000 Lille, France\label{aff130}
\and
ICSC - Centro Nazionale di Ricerca in High Performance Computing, Big Data e Quantum Computing, Via Magnanelli 2, Bologna, Italy\label{aff131}
\and
Instituto de F\'isica Te\'orica UAM-CSIC, Campus de Cantoblanco, 28049 Madrid, Spain\label{aff132}
\and
CERCA/ISO, Department of Physics, Case Western Reserve University, 10900 Euclid Avenue, Cleveland, OH 44106, USA\label{aff133}
\and
Technical University of Munich, TUM School of Natural Sciences, Physics Department, James-Franck-Str.~1, 85748 Garching, Germany\label{aff134}
\and
Max-Planck-Institut f\"ur Astrophysik, Karl-Schwarzschild-Str.~1, 85748 Garching, Germany\label{aff135}
\and
Laboratoire Univers et Th\'eorie, Observatoire de Paris, Universit\'e PSL, Universit\'e Paris Cit\'e, CNRS, 92190 Meudon, France\label{aff136}
\and
Departamento de F{\'\i}sica Fundamental. Universidad de Salamanca. Plaza de la Merced s/n. 37008 Salamanca, Spain\label{aff137}
\and
Universit\'e de Strasbourg, CNRS, Observatoire astronomique de Strasbourg, UMR 7550, 67000 Strasbourg, France\label{aff138}
\and
Center for Data-Driven Discovery, Kavli IPMU (WPI), UTIAS, The University of Tokyo, Kashiwa, Chiba 277-8583, Japan\label{aff139}
\and
Ludwig-Maximilians-University, Schellingstrasse 4, 80799 Munich, Germany\label{aff140}
\and
Max-Planck-Institut f\"ur Physik, Boltzmannstr. 8, 85748 Garching, Germany\label{aff141}
\and
Dipartimento di Fisica - Sezione di Astronomia, Universit\`a di Trieste, Via Tiepolo 11, 34131 Trieste, Italy\label{aff142}
\and
California Institute of Technology, 1200 E California Blvd, Pasadena, CA 91125, USA\label{aff143}
\and
Department of Physics \& Astronomy, University of California Irvine, Irvine CA 92697, USA\label{aff144}
\and
Department of Mathematics and Physics E. De Giorgi, University of Salento, Via per Arnesano, CP-I93, 73100, Lecce, Italy\label{aff145}
\and
INFN, Sezione di Lecce, Via per Arnesano, CP-193, 73100, Lecce, Italy\label{aff146}
\and
INAF-Sezione di Lecce, c/o Dipartimento Matematica e Fisica, Via per Arnesano, 73100, Lecce, Italy\label{aff147}
\and
Departamento F\'isica Aplicada, Universidad Polit\'ecnica de Cartagena, Campus Muralla del Mar, 30202 Cartagena, Murcia, Spain\label{aff148}
\and
Instituto de F\'isica de Cantabria, Edificio Juan Jord\'a, Avenida de los Castros, 39005 Santander, Spain\label{aff149}
\and
CEA Saclay, DFR/IRFU, Service d'Astrophysique, Bat. 709, 91191 Gif-sur-Yvette, France\label{aff150}
\and
Institute of Cosmology and Gravitation, University of Portsmouth, Portsmouth PO1 3FX, UK\label{aff151}
\and
Department of Computer Science, Aalto University, PO Box 15400, Espoo, FI-00 076, Finland\label{aff152}
\and
Instituto de Astrof\'\i sica de Canarias, c/ Via Lactea s/n, La Laguna 38200, Spain. Departamento de Astrof\'\i sica de la Universidad de La Laguna, Avda. Francisco Sanchez, La Laguna, 38200, Spain\label{aff153}
\and
Ruhr University Bochum, Faculty of Physics and Astronomy, Astronomical Institute (AIRUB), German Centre for Cosmological Lensing (GCCL), 44780 Bochum, Germany\label{aff154}
\and
Department of Physics and Astronomy, Vesilinnantie 5, 20014 University of Turku, Finland\label{aff155}
\and
Serco for European Space Agency (ESA), Camino bajo del Castillo, s/n, Urbanizacion Villafranca del Castillo, Villanueva de la Ca\~nada, 28692 Madrid, Spain\label{aff156}
\and
ARC Centre of Excellence for Dark Matter Particle Physics, Melbourne, Australia\label{aff157}
\and
Centre for Astrophysics \& Supercomputing, Swinburne University of Technology,  Hawthorn, Victoria 3122, Australia\label{aff158}
\and
Department of Physics and Astronomy, University of the Western Cape, Bellville, Cape Town, 7535, South Africa\label{aff159}
\and
DAMTP, Centre for Mathematical Sciences, Wilberforce Road, Cambridge CB3 0WA, UK\label{aff160}
\and
Kavli Institute for Cosmology Cambridge, Madingley Road, Cambridge, CB3 0HA, UK\label{aff161}
\and
Department of Astrophysics, University of Zurich, Winterthurerstrasse 190, 8057 Zurich, Switzerland\label{aff162}
\and
Department of Physics, Centre for Extragalactic Astronomy, Durham University, South Road, Durham, DH1 3LE, UK\label{aff163}
\and
IRFU, CEA, Universit\'e Paris-Saclay 91191 Gif-sur-Yvette Cedex, France\label{aff164}
\and
Oskar Klein Centre for Cosmoparticle Physics, Department of Physics, Stockholm University, Stockholm, SE-106 91, Sweden\label{aff165}
\and
Astrophysics Group, Blackett Laboratory, Imperial College London, London SW7 2AZ, UK\label{aff166}
\and
Univ. Grenoble Alpes, CNRS, Grenoble INP, LPSC-IN2P3, 53, Avenue des Martyrs, 38000, Grenoble, France\label{aff167}
\and
INAF-Osservatorio Astrofisico di Arcetri, Largo E. Fermi 5, 50125, Firenze, Italy\label{aff168}
\and
Dipartimento di Fisica, Sapienza Universit\`a di Roma, Piazzale Aldo Moro 2, 00185 Roma, Italy\label{aff169}
\and
Centro de Astrof\'{\i}sica da Universidade do Porto, Rua das Estrelas, 4150-762 Porto, Portugal\label{aff170}
\and
HE Space for European Space Agency (ESA), Camino bajo del Castillo, s/n, Urbanizacion Villafranca del Castillo, Villanueva de la Ca\~nada, 28692 Madrid, Spain\label{aff171}
\and
Department of Astrophysical Sciences, Peyton Hall, Princeton University, Princeton, NJ 08544, USA\label{aff172}
\and
Theoretical astrophysics, Department of Physics and Astronomy, Uppsala University, Box 515, 751 20 Uppsala, Sweden\label{aff173}
\and
Minnesota Institute for Astrophysics, University of Minnesota, 116 Church St SE, Minneapolis, MN 55455, USA\label{aff174}
\and
Mathematical Institute, University of Leiden, Einsteinweg 55, 2333 CA Leiden, The Netherlands\label{aff175}
\and
Institute of Astronomy, University of Cambridge, Madingley Road, Cambridge CB3 0HA, UK\label{aff176}
\and
Space physics and astronomy research unit, University of Oulu, Pentti Kaiteran katu 1, FI-90014 Oulu, Finland\label{aff177}
\and
Center for Computational Astrophysics, Flatiron Institute, 162 5th Avenue, 10010, New York, NY, USA\label{aff178}
\and
Department of Physics and Astronomy, University of British Columbia, Vancouver, BC V6T 1Z1, Canada\label{aff179}}    